\documentclass[aps,pra,twocolumn,superscriptaddress]{revtex4-1}
\usepackage{graphicx} %Loading the package
\graphicspath {{figs/}}
\usepackage{lipsum}
\usepackage{xcolor}
\usepackage{amsmath}
\usepackage{amsfonts}
\usepackage{textcomp}
\usepackage{gensymb}
\usepackage{lineno}
\usepackage{float}
\usepackage{soul} % for strikethrough 
\usepackage{hyperref}
\hypersetup{final}% for hyperlinks within text & urls in references

\emergencystretch=\maxdimen
\begin{document}
\raggedbottom

\newcommand{\todo}[1]{{\color[rgb]{1.0,0.0,0.0}#1}} % gaps to be filled in red
\newcommand{\yg}[1]{{\color[rgb]{0.0,0.0,1.0}#1}} % Yvonne Comments in blue
\newcommand{\kn}[1]{{\color[rgb]{0.2,0.7,0.2}#1}} % KJ comments in blue
\newcommand{\aj}[1]{{\color[rgb]{0.0,0.7,0.5}#1}} % AJ comments in blue
\newcommand{\cut}[1]{\st{#1}} % Use this line to show deleted text as struck out

\title{Quantum Information Processing with Bosonic Qubits in Circuit QED}
\author{Atharv Joshi}
\affiliation{Centre for Quantum Technologies, National University of Singapore}
\author{Kyungjoo Noh}\thanks{This review paper was written before K. Noh joined AWS Center for Quantum Computing.}
\affiliation{AWS Center for Quantum Computing, Pasadena, CA, 91125, USA}
\author{Yvonne Y. Gao}
\email[Corresponding author: ]{yvonne.gao@nus.edu.sg}
\affiliation{Centre for Quantum Technologies, National University of Singapore}

%\author{Atharv Joshi$^1$, Kyungjoo Noh$^{2,\dagger}$, and Yvonne Y. Gao$^{1,*}$}
%\address{$^1$ Centre for Quantum Technologies, National Univeristy of Singapore\\
%        $^2$ AWS Center for Quantum Computing, Pasadena, CA, 91125, USA\\
%        $^{\dagger}$ This review paper was written before K.N. joined AWS.\\
%        $^*$ \textbf{Email}: yvonne.gao@nus.edu.sg}

\begin{abstract}

The unique features of quantum theory offer a powerful new paradigm for information processing. Translating these mathematical abstractions into useful algorithms and applications requires quantum systems with significant complexity and sufficiently low error rates. Such quantum systems must be made from robust hardware that can coherently store, process, and extract the encoded information, as well as possess effective quantum error correction (QEC) protocols to detect and correct errors. Circuit quantum electrodynamics (cQED) provides a promising hardware platform for implementing robust quantum devices. In particular, bosonic encodings in cQED that use multi-photon states of superconducting cavities to encode information have shown success in realizing hardware-efficient QEC. Here, we review recent developments in the theory and implementation of quantum error correction with bosonic codes and report the progress made towards realizing fault-tolerant quantum information processing with cQED devices. 
\end{abstract}

%\pacs{}
\maketitle
%\ioptwocol
\tableofcontents

\vspace{1in} % to begin intro on new column

\section{Introduction}

A quantum computer harnesses unique features of quantum theory, such as superposition and entanglement, to tackle classically challenging tasks. To perform faithful computation, quantum information must be protected against errors due to decoherence mechanisms and operational imperfections. While these errors are relatively insignificant individually, they can quickly accumulate to completely scramble the information. 

To protect quantum information from scrambling, the theoretical frameworks of quantum error correction (QEC) ~\cite{shor1995_scheme, steane1996_error} and fault-tolerant quantum computation~\cite{preskill1998_reliable} were developed in the early days of quantum computing. Essentially, these frameworks devise encodings which map a collection of physical elements onto a single `logical' bit of quantum information. Such a logical qubit is endowed with cleverly chosen symmetry properties that allow us to extract error syndromes and enact error correction without disturbing the encoded information.

An important metric for evaluating the effectiveness of QEC implementations is the break-even point, which is achieved when the lifetime of a logical qubit exceeds that of the best single physical element in the system. Achieving the break-even point entails that additional physical elements and operations introduced to a QEC process do not cause more degradation than the protection they afford. Hence, reaching the break-even point is a critical pre-requisite for implementing fault-tolerant gates and eventually performing robust quantum information processing on a large scale.

In the conventional approach to QEC, the physical elements are realized by discrete two-level systems. In this approach, even a simple QEC scheme designed to correct single errors, such as the Steane code~\cite{steane1996_error}, requires tens of two-level systems, ancillary qubits, and measurement elements. Constructing physical devices that contain these many interconnected elements can be a significant engineering challenge. More crucially, having many interconnected elements often degrades the device performance and introduces new uncorrectable errors such as cross-talk due to undesired couplings between the elements. Over the last decade, many proof-of-principle demonstrations of QEC schemes have been realized with encoding schemes based on two-level systems~\cite{chiaverini2004_realization, reed2012_realization, taminiau2014_universal, waldherr2014_quantum, linke2017_faulttolerant, andersen2020_repeated, erhard2021_entangling}. However, given the practical challenges described above, these demonstrations have not deterministically extended the performance of the logical qubits beyond that of the best available physical qubit in the system.

A promising alternative with the potential to realize robust universal quantum computing with effective QEC beyond the break-even point involves encoding logical qubits in continuous variables~\cite{braunstein1998_quantum, bartlett2002_universal, bartlett2002_quantum}. In particular, superconducting microwave cavities coupled to one or more anharmonic elements in the circuit quantum electrodynamics (cQED) architecture provide a valuable resource for the hardware-efficient encoding of logical qubits~\cite{devoret2013_superconducting, blais2020_circuit}. These cavities have a large Hilbert space for encoding information in multi-photon states compactly and thus form a logical qubit in a single piece of hardware. The anharmonic element, typically in the form of a transmon and henceforth referred to as the `ancilla', provides the necessary non-linearity to control and measure cavity states. This strategy of using multi-photon states of superconducting cavities to encode logical information is also known as bosonic codes. Implementations of bosonic codes in cQED have thus far not only demonstrated QEC at the break-even point~\cite{ofek2016_extending}, but also robust operations~\cite{heeres2017_implementing, chou2018_deterministic, gao2019_entanglement, xu2020_demonstration, reinhold2020_error-corrected} and fault-tolerant measurement of error syndromes~\cite{rosenblum2018_fault-tolerant}, thus making rapid progress in recent years. 

In this article, we review the recent developments in  bosonic codes in the cQED setting. In particular, we highlight the progress made in demonstrating effective QEC and information processing with logical elements implemented using bosonic codes. These recent works provide compelling evidence for the vast potential of bosonic codes in cQED as the fundamental building blocks for robust universal quantum computing. 

\subsection{Organization of the article}
In Sec.~\ref{sec:concepts}, we begin by outlining the basic principles of QEC as well as bosonic codes. Here, we highlight their nature by comparing a bosonic code with a multi-qubit code in the presence of similar errors. In Sec.~\ref{sec:codes}, we present various bosonic encoding schemes proposed in literature and compare their respective strengths and limitations. In particular, we wish to emphasize  crucial considerations for constructing these codes and evaluating their performance in the presence of naturally-occurring errors.

In Sec.~\ref{sec:cavities}, we introduce the key hardware building blocks required for cQED implementations of bosonic encoding schemes. In this section, we also consolidate the progress made in improving the intrinsic  quality factors of superconducting microwave cavities over the last decade. Subsequently, in Sec.~\ref{section: realization}, we explore the latest developments in implementing robust universal control on bosonic qubits encoded in superconducting cavities, both in terms of single-mode gates as well as novel two-mode operations. We then describe the different strategies for detecting and correcting quantum errors on bosonic logical qubits encoded in superconducting cavities. 

In Sec.~\ref{sec:FT}, we discuss the concept of fault-tolerance and how that might be realized with protected bosonic qubits. Here, we also feature some novel schemes that concatenate bosonic codes with other QEC codes to protect against quantum errors more comprehensively. Finally, in Sec.~\ref{sec:future}, we provide some perspectives for achieving QEC on a larger scale. We conclude by remarking on the appeal of the modular architecture, which offers a promising path for practical and robust quantum information processing with individually protected bosonic logical elements. 

\section{Concepts of Bosonic Quantum Error Correction}\label{sec:concepts}

The general principle of QEC is to encode logical quantum information redundantly in a large Hilbert space with certain symmetry properties, which can be used to detect errors. In particular, logical code states are designed such that they can be mapped onto orthogonal subspaces under distinct errors. Crucially, the logical information can be recovered faithfully only if the mapping between the logical and the error states does not distort the code words. 

Mathematically, these requirements can be succinctly described by the Knill-Laflamme condition~\cite{knill1997_theory}, which states that an error-correcting code $\mathcal{C}$ can correct any error operators in the span of an error set \mbox{$\mathcal{E} \equiv \lbrace \hat{E}_{1},\cdots,\hat{E}_{|\mathcal{E}|} \rbrace$} if and only if it satisfies:
\begin{linenomath}
\begin{gather}
    \hat{P}_{\mathcal{C}} \hat{E}_{\ell}^{\dagger}\hat{E}_{\ell'} \hat{P}_{\mathcal{C}} = \alpha_{\ell\ell'}\hat{P}_{\mathcal{C}} %\textrm{ for all }\ell,\ell'\in \lbrace 1,\cdots,|\mathcal{E}| \rbrace, 
\end{gather}
\end{linenomath}
for all $\ell,\ell'\in \lbrace 1,\cdots,|\mathcal{E}| \rbrace$, where $|\mathcal{E}|$ is the size of the error set, $\hat{P}_{\mathcal{C}}$ is the projection operator to the code space $\mathcal{C}$, and $\alpha_{\ell\ell'}$ are matrix elements of a Hermitian and positive semi-definite matrix. A derivation of the Knill-Laflamme condition is given in Ref.~\cite{nielsen2000_quantum}.

Bosonic codes achieve these requirements by cleverly configuring the excitations in a harmonic oscillator mode. For instance, in the cQED architecture, information is encoded in multi-photon states of superconducting microwave cavities. We will illustrate how bosonic QEC codes work by describing the simplest binomial code~\cite{michael2016_new}, also known as the `kitten' code. We will then compare this code with its multi-qubit cousin, the `4-qubit code' (or the $[[4,1,2]]$ code or the distance-$2$ surface code)~\cite{leung1997_approximate}, in the presence of similar errors in the cQED architecture. This comparison, adapted from Ref.~\cite{girvin2019_quantum}, emphasizes the hardware-efficient nature of the bosonic QEC approach.

\begin{figure*}[t!]
    \centering
    \includegraphics[scale=1]{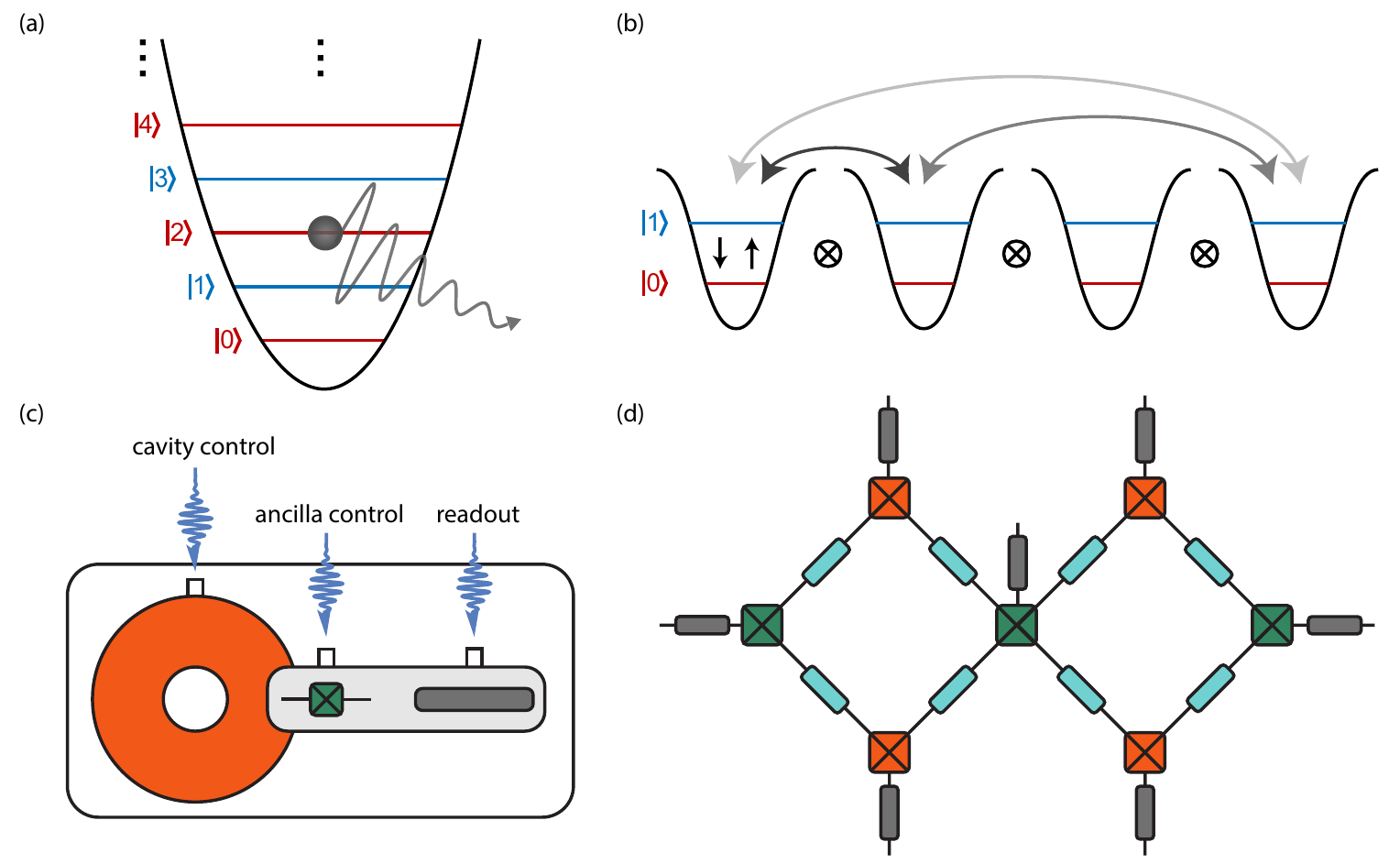}
    \caption[width=\textwidth]{\textbf{Hardware comparison between the kitten code and the 4-qubit code.} (a) Encoding a single logical qubit using multiple energy levels of a single harmonic oscillator, where the only dominant error channel is photon loss. (b) The 4-qubit scheme based on a collection of two-level systems. Individually, each two-level system can experience errors due to spontaneous absorption and emission. Moreover, additional errors can be introduced due to undesired couplings between these systems. (c) The typical hardware for implementing a bosonic logical qubit. The logical information is encoded in a single superconducting cavity (orange) in the kitten code, which can be fully controlled by a single non-linear ancilla (green) and read out via another cavity (gray). (d) Example of a system designed to implement the 4-qubit code, where the logical information is spread out across four data qubits (orange) checked by three ancillary ones (green). The interactions between each two-level system are mediated by coupling cavities (teal), and their respective states are read out via seven planar cavities (gray).
    } \label{fig:hardware}
\end{figure*}

The kitten code is designed to correct only single photon loss events, or $\hat{a}$, which are the dominant error channel in superconducting cavities.  This scheme encodes logical information in the even photon number parity subspace of a harmonic oscillator (Fig.~\ref{fig:hardware}a):
\begin{linenomath}
\begin{align}
    |0_{L}\rangle = \frac{1}{\sqrt{2}}( |0\rangle + |4\rangle ) , \quad |1_{L}\rangle = |2\rangle. \label{eq:bin N1S1}
\end{align}
\end{linenomath}
With these code words, a single photon loss event maps an even parity logical state $|\psi_{L}\rangle = \alpha|0_{L}\rangle + \beta|1_{L}\rangle$ to an odd parity error state $\hat{a}|\psi_{L}\rangle = \sqrt{2}(\alpha|3\rangle+\beta|1\rangle)$. Because of the parity difference, the code space and the error space are mutually orthogonal. Therefore, single photon loss events can be detected by measuring the photon number parity operator $\hat{\Pi}_{2} = (-1)^{\hat{n}}$. Importantly, since the error states $\hat{a}|0_{L}\rangle$ and $\hat{a}|1_{L}\rangle$ have the same normalization constant, 
\begin{linenomath}
\begin{align}
    \langle 0_{L}| \hat{a}^{\dagger}\hat{a}|0_{L}\rangle = \langle 1_{L}| \hat{a}^{\dagger}\hat{a}|1_{L}\rangle = 2,
\end{align}
\end{linenomath}
or equivalently, the logical states $|0_{L}\rangle$ and $|1_{L}\rangle$ have the same average photon number, a single photon loss event does not distort the encoded information. In other words, by mapping the normalized error states $|3\rangle$ and $|1\rangle$ back to the original code states $|0_{L}\rangle$ and $|1_{L}\rangle$, we can recover the input logical state up to an overall normalization constant: 
\begin{linenomath}
\begin{align}
    \hat{a}|\psi_{L}\rangle &= \sqrt{2}(\alpha |3\rangle + \beta |1\rangle) \nonumber \\ &\xrightarrow{\textrm{recovery}} \sqrt{2}(\alpha |0_{L}\rangle + \beta |1_{L}\rangle ) \propto |\psi_{L}\rangle.
\end{align}
\end{linenomath}

Note that we cannot faithfully recover logical information if the logical states have different average photon numbers. For instance,  with $|0_{L}\rangle = \frac{1}{2}|0\rangle + \frac{\sqrt{3}}{2}|4\rangle$, which has $3$ photons on average, as opposed to $2$ photons for $|1_{L}\rangle = |2\rangle$, a single photon loss and recovery event yields a state which is not proportional to $|\psi_{L}\rangle$: 
\begin{linenomath}
\begin{align}
    \hat{a}|\psi_{L}\rangle &= \sqrt{3}\alpha|3\rangle + \sqrt{2}\beta|1\rangle \nonumber \\ &\xrightarrow{\textrm{recovery}}    \sqrt{3}\alpha|0_{L}\rangle + \sqrt{2}\beta|1_{L}\rangle. \label{eq:adjusted weight}
\end{align}
\end{linenomath}
Specifically, the error and recovery process distorts the relative phase between $|0_{L}\rangle$ and $|1_{L}\rangle$. An intuitive way to understand such a phase distortion is by considering the role of the environment. When $|0_{L}\rangle$ and $|1_{L}\rangle$ have different average photon numbers, the environment gains partial information on whether $|\psi_{L}\rangle$ was in $|0_{L}\rangle$ or in $|1_{L}\rangle$. In the example above, $|0_{L}\rangle$ has more photons than $|1_{L}\rangle$ and hence has a higher probability of losing a photon. When one photon is lost, the environment knows that the encoded state was more likely to be $|0_{L}\rangle$ than $|1_{L}\rangle$ and the weights in Eq.\ \eqref{eq:adjusted weight} are therefore adjusted accordingly. Alternatively, we can say that the environment performs a weak measurement on the logical state in the $|0_{L}/1_{L}\rangle$ basis, thus leading to the dephasing of $|\psi_{L}\rangle$ in the $|0_{L}/1_{L}\rangle$ basis.

Hence, while the even parity structure allows the detection of single photon loss events, it does not guarantee the recoverability of the logical information without any distortion. Faithful recovery is ensured by selecting $|0_{L}\rangle$ and $|1_{L}\rangle$ to have the same average photon number. In the case of the kitten code, average photon numbers are matched by choosing equal coefficients for the vacuum and the four photon state components in the logical zero state. 

More generally, a code $\mathcal{C}$ can correct single photon loss events if it satisfies the Knill-Laflamme condition for the error set $\lbrace \hat{I},\hat{a} \rbrace$. That is, the projection operator to the code space should satisfy $\hat{P}_{\mathcal{C}} \hat{x} \hat{P}_{\mathcal{C}} \propto \hat{P}_{\mathcal{C}}$ for all \mbox{$\hat{x}\in \lbrace \hat{I},\hat{a},\hat{a}^{\dagger},\hat{a}^{\dagger}\hat{a} \rbrace$}. The first condition with $\hat{x}=\hat{I}$ is trivially satisfied for any code. For even parity codes, which are composed of logical states of even photon number parity, the second and the third conditions with $\hat{x} = \hat{a} $ and $\hat{x} = \hat{a}^{\dagger}$ are satisfied due to the parity structure. This implies that even parity codes are capable of detecting, but not necessarily correcting, any single photon loss and gain events. Single photon loss events can be corrected if the fourth condition is met, that is, $\hat{P}_{\mathcal{C}} \hat{a}^{\dagger}\hat{a} \hat{P}_{\mathcal{C}} \propto \hat{P}_{\mathcal{C}}$, or equivalently, if all logical states have the same average photon number.    

Now, using an example inspired by Ref.~\cite{girvin2019_quantum}, we compare the kitten code with its multi-qubit cousin, the 4-qubit code, whose logical states consist of 4 distinct two-level systems (Fig.~\ref{fig:hardware}b): 
\begin{linenomath}
\begin{align}
    |0_{L}\rangle = \frac{1}{\sqrt{2}} ( |0000\rangle + |1111\rangle ), \nonumber \\ 
    |1_{L}\rangle = \frac{1}{\sqrt{2}} ( |0101\rangle + |1010\rangle ).
\end{align}
\end{linenomath}
The $4$-qubit code is stabilized by three stabilizers: \mbox{$\hat{S}_{1} = \hat{Z}_{1}\hat{Z}_{3}$}, \mbox{$\hat{S}_{2} = \hat{Z}_{2}\hat{Z}_{4}$}, and \mbox{$\hat{S}_{3} = \hat{X}_{1}\hat{X}_{2}\hat{X}_{3}\hat{X}_{4}$}. This scheme is capable of detecting any arbitrary single-qubit errors. Moreover, it can correct single excitation loss errors,  $\lbrace\hat{\sigma}^{-}_{1},\hat{\sigma}^{-}_{2},\hat{\sigma}^{-}_{3},\hat{\sigma}^{-}_{4}\rbrace$, via approximate QEC~\cite{leung1997_approximate}, which happens when the Knill-Laflamme condition is approximately satisfied only to a certain low order in the error parameters. This capability is comparable to the protection afforded by the kitten code against single photon loss errors.

Despite their comparable error-correcting capability, the $4$-qubit code and the kitten code incur significantly different hardware overheads. A cQED implementation of the kitten code (Fig.~\ref{fig:hardware}c) requires a single bosonic mode to store logical information, a single ancilla (typically a transmon) to measure and control the cavity state, and a single readout cavity mode to measure the ancilla state. In contrast, a cQED realization of the $4$-qubit code (Fig.~\ref{fig:hardware}d) uses $4$ data qubits to encode logical information, $3$ ancillae to measure the three stabilizers, and additional cavity modes to connect and measure all $7$ physical qubits. Apart from the pure complexity of realizing such a device, the presence of additional elements introduces other error channels such as cross-talk arising from spurious couplings between the physical qubits. While these effects can be calibrated and mitigated on a small scale with clever techniques~\cite{sheldon2016_procedure, rol2019_fast}, they can quickly become intractable for more complex devices. This comparison thus illustrates the advantage of using the multiple levels of a single bosonic mode over using multiple two-level systems as a redundant resource for QEC.

While we focus on bosonic codes that encode a qubit in a single bosonic mode, there are proposals for encoding a qubit in many bosonic modes, namely permutation-invariant codes \cite{ouyang2014_permutationinvariant, ouyang2016_permutationinvariant, ouyang2017_permutationinvariant, ouyang2020_permutationinvariant}. These codes are tailored for excitation loss errors and generalize the simple 4-qubit code. Other codes of a similar nature have also been studied in Refs.~\cite{wasilewski2007_protecting, kapit2016_hardwareefficient, bergmann2016_quantum, kapit2018_errortransparent}.

\section{Performance of Bosonic Codes for Loss and Dephasing Errors}\label{sec:codes}
In general, bosonic modes in cQED systems typically undergo both photon loss and dephasing errors. Photon loss is considered to be the dominant error-channel. The rate of photon loss is determined by the internal quality factor ($Q_{\mathrm{int}}$) of the superconducting cavity. Intrinsic dephasing is usually insignificant for such cavities~\cite{reagor2013_reaching}. 

However, as the cavity is dispersively coupled to a non-linear ancilla, if the ancilla experiences undesired absorption ($\Gamma_{\uparrow}$) or emission ($\Gamma_{\downarrow}$) of excitations due to stray radiation, then the encoded logical information undergoes induced dephasing. The rate of this dephasing depends on $\Gamma_{\uparrow}$, $\Gamma_{\downarrow}$, and the coupling strength $\chi$. In general, a transition of the ancilla state leads to a rotation on the logical qubit by an angle $\sim \chi t$, where $t$ is the time the ancilla spends in the resulting state after the transition. In the limit where this rotation is small, we can apply the Markovian approximation and use the following Lindblad master equation to describe the evolution of the encoded qubit:
% \begin{linenomath}
\begin{align}
    \frac{d\hat{\rho}(t)}{dt} = \big{(} \kappa \mathcal{D}[\hat{a}] + \kappa_{\phi}\mathcal{D}[\hat{a}^{\dagger}\hat{a}] \big{)} \hat{\rho}(t),
\end{align}
% \end{linenomath}
where $\kappa$ and $\kappa_{\phi}$ are the photon loss and dephasing rates respectively, and \mbox{$\mathcal{D}[\hat{A}](\hat{\rho})\equiv \hat{A}\hat{\rho}\hat{A}^{\dagger}-\frac{1}{2}\lbrace \hat{A}^{\dagger}\hat{A},\hat{\rho} \rbrace$} is the dissipation superoperator. Note that loss and dephasing errors are generated by the jump operators $\hat{a}$ and $\hat{n} = \hat{a}^{\dagger}\hat{a}$ respectively. We say that a system is loss dominated if $\kappa\gg \kappa_{\phi}$ and dephasing dominated if $\kappa \ll \kappa_{\phi}$. 

\begin{figure*}[t!]
    \centering
  %  \captionsetup{width=\textwidth}
    \includegraphics[scale=1]{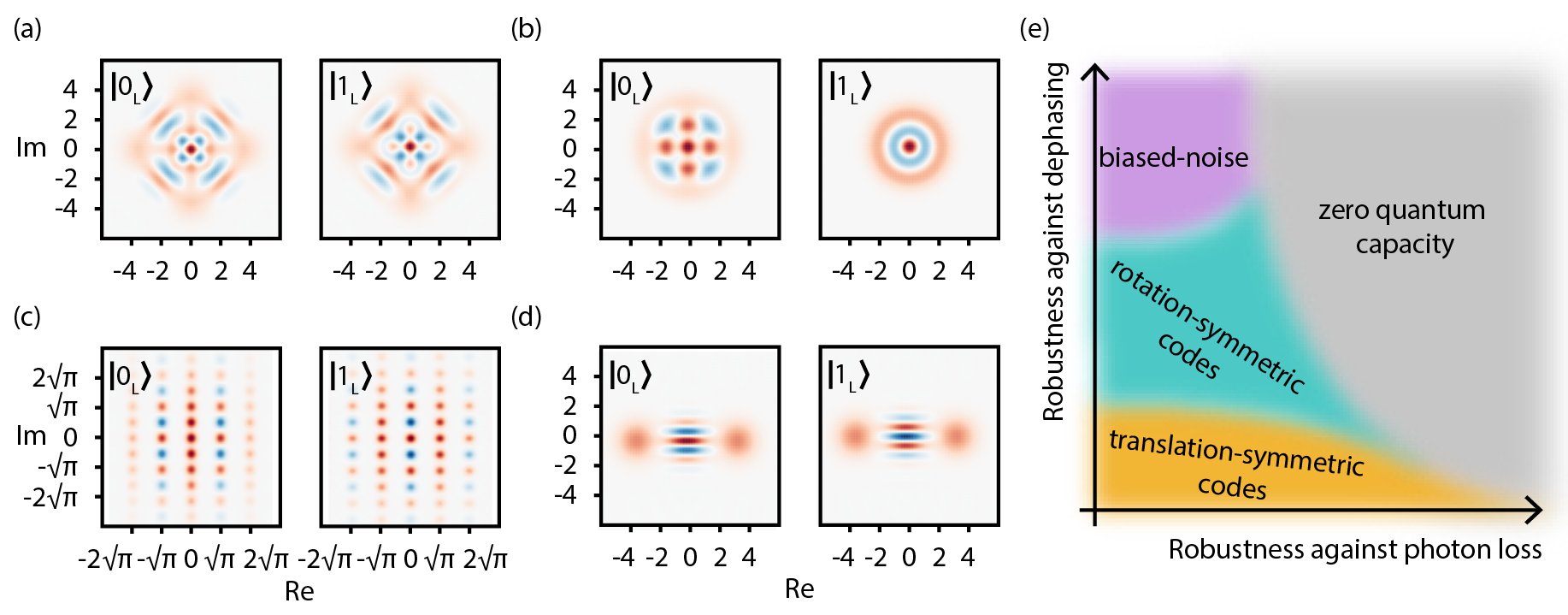}
    \caption[width=\textwidth]{\textbf{Visualization and performance of different bosonic codes.} (a) - (d) Wigner functions of the code words for the four-component cat, binomial kitten, GKP square lattice, and two-component cat code, respectively. (e) A qualitative description of the robustness of various classes of bosonic codes against dephasing and photon loss errors. Note that for the four-component cat code and the binomial kitten code, the logical states in the $|0/1_{L}\rangle$ basis are invariant under a $90\degree$ rotation, up to a global phase. However, a general code state, that is, an arbitrary superposition of $|0_{L}\rangle$ and $|1_{L}\rangle$), is invariant only under the $180\degree$ rotation.} \label{fig:codes}
\end{figure*} 

Importantly, in typical experimental regimes, dephasing errors induced by transitions of the ancilla state are generally non-Markovian. This limits the performance of some current implementations~\cite{lescanne2020_exponential} of bosonic QEC codes.

% \yg{It is important to note that in typical experimental regimes, the dephasing errors induced by transitions of the ancilla state are generally non-Markovian. This poses an outstanding challenge that limits the performance of some current implementations~\cite{lescanne2020_exponential} of bosonic QEC codes.}

In this section, we review the various single-mode bosonic codes, namely, cat (Fig.~\ref{fig:codes}a) and binomial (Fig.~\ref{fig:codes}b) codes (rotation-symmetric),  Gottesman-Kitaev-Preskill (GKP)  codes (translation-symmetric) (Fig.~\ref{fig:codes}c), and two-component cat codes (biased-noise bosonic qubits) (Fig.~\ref{fig:codes}d), and discuss their error-correcting capability against both photon loss and dephasing errors.    

\subsection{Rotation-symmetric codes: Binomial and cat codes}\label{subsection:rotation symmetric codes}
The class of rotation-symmetric bosonic codes~\cite{grimsmo2020_quantum} refers to encodings that remain invariant under a set of discrete rotations in phase space. Encoding schemes with rotation-symmetry, such as the cat and binomial codes, are stabilized by a photon number super-parity operator \mbox{$\hat{\Pi}_{N} = e^{i(2\pi/N)\hat{n}}$ ($N \in \lbrace 2,3,\cdots \rbrace$)}, which is equivalent to the $360/N\degree$ rotation operator. As such, these codes are, by design, capable of detecting $N-1$ photon loss events.

Furthermore, these rotation-symmetric codes can also be made robust against dephasing errors in addition to photon loss errors~\cite{ouyang2020_tradeoffs}. For instance, the kitten code may be modified to have the following logical states:
\begin{linenomath}
\begin{align}
    |0_{L}\rangle &= \frac{1}{2}( |0\rangle + \sqrt{3}|4\rangle ), \nonumber \\
    |1_{L}\rangle &= \frac{1}{2}( \sqrt{3}|2\rangle + |6\rangle ).  \label{eq:bin N2S1}
\end{align}
\end{linenomath}
This version of the binomial code is also stabilized by the parity operator and is invariant under the $180\degree$ rotation, thus making it robust against single photon loss errors. However, the modified code is now also higher in energy as its logical states have $3$ photons on average, as opposed to $2$ in the case of the kitten code. This additional redundancy makes the modified binomial code robust against single dephasing events as well. In other words, the logical states of the modified binomial code satisfy the Knill-Laflamme condition for an extended error set \mbox{$\lbrace \hat{I},\hat{a},\hat{a}^{\dagger}\hat{a} \rbrace$ where $\hat{a}^{\dagger}\hat{a}$} describes single dephasing errors in the cavity. This means that the two logical code words also have the same second moment of the photon number probability distribution besides having the same average photon number. Note that when more moments of the photon number probability distribution are the same for the two code words, distinguishing them becomes more difficult for the environment, which enhances the error-correcting capability of the code.

Another way to generalize rotation-symmetric codes is by considering different rotation angles. For instance, we can design codes that are invariant under a $120\degree$ rotation instead of a $180\degree$ rotation and are thus stabilized by the super-parity modulo $3$ operator \mbox{$\hat{\Pi}_{3} = e^{ i (2\pi /3) \hat{n} }$}. By taking advantage of a larger spacing in the photon number basis, $120\degree$-rotation-symmetric codes can detect two-photon loss events as well as single photon loss events, thus providing enhanced protection to the logical information. For instance, a variant of the binomial code with the logical states 
\begin{linenomath}
\begin{align}
    |0_{L}\rangle &= \frac{1}{2} ( |0\rangle + \sqrt{3} |6\rangle ), \nonumber \\
    |1_{L}\rangle &= \frac{1}{2}(\sqrt{3}|3\rangle + |9\rangle), \label{eq:bin N2S2}
\end{align}
\end{linenomath}
is invariant under the $120\degree$ rotation, or equivalently, has photon numbers that are integer multiples of $3$. Moreover, the particular coefficients associated with each photon number state in the code words, which are derived from the binomial coefficients (hence the name `binomial' code), ensure that the code satisfies the Knill-Laflamme condition for an error set \mbox{$\lbrace \hat{I},\hat{a},\hat{a}^{2},\hat{a}^{\dagger}\hat{a} \rbrace$}. This indicates that the code is robust against both single and two-photon loss events, as well as single dephasing errors. The binomial code can be further generalized to protect against higher-order effects of loss and dephasing errors~\cite{michael2016_new} as well as used for autonomous QEC~\cite{lihm2018_implementation}.

Cat codes are another important example of rotation-symmetric bosonic codes. The four-component cat codes (or 4-cat codes), which are composed of four coherent states $|\pm\alpha\rangle$, $|\pm i\alpha\rangle$, are the simplest variants of the cat codes that are robust against photon loss errors~\cite{leghtas2013_hardware-efficient, mirrahimi2014_dynamically}. In this encoding, the logical states are defined by superpositions of coherent states:
\begin{linenomath}
\begin{align}
    |0_{L}\rangle &\propto |\alpha\rangle + |i\alpha\rangle + |-\alpha\rangle + |-i\alpha\rangle 
    \nonumber \\
    &\propto \sum_{n=0}^{\infty}\frac{\alpha^{4n}}{\sqrt{ (4n)! }}|4n\rangle,  
    \nonumber \\
    |1_{L}\rangle &\propto |\alpha\rangle - |i\alpha\rangle + |-\alpha\rangle - |-i\alpha\rangle 
    \nonumber\\
    &\propto \sum_{n=0}^{\infty}\frac{\alpha^{4n+2}}{\sqrt{ (4n+2)! }}|4n+2\rangle . 
\end{align}
\end{linenomath}
Note that the amplitude of the coherent states $|\alpha|$ determines the size of the cat code. Similar to the kitten code (Eq.~\eqref{eq:bin N1S1}), the logical states of the 4-cat code have an even number of photons. Thus, the 4-cat code is invariant under the $180\degree$ rotation and is able to detect single photon loss events. Here, we reinforce that the parity (or rotation-symmetry) alone does not ensure the recoverability of logical information against single photon loss errors, which is only guaranteed when the Knill-Laflamme condition is satisfied for the error set \mbox{$\lbrace \hat{I},\hat{a} \rbrace$}. For even parity codes, as explained in Sec~\ref{sec:concepts}, recoverability requires the two logical states to have the same number of photons on average. 

For large cat codes with $|\alpha|\gg 1$, the average photon number is approximately given by \mbox{$\bar{n} \simeq |\alpha|^{2}$} for both logical states, and the Knill-Laflamme condition is approximately fulfilled. More importantly, the average photon numbers of the two logical states are exactly the same for certain values of $|\alpha|$ (also known as `sweet spots'~\cite{li2017_cat}) such that:
\begin{linenomath}
\begin{align}
    \tan|\alpha|^{2} = - \tanh |\alpha|^{2}. 
\end{align}
\end{linenomath}
The smallest such $|\alpha|$ is given by $|\alpha| = 1.538$, which corresponds to the average photon number $\bar{n} = 2.324$. By increasing the size of the cat codes, we can construct logical states that are robust against both single photon loss and dephasing errors. In particular, cat codes with a large average photon number $|\alpha|^{2}$ satisfy the Knill-Laflamme condition for the dephasing error set $\lbrace \hat{I}, \hat{n}, \hat{n}^{2}, \cdots \rbrace$ approximately modulo an inaccuracy that scales as $e^{-2|\alpha|^{2}}$ \cite{albert2018_performance}. However, an extra error-correcting mechanism, such as a multi-photon engineered dissipation \cite{mirrahimi2014_dynamically}, is required to exploit the intrinsic error-correcting capability of the cat codes against dephasing errors. Moreover, increasing the number of coherent state components in the cat code introduces further protection~\cite{albert2016_holonomic,bergmann2016_quantum}. For instance, with six coherent state components, the code words become $120\degree$-rotation-symmetric and can thus detect up to two-photon loss events. Generalizations of cat codes and analyses of their performance can be found in Ref.~\cite{li2017_cat}, while a multi-mode generalization is available in Ref.~\cite{albert2019_pair-cat}.   

\subsection{Translation-symmetric codes: GKP codes}\label{subsection:translation symmetric codes}

Another class of bosonic codes are translation-symmetric, with the Gottesman-Kitaev-Preskill (GKP) codes~\cite{gottesman2001_encoding} being a prominent example. The simplest variant of the GKP codes is the square lattice GKP code, which encodes a logical qubit in the phase space of a harmonic oscillator stabilized by two commuting displacement operators:
\begin{linenomath}
\begin{align}
    \hat{S}_{q} &\equiv \exp[i2\sqrt{\pi}\hat{q}] = \hat{D}(i\sqrt{2\pi}), \nonumber \\ 
    \hat{S}_{p} &\equiv \exp[-i2\sqrt{\pi}\hat{p}] = \hat{D}(\sqrt{2\pi}),
    \label{eq:square gkp stabilizers}
\end{align}
\end{linenomath}
where $\hat{q}$ and $\hat{p}$ are the position and momentum operators respectively. 

A key motivation for choosing these stabilizers is to circumvent the Heisenberg uncertainty principle, which dictates that the position and momentum operators cannot be measured simultaneously as they do not commute. Since the two displacement operators in Eq.~\eqref{eq:square gkp stabilizers} commute with each other, we can measure them simultaneously. Measuring the displacement operators $\exp[i2\sqrt{\pi}\hat{q}]$ and $\exp[-i2\sqrt{\pi}\hat{p}]$ is equivalent to measuring their phases $2\sqrt{\pi}\hat{q}$ and $-2\sqrt{\pi}\hat{p}$ modulo $2\pi$, which is in turn the same as simultaneously measuring $\hat{q}$ and $\hat{p}$ modulo $\sqrt{\pi}$. Therefore, the square lattice GKP code is, by design, capable of addressing two non-commuting quadrature operators by simultaneously measuring them within a unit cell of a square lattice. The uncertainty now lies in the fact that we do not know which unit cell the state is in.

The two logical states of the square lattice GKP code are explicitly given by:
\begin{linenomath}
\begin{align}
    |0_{L}\rangle &\propto \sum_{n \in \mathbb{Z}} 
    |\hat{q} = (2n) \sqrt{\pi} \rangle \propto \sum_{n\in\mathbb{Z}} |\hat{p} = n\sqrt{\pi} \rangle ,
    \nonumber \\
    |1_{L}\rangle &\propto \sum_{n\in\mathbb{Z}} |\hat{q} = (2n+1)\sqrt{\pi} \rangle \propto \sum_{n\in\mathbb{Z}} (-1)^{n}|\hat{p} = n\sqrt{\pi} \rangle, 
\end{align}
\end{linenomath}
and satisfy \mbox{$\hat{q} = \hat{p} = 0$} modulo $\sqrt{\pi}$, thus clearly illustrating that a periodic simultaneous quadrature measurement is indeed possible if the spacing is chosen appropriately. Additionally, the code states are invariant under discrete translations of length $2\sqrt{\pi}$ in both the position and momentum directions, which makes the code symmetric under translations. 

Conceptually, ideal GKP code states consist of infinitely many infinitely squeezed states where each component is described by a Dirac Delta function. However, precisely implementing these ideal states is not feasible in realistic quantum systems. In practice, only approximate GKP states can be realized where each position or momentum eigenstate is replaced by a finitely squeezed state and large position and momentum components are suppressed by a Gaussian envelope~\cite{gottesman2001_encoding, terhal2016_encoding, matsuura2019_equivalence}. Many proposals to realize approximate GKP states in various physical platforms~\cite{gottesman2001_encoding, travaglione2002_preparing, pirandola2004_constructing, pirandola2006_generating, vasconcelos2010_alloptical, terhal2016_encoding, motes2017_encoding, weigand2018_generating, arrazola2019_machine, shi2019_fault-tolerant, su2019_conversion, eaton2019_nongaussian, hastrup2019_measurementfree, weigand2020_realizing, hastrup2020_improved, royer2020_stabilization}  and ways to simulate approximate states efficiently~\cite{wan2019_memoryassisted, tzitrin2020_progress, terhal2020_scalable, pantaleoni2020_modular, walshe2020_continuousvariable, mensen2020_phasespace} have been explored in the field. The quality of such states can be characterized by the degree of squeezing in both the position and momentum quadratures. As the squeezing, and hence the average photon number, increases, the approximate state will converge towards an ideal code state. Recently, approximate states of squeezing \mbox{$5.5$ -- $9.5$ dB} have been realized in trapped ion~\cite{fluhmann2019_encoding, deNeeve2020_error} and cQED \cite{campagne-ibarcq2019_quantum} systems. The cQED realization is discussed further in Sec.~\ref{section: realization}.

With a discrete translation-symmetry, GKP codes are naturally robust against random displacement errors as long as the size of the displacement is small compared to the separation between distinct logical states. For instance, the square lattice GKP code is robust against any displacements of size less than $\sqrt{\pi}/2$ as they can be identified via the quadrature measurements modulo $\sqrt{\pi}$ and then countered accordingly. For moderately squeezed approximate GKP states that contain a small number of photons, photon loss errors can be decomposed as small shift errors and therefore can be effectively addressed by the code~\cite{gottesman2001_encoding,terhal2016_encoding}. On the other hand, for large approximate GKP states that are highly squeezed, even a tiny fraction of photon loss results in large shift errors which cannot be corrected by the code. Thus, naively using the standard GKP error correction protocol to decode does not work for large GKP states under photon loss errors. Nevertheless, studies have observed that if an optimal decoding scheme is adopted, excellent performance against photon loss errors can be achieved even with large GKP states~\cite{albert2018_performance, noh2019_quantum}. This improvement happens because photon loss errors can be converted into random displacement errors via amplification. This implies that for highly squeezed large GKP codes, a suitable decoding strategy is to first amplify the contracted states and then correct the resulting random shift errors by measuring the quadrature operators modulo $\sqrt{\pi}$~\cite{gottesman2003_secure}. We note that at present, no analogous techniques are known for dephasing errors. Thus, highly squeezed GKP codes are not robust against dephasing errors since even a small random rotation can result in large shift errors~\cite{terhal2016_encoding}.      

While the preparation of GKP states is challenging, implementing logical operations on GKP states is relatively straightforward. Any logical Pauli or Clifford operation on GKP states can be realized by a displacement or Gaussian operation (via a linear drive or a bilinear coupling). Moreover, magic states~\cite{bravyi2005_universal} encoded in the GKP code, which are necessary for implementing non-Clifford operations, can be prepared with only Gaussian operations and GKP states~\cite{terhal2016_encoding, baragiola2019_allgaussian}. Thus, the preparation of code words is the only required non-Gaussian operation for performing universal quantum computation with the GKP code. Non-Clifford operations can be directly enacted on GKP qubits~\cite{gottesman2001_encoding}, although the cubic phase gate suggested in the original proposal has been recently shown to perform poorly for this purpose~\cite{hastrup2020_cubic}.

Furthermore, GKP states can be defined over lattices other than the square lattice. For instance, hexagonal GKP codes can correct any shift errors of size less than \mbox{$(2/\sqrt{3})^{1/2} \sqrt{\pi}/2 \simeq 1.07\sqrt{\pi}/2$}, which is larger than the size of shifts that are correctable by the square lattice GKP code~\cite{gottesman2001_encoding, noh2019_quantum}. A recent work has also explored the rectangular GKP code~\cite{hanggli2020_enhanced}. More generally, a multi-mode GKP code can be defined over any symplectic lattice~\cite{harrington2001_achievable}. Lastly, the GKP code may also be used for building robust quantum repeaters for long-distance quantum communication~\cite{rozpedek2020_quantum, fukui2020_alloptical}.

\subsection{Biased-noise bosonic qubits: Two-component cat codes}\label{subsection:biased noise qubits}

Recently, there has been growing interest in biased-noise bosonic qubits, where a quantum system is engineered to have one type of error occur with a much higher probability than other types of errors. This noise bias can simplify the next layer of error correction~\cite{aliferis2008_faulttolerant}. For instance, we can design a biased-noise code that suppresses bit-flips and then correct the dominant phase-flip errors by using repetition codes~\cite{guillaud2019_repetition, guillaud2020_error}. Alternatively, we can also tailor the surface code to leverage the advantages of biased-noise models to increase fault-tolerance thresholds and reduce resource overheads~\cite{tuckett2018-ultrahigh, tuckett2019_tailoring, tuckett2020_faulttolerant, chamberland2020_building}.

A promising candidate for biased-noise bosonic qubits is the two-component cat code~\cite{cochrane1999_macroscopically, jeong2002_efficient, ralph2003_quantum, glancy2004_transmission, lund2008_faulttolerant}, or 2-cat code, whose logical code words are given by: 
\begin{linenomath}
\begin{align}
    |+_{L}\rangle &\propto |\alpha\rangle + |-\alpha\rangle \propto \sum_{n=0}^{\infty} \frac{\alpha^{2n}}{\sqrt{ (2n)! }}|2n\rangle , 
    \nonumber \\
    |-_{L}\rangle &\propto |\alpha\rangle - |-\alpha\rangle \propto \sum_{n=0}^{\infty} \frac{\alpha^{2n+1}}{\sqrt{ (2n+1)! }}|2n+1\rangle . 
\end{align}
\end{linenomath}
If $\alpha$ is large enough, 2-cat codes are capable of correcting dephasing errors. One way to implement 2-cat codes is by autonomously stabilizing them via an engineered dissipation of the form \mbox{$\kappa_{2}\mathcal{D}[\hat{a}^{2}-\alpha^{2}]$}~\cite{mirrahimi2014_dynamically}. This engineered two-photon dissipation can exponentially suppress the logical bit-flip error in the code states due to dephasing in superconducting cavities \mbox{($\kappa_{\phi}\mathcal{D}[\hat{a}^{\dagger}\hat{a}]$)}, i.e., 
\begin{linenomath}
\begin{align}
    \gamma_{\textrm{bit-flip}} \simeq 2\kappa_{\phi}|\alpha|^{2}e^{-2|\alpha|^{2}} \,\,\, \textrm{if}\,\,\, \kappa_{\phi}\ll \kappa_{2}, 
\end{align}
\end{linenomath}
where $|\alpha|^{2}$ is the size of the cat code, $\kappa_{2}$ is the two-photon dissipation rate, and $\kappa_{\phi}$ is the dephasing rate. 

However, the dominant error source in superconducting cavities is photon loss. While dephasing of the cavity does not change the photon number parity, a single photon loss can directly flip the photon number parity of the state. In other words, a single photon loss maps an even parity state $|+_{L}\rangle$ to an odd parity state $|-_{L}\rangle$, thus causing a phase-flip error. As such, phase-flip errors due to single photon loss cannot be mitigated by the 2-cat code. Nevertheless, bit-flip errors due to single photon loss can be suppressed exponentially in the photon number $\alpha^{2}$ provided that the two-photon dissipation is strong enough. Once bit-flip errors are countered, the next layer of QEC will only need to tackle the phase-flip errors, which can be induced by spurious transitions of the ancilla states. In particular, in the stabilized 2-cat codes, thermal excitation of the ancilla during idle times can result in a significant rotation of the cavity state and complete dephasing of the encoded qubit. Currently, such events cannot be corrected by the code and are observed to be a limiting factor for the logical lifetime of the encoded qubit~\cite{lescanne2020_exponential}.

Biased-noise 2-cat qubits can also be realized by using an engineered Hamiltonian \mbox{$\hat{H} = -K(\hat{a}^{\dagger}-(\alpha^{*})^{2})( \hat{a}^{2}-\alpha^{2} )$}, which has two coherent states $|\pm\alpha\rangle$ as degenerate eigenstates~\cite{puri2017_engineering, puri2019_stabilized, puri2020_bias-preserving}. A crucial general consideration for these biased-noise bosonic codes is that the asymmetry in the noise must be preserved during the implementation of gates. The theoretical framework to achieve a universal gate set on the 2-cat codes in a bias-preserving manner has been proposed recently~\cite{guillaud2019_repetition, puri2020_bias-preserving}.

\subsection{Comparison of various bosonic codes for loss and dephasing errors}\label{subsection:code comparision}

In summary, rotation-symmetric codes (e.g., 4-cat codes and binomial codes) are robust against both photon loss and dephasing errors. Translation-symmetric codes such as the GKP codes can be made highly robust against photon loss errors but are susceptible to dephasing errors. Thus, translation-symmetric codes are suited for loss dominated systems. In contrast, 2-cat codes can correct dephasing errors well if they are stabilized by an engineered two-photon dissipation, but are not capable of correcting photon loss errors. Hence, 2-cat codes are naturally suited for dephasing dominated systems. However, as discussed in Sec.~\ref{subsection:biased noise qubits}, 2-cat codes can also be useful in the loss dominated regime as their large noise bias can simplify any higher-level error correction schemes. In Fig.~\ref{fig:codes}(e), we provide a qualitative schematic that represents different regimes of photon loss and dephasing where each code is designed to perform well. Note that if the loss and dephasing error probabilities are too high, the quantum capacity~\cite{lloyd1997_capacity, devetak2005_private} of the corresponding quantum channel will vanish. When this happens, encoding logical quantum information in a reliable way becomes impossible even with an optimal QEC code~\cite{holevo2001_evaluating, wolf2007_quantum, wilde2012_quantum, wilde2018_energyconstrained, arqand2020_quantum}. 

While photon loss is the dominant error channel in typical cQED implementations, excitation gain events do also occur, although at a much lower rate. In general, codes robust against photon loss errors tend to be robust against photon gain as well. For rotation-symmetric codes, parity (or super-parity) measurements can detect both photon loss and photon gain events. In the case of translation-symmetric GKP codes, photon gain can be converted into a random shift error by applying a suitable photon loss channel and hence be monitored with the modular quadrature measurement. On the other hand, since the 2-cat codes are not robust against photon loss, they are consequently susceptible to photon gain too, which results in phase-flip errors. Nevertheless, in the presence of engineered dissipation, the stabilized 2-cat codes can still preserve their noise bias under both photon loss and gain errors.

Until now, we have focused only on the intrinsic error-correcting capability of various bosonic codes without considering practical imperfections. In the following sections, we will discuss the implementation of bosonic QEC in cQED systems and examine errors that occur in practical situations.

\section{The {cQED} Hardware for Bosonic Codes}
\label{sec:cavities}

In the preceding sections, we have discussed the concepts and merits of various bosonic QEC codes. These ideas are brought to reality by developing robust quantum hardware with both coherent harmonic modes for encoding information as well as non-linear ancillae for effective control and tomography of the encoded information. Such systems have been realized with the motional degree of freedom in trapped ions~\cite{fluhmann2019_encoding, deNeeve2020_error}, electromagnetic fields of microwave~\cite{deleglise2008_reconstruction} and optical cavities~\cite{hacker2019_deterministic}, Rydberg atom arrays~\cite{omran2019_generation}, and flying photons~\cite{ourjoumtsev2006_generating}, and there are proposals that use other physical systems~\cite{hou2016_generation, bulutay2017_cat-state} as well. Among the various platforms, the cQED architecture consisting of superconducting microwave cavities and Josephson junction-based non-linear ancillae have enabled many prominent experimental milestones towards realizing QEC using bosonic codes. In this section, we will introduce the key hardware building blocks necessary for the successful realization of bosonic codes in cQED.

\subsection{Components of the cQED architecture}

Circuit quantum electrodynamics (cQED) explores light-matter interactions by confining quantized electromagnetic fields in precisely engineered compositions of superconducting inductors and capacitors~\cite{blais2020_circuit}. These superconducting circuit elements can be tailor-made by conventional fabrication techniques~\cite{frunzio2005_fabrication} and controlled by commercially available microwave electronics and dilution refrigerators~\cite{schoelkopf2008_wiring} to access strong coupling regimes~\cite{wallraff2004_strong, schuster2007_resolving, niemczyk2010_circuit}. These characteristics make cQED a compelling platform for universal quantum computation, as noted by several review articles~\cite{girvin2009_circuit, ladd2010_quantum, devoret2013_superconducting, wendin2017_quantum, krantz2019_quantum, kjaergaard2019_superconducting, blais2020_quantum, cai2020_bosonic}.

Quantum devices in the cQED framework are typically built by coupling two components, a linear oscillator mode and an anharmonic mode, in different configurations~\cite{blais2020_circuit}. The anharmonic modes are discrete few-level systems that can be implemented using Josephson junctions. They can be designed to interact with one or more harmonic modes, akin to atoms in an optical field in the cavity QED framework~\cite{haroche2006_exploring}. Unlike naturally occurring atoms, the parameters of these anharmonic oscillators, or artificial atoms, can be precisely engineered. For instance, they may be made to have a fixed resonance frequency in devices with a single Josephson junction or have tunable frequencies by integrating multiple junctions in the presence of an external magnetic flux. The lowest two or three energy levels of these artificial atoms, typically in the form of transmons~\cite{koch2007_charge-insensitive, schreier2008_suppressing}, can be used to effectively encode and process quantum information, as demonstrated by successful realizations of various NISQ era~\cite{preskill2018_quantum} processors~\cite{xiang2017_experimental, otterbach2017_unsupervised, kandala2017_hardware, arute2019_quantum, jurcevic2020_demonstration, google2020_hartree-fock, lacroix2020_improving}.

Linear oscillators are typically realized in cQED by superconducting microwave cavities. Commonly used architectures include coplanar waveguide (CPW)~\cite{wallraff2004_strong}, three-dimensional (3D) rectangular~\cite{paik2011_observation} and cylindrical co-axial~\cite{reagor2016_quantum}, and micromachined~\cite{brecht2015_demonstration} cavities. These quantum harmonic oscillators have well-defined but degenerate energy transitions. Therefore, to selectively address their transitions, we must introduce some non-linearities in them. In cQED, non-linearities are introduced by coupling superconducting cavities to artificial atoms in either the resonant or dispersive regimes. 

Under resonant coupling, the transition frequencies of the artificial atom and the cavity coincide, which allows the direct exchange of energy from one mode to the other. In this configuration, cavities can act as an on-demand single photon source~\cite{houck2007_generating} or a quantum bus that mediates operations between two isolated artificial atoms by sequentially interacting with each of them~\cite{majer2007_coupling, sillanpaa2007_coherent, dicarlo2009_demonstration}. 
Dispersive coupling is achieved by detuning the frequencies of the cavity, $\omega_{a}$, and the artificial atom, $\omega_{b}$, such that the detuning is much larger than the direct interaction strength between them. In this regime, there is no resonant energy exchange between the modes. Instead, the coupling translates into a state-dependent frequency shift, which can be described by the following Hamiltonian:
\begin{linenomath}
\begin{align}
    \hat{H} &= \omega_{a}\hat{a}^{\dagger}\hat{a} + \omega_{b}\hat{b}^{\dagger}\hat{b} - \chi_{ab}\hat{b}^{\dagger}\hat{b}\hat{a}^{\dagger}\hat{a} - \frac{\alpha}{2}\hat{b}^{\dagger2}\hat{b}^2 - \frac{K}{2}\hat{a}^{\dagger2}\hat{a}^2,
    \label{eq:dispersive_hamiltonian}
\end{align}
\end{linenomath}
where $\hat{a},\hat{b}$ are the annihilation operators associated with cavity and artificial atom respectively, $\chi_{ab}$ is the dispersive coupling strength between them, $\alpha$ is the anharmonicity of the artificial atom, and $K$ is the non-linearity of the cavity inherited from the atom. Eq.~\eqref{eq:dispersive_hamiltonian} further illustrates that the coupling between the two modes is symmetric. In other words, the frequency of the cavity shifts conditioned on the state of the artificial atom, and vice versa. 

Superconducting cavities dispersively coupled to an artificial atom constitute a highly versatile tool that can be employed to fulfill many different roles for quantum information processing. For instance, cavities that are strongly coupled to a transmission line are useful for the efficient readout of the quantum state of the artificial atom~\cite{wallraff2004_strong, wallraff2005_approaching, johansson2006_vacuum}. Conversely, cavities weakly coupled to the environment can be used as quantum memories for storing information coherently~\cite{reagor2016_quantum}. Moreover, when the linewidth of the cavity is narrow compared to the dispersive shift $\chi_{ab}$, we can selectively address the individual energy levels of the cavity via the artificial atom~\cite{schuster2007_resolving}. In this case, the artificial atoms act as non-linear ancillae whose role is to enable conditional operations and perform efficient tomography of the cavity state~\cite{blais2004_cavity}. This configuration where multi-photon states of the cavity encode logical information and the ancilla affords universal control~\cite{krastanov2015_universal} has become an increasingly prevalent choice for implementing bosonic QEC.

\subsection{Coherence of superconducting cavities}\label{subsection:oscillators}

Superconducting microwave cavities may be realized in several geometries, with each having their respective advantages. Typically, cavities are constructed in two main architectures, which are 2-dimensional (2D) and 3-dimensional (3D), based on the dimensionality of the electric field distribution. A third possibility combines the advantages of both the 2D and 3D designs to realize a compact and highly coherent `2.5D' cavity structure, for instance, using micromachining techniques~\cite{brecht2015_demonstration, brecht2017_micromachined}.

\begin{figure*}[t!]
    \centering
    \includegraphics[scale=1]{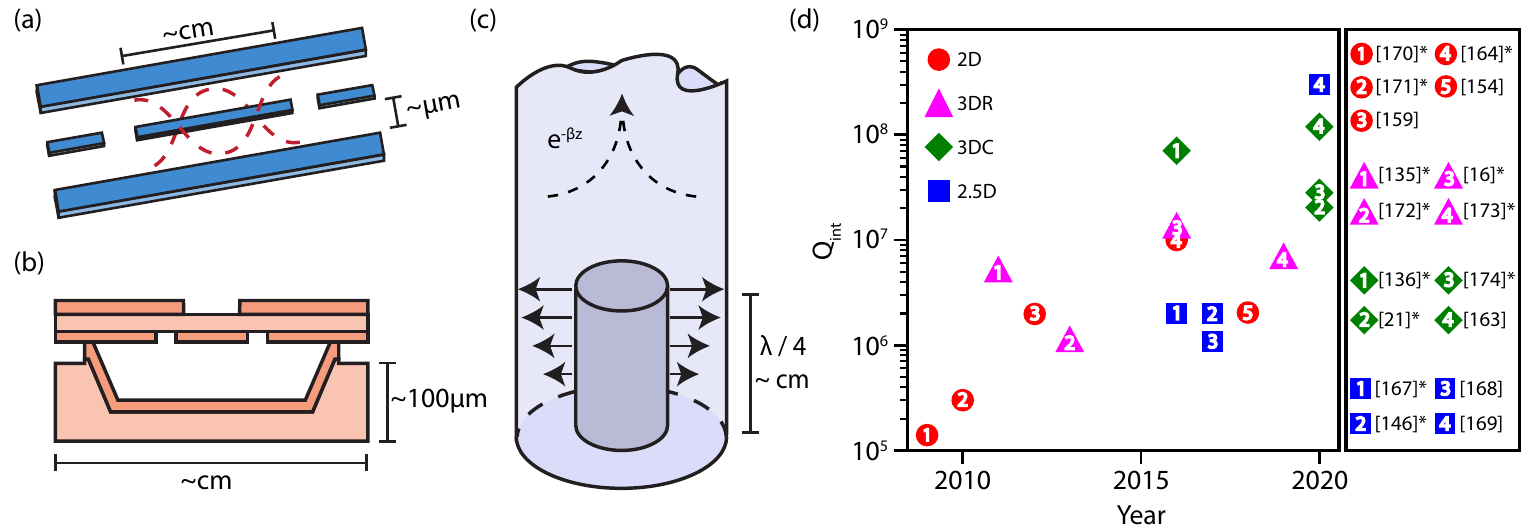}
    \caption[width=\textwidth]{\textbf{Designs of superconducting cavities and their coherence} (a)-(c) Illustrations of 2D CPW, 2.5D micromachined, and 3D co-axial cavities respectively. (d) A selected set of $Q_{\mathrm{int}}$ for four commonly used cavity designs in cQED extracted from the literature. Overall, the coherence properties of superconducting cavities have been steadily increasing over the last decade, with the 3D co-axial cavities currently being the main architecture for realizing bosonic QEC.  The references for each cavity design, in chronological order, are: 2D cavities: Refs. \cite{hofheinz2009_synthesizing}, \cite{leek2010_cavity}, \cite{megrant2012_planar}, \cite{axline2016_architecture}, \cite{calusine2018_analysis}; 3D rectangular cavities (3DR): Refs. \cite{paik2011_observation}, \cite{vlastakis2013_deterministically}, \cite{ofek2016_extending}, \cite{ hu2019_quantum}; 3D cylindrical co-axial cavities (3DC): Refs.  \cite{reagor2016_quantum}, \cite{reinhold2020_error-corrected}, \cite{ma2020_error-transparent}, \cite{kudra2020_high}; 2.5D cavities: Refs.~\cite{minev2016_planar}, \cite{brecht2017_micromachined}, \cite{zoepfl2017_characterization}, \cite{lei2020_high}. We have highlighted studies that successfully integrated one or more non-linear ancillae with an asterisk (*).}
    \label{fig:cavities}
\end{figure*}

In the 2D architecture, such as the coplanar waveguide (CPW), the cavity is defined by gaps between circuit elements printed on a substrate which is typically made from silicon or sapphire (Fig.~\ref{fig:cavities}a). In such planar structures, the energy is mostly stored in the substrate, surfaces, and interfaces, all of which suffer from losses due to spurious two-level systems in the resonator dielectrics~\cite{martinis2005_decoherence, oconnell2008_microwave, gao2008_experimental, muller2019_towards}. Hence, the $Q_{\mathrm{int}}$ of 2D cavities is currently limited to \mbox{$\sim {10^5 - 10^6}$} but can potentially be improved with more sophisticated cavity design~\cite{sage2011_study, vissers2012_reduced, geerlings2012_improving, calusine2018_analysis}, materials selection~\cite{wang2009_improving, barends2010_minimal, vissers2010_low, place2020_new}, and surface treatment~\cite{megrant2012_planar, sandberg2012_etch, bruno2015_reducing, melville2020_comparison}. Despite their limited coherence properties, 2D cavities are widely featured in cQED, and especially in NISQ processors, as they have a small footprint and a straightforward fabrication process. Furthermore, in these processors, these cavities are typically used as readout or bus modes, which do not require long coherence times. 

In contrast, 3D cavities (Fig.~\ref{fig:cavities}c) achieve a higher $Q_{\mathrm{int}}$ of \mbox{$\sim {10^7 - 10^8}$}, at the cost of a much larger footprint than their 2D counterparts, by storing energy in the vacuum between the walls of a superconducting box. Among 3D designs, 3D co-axial cavities machined out of high-purity aluminum \mbox{($\simeq {99.999}\%$)} have shown a $Q_{\mathrm{int}}$ as high as \mbox{$1.1 \times 10^8$~\cite{kudra2020_high}}. This is achieved by significantly suppressing the dissipation due to current flowing along the seams of the superconducting walls and imperfections on the inner surfaces of the structure. Additionally, this design is particularly effective as a platform for implementing bosonic codes as one or more ancillae and readout modes can be conveniently integrated into the platform~\cite{axline2016_architecture}.

One strategy to take advantage of both the long lifetimes of 3D designs and the small footprint and scalable fabrication of 2D geometries is to construct compact 2.5D cavities~\cite{minev2013_planar, brecht2015_demonstration, brecht2016_multilayer, minev2016_planar,  axline2016_architecture, brecht2017_micromachined, zoepfl2017_characterization}. In particular, the micromachining technique provides a promising method to fabricate lithographically defined 2.5D cavities etched out of silicon wafers~\cite{brecht2015_demonstration}. In this configuration, the energy can be stored primarily in the vacuum but the depth of the cavity is much smaller compared to the other dimensions (Fig.~\ref{fig:cavities}b). Here, the key challenge is to realize a high-quality contact between the top wall and the etched region. With carefully optimized indium bump-bonding methods, internal quality factors surpassing 300 million have recently been achieved in the 2.5D architecture~\cite{lei2020_high}. Moreover, the integration of a transmon ancilla in this design has also been demonstrated~\cite{brecht2017_micromachined}, thus making these 2.5D cavities promising candidates for realizing large-scale quantum devices based on bosonic modes. 

Regardless of the architecture, the effective implementation of bosonic QEC schemes requires a careful balance between the need for isolation and coherence as well as the ability to effectively manipulate and characterize the cavity. The various loss channels~\cite{oliver2013_materials, mcrae2020_materials} as well as lossy interfaces~\cite{wenner2011_surface, wang2015_surface, woods2019_determining} of superconducting microwave cavities have been extensively studied and their intrinsic coherence properties have been improving significantly over the last 15 years~\cite{mcrae2020_materials}. In Fig.~\ref{fig:cavities}d, we compile a non-exhaustive summary of the internal quality factors of cavities demonstrated in various geometries over the last decade. Besides enhancing the $Q_{\mathrm{int}}$ of the cavities, integrating them with ancillary mode(s) is crucial for realizing bosonic logical qubits. However, note that introducing an ancilla results in the degradation of the $Q_{\mathrm{int}}$, as the best ancilla coherence times \mbox{($\sim {50 - 100} {\mu}$s)} are typically about $10 - 20$ times lower than those of the state-of-the-art superconducting cavities. Hence, while comparing the performance of the cavities in Fig.~\ref{fig:cavities}d, we have only included demonstrations that are compatible with being coupled to non-linear ancillae in the cQED architecture. From the figure, the 3D co-axial cavities emerge as the leading design currently used to realize bosonic qubits. 

\section{Realization of Bosonic Logical Qubits}\label{section: realization}

The remarkable improvements in the performance of cQED hardware components highlighted in Sec.~\ref{sec:cavities} have made realizing protected logical qubits using bosonic codes a realistic goal. Studies that encode a single logical mode and protect it against dominant error channels have been reported for the four-component cat~\cite{ofek2016_extending}, binomial kitten~\cite{hu2019_quantum}, and square and hexagonal GKP~\cite{campagne-ibarcq2019_quantum} codes. Moreover, robust operations on both single~\cite{heeres2017_implementing, ma2020_error-transparent, reinhold2020_error-corrected} and two bosonic modes~\cite{gao2018_programmable, gao2019_entanglement, xu2020_demonstration} have also been explored, thus paving the way towards building a fault-tolerant universal quantum computer based on bosonic logical qubits. In this section, we highlight recent efforts to implement universal control of as well as error correction protocols with bosonic modes.

\subsection{Operations on single bosonic modes}\label{subsection:operations_single}

For processing quantum information encoded in multi-photon states of superconducting cavities, we must be able to perform effective operations on and characterization of the cavity states. The only operation available for a standalone cavity mode is a displacement, \mbox{$\hat{D}(\alpha) = e^{\alpha \hat{a}^{\dagger} - \alpha^{*}\hat{a}}$}, which displaces the position and/or momentum of the harmonic oscillator depending on the value of the complex number $\alpha$. Real values of $\alpha$ correspond to pure position displacements, while imaginary values of $\alpha$ correspond to pure momentum ones~\cite{glauber1963_coherent}. Displacements can only result in the generation of coherent states from vacuum without the possibility to selectively address individual photon number states in the cavity. Therefore, non-trivial operations on these bosonic logical qubits are implemented by dispersively coupling the bosonic mode to a non-linear ancilla in combination with simple displacements. 

\begin{figure*}[t!]
    \centering
    \includegraphics[scale=1]{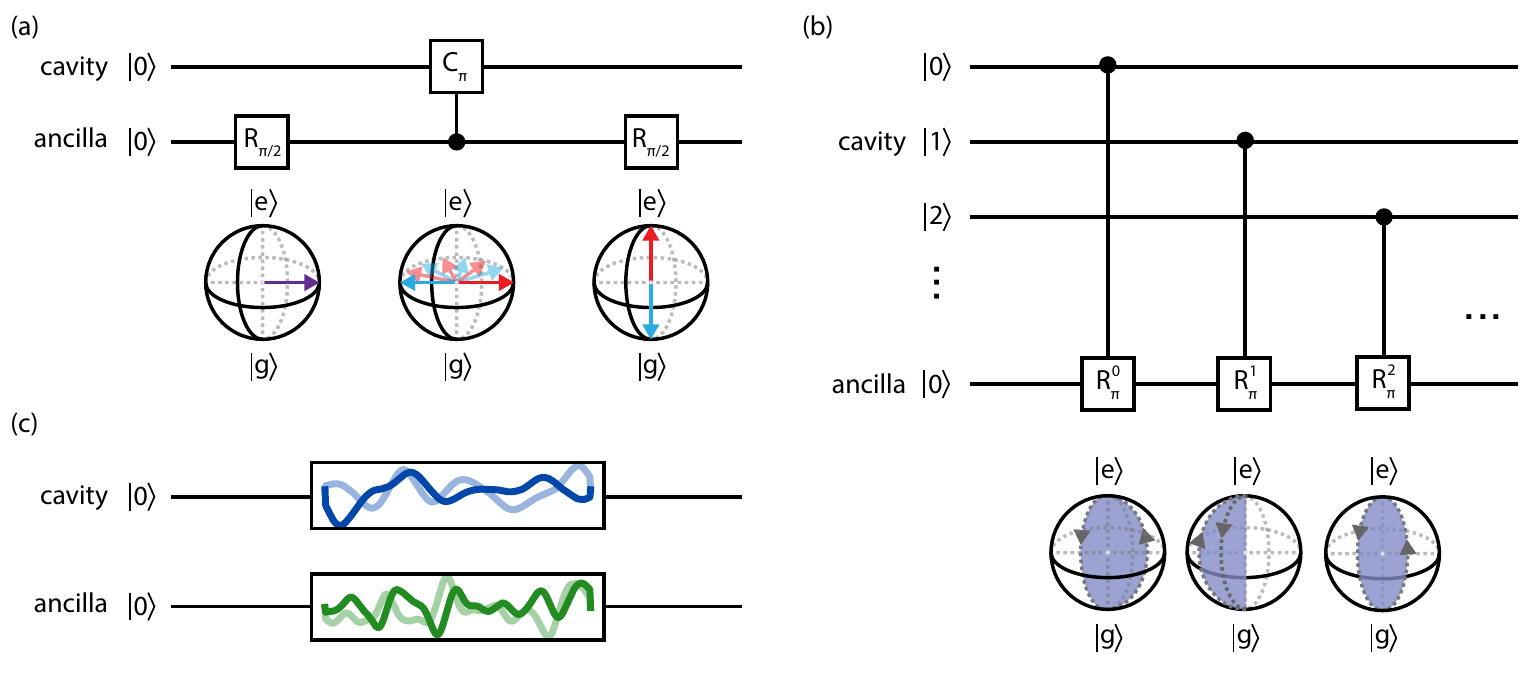}
    \caption[width=\textwidth]{\textbf{Operations on single bosonic modes.} (a) The sequence for mapping the photon number parity of a bosonic mode to the states of its ancilla. (b) Creation of arbitrary quantum states via repeat SNAP operations on the bosonic mode. (c) Implementing universal control on a single bosonic mode by concurrently driving the cavity and its ancilla with numerically optimized pulses.
    } \label{fig:operations}
\end{figure*}

A key capability enabled by this natural dispersive coupling is a controlled phase shift (CPS). CPS is a unitary operation that imparts a well-defined ancilla state-dependent phase on arbitrary cavity states, and is governed by \mbox{$\hat{U}_{\mathrm{CPS}}(t) = |g\rangle\langle g|\otimes\hat{I} + |e\rangle\langle e|\otimes e^{i\hat{n}\chi t}$}, where $\hat{n}$ is the cavity photon number operator, $\chi$ is the dispersive coupling strength, and $t$ is the evolution time. With this unitary, we can efficiently implement conditional phase operations by simply adjusting the evolution time. In particular, when \mbox{$t = \pi/\chi$}, all the odd photon number states acquire an overall $\pi$-phase while the even states get none. This allows us to effectively map the photon number parity of the bosonic mode onto the state of the ancilla (Fig.~\ref{fig:operations}a). 

Note that the time required for the parity-mapping operation scales inversely with the dispersive coupling strength, $\chi$. Naively, one might want to minimize the operation time by engineering a large $\chi$. However, increasing $\chi$ can also result in a stronger inherited non-linearity in the cavity (also known as the Kerr effect), which distorts the encoded information. Therefore, the coupling strength between the cavity and its ancilla, as well as the type of the non-linear ancillary mode, such as the transmon or the SNAIL design~\cite{frattini2017_3-wave}, are usually carefully optimized for each system to produce the desired Hamiltonian configuration.

With $\hat{U}_{\mathrm{CPS}}$, we can deterministically create complex bosonic encodings using analytically designed protocols such as the qcMAP operation, which maps an arbitrary qubit state onto a superposition of coherent states in the cavity~\cite{leghtas2013_deterministic}. This scheme is useful for preparing 4-cat states in both single~\cite{vlastakis2013_deterministically} and multiple~\cite{wang2016_schrodinger} cavities. Furthermore,  \mbox{$\hat{U}_{\mathrm{CPS}}(\pi/\chi)$} is also employed as the parity-mapping operation which is crucial for the characterization and tomography of the encoded bosonic qubits and the engineered gates on these qubits. 

In general, the quantum states encoded in a cavity can be fully characterized by probing the cavity's quasi-probability distributions, which is commonly achieved by performing Wigner tomography. The Wigner function can be defined as the expectation value of the displaced photon number parity operator, \mbox{$W(\beta) = \frac{2}{\pi} \mathrm{Tr}[\hat{D}(\beta)^\dagger \rho \hat{D}(\beta) \hat{P}]$}. In cQED, the Wigner functions of arbitrary quantum states can be measured precisely with a well-defined sequence that uses only the cavity displacement, ancilla rotation, and controlled phase shift operations~\cite{blais2020_circuit}. From the results of the Wigner tomography, we can reconstruct the full density matrix and characterize the action enacted on the cavity states in either the Pauli Transfer Matrix~\cite{chow2012_universal} or the process matrix~\cite{nielsen2000_quantum}.

Another crucial operation arising from the natural dispersive coupling is the non-linear selective number-dependent arbitrary phase (SNAP) gate~\cite{krastanov2015_universal, heeres2015_cavity}. A SNAP gate (Fig.~\ref{fig:operations}b), defined as \mbox{$\hat{S}_{n}(\theta_n)= e^{i\theta_n(|n\rangle\langle n|)}$}, selectively imparts a phase $\theta_n$ to the number state $|n\rangle$. Due to the energy-preserving nature of this operation, we can simultaneously perform $\hat{S}_{n}(\theta_n)$ on multiple number states. By numerically optimizing the linear displacement gates and the phases applied to each photon number state in the cavity, we can effectively cancel out the undesired Fock components via destructive interference to obtain the intended target state. 

While schemes like qcMAP and SNAP are sufficient to realize universal control on the cavity state, they quickly become impractical for handling more complex bosonic states. For instance, an operation on $n$ photons requires $O(n^2)$ gates using the SNAP protocol. To address this challenge, a fully numerical approach using optimal control theory (OCT) has been developed and widely adopted in recent years. The OCT framework provides an efficient general-purpose tool to implement arbitrary operations. In particular, the Gradient Ascent Pulse Engineering (GRAPE) method~\cite{khaneja2005_optimal, defouquieres2011_second} has been successfully deployed in other physical systems~\cite{dolde2014_high-fidelity, anderson2015_accurate} to implement robust quantum control. By constructing an accurate model of the time-dependent Hamiltonian of the system in the presence of arbitrary control fields, we can apply this technique to cQED systems to realize high-fidelity universal gate sets on any bosonic qubit encoded in cavities, as demonstrated in Ref.~\cite{heeres2017_implementing}. A typical example of a set of pulses obtained through the
gradient-based OCT framework is shown in (Fig.~\ref{fig:operations}c).

More recently, various concepts from classical machine learning, such as automatic differentiation~\cite{abdelhafez2019_gradient-based} and reinforcement learning~\cite{niu2019_universal, zhang2019_reinforcement}, have been applied to enhance the efficiency of the numerical optimization. Crucially, the success of these techniques does not only rely on the robustness of the algorithms, but also depends on the choice of boundary conditions. Knowledge of these boundary conditions requires comprehensive investigations of the physics of the quantum system and the practical constraints of the control and measurement apparatus. 

\subsection{Operations on multiple bosonic modes}\label{subsection:operations_multiple}

Apart from robust single-mode operations, universal quantum computation using bosonic qubits also requires at least one entangling gate between two modes. Realizing such an operation can be challenging due to the lack of a natural coupling between cavities. Moreover, the individual coherence of each bosonic qubit must be maintained while maximizing the rate of interactions between them. One promising strategy to tackle this issue is to use the non-linear frequency conversion capability of the Josephson junction to provide a driven coupling between two otherwise isolated cavities~\cite{bergeal2010_phase-preserving}. Such operations are fully activated by external microwave drives which can be tuned on and off on-demand without modifying the hardware. This arrangement ensures that the individual bosonic modes remain well-isolated during idle times and undergo the engineered interaction only when an operation is enacted. 

Using this strategy, a CNOT gate was the first logical gate enacted on two bosonic qubits~\cite{rosenblum2018_cnot}. This gate is facilitated by a parametrically-driven sideband transition between the ancilla and the control mode together with a carefully chosen conditional phase gate between the ancilla and the target cavity. By achieving a gate fidelity above 98\%, this study showcases the potential of such engineered quantum gates between bosonic modes. More recently, a controlled-phase gate has been demonstrated between two binomial logical qubits in Ref.~\cite{xu2020_demonstration}. Here, the microwave drives are tailored to induce a geometric phase that depends on the joint state of the two bosonic modes. However, these two types of operations are both customized for a selective set of code words and do not yet generalize readily to other bosonic encoding schemes. 

A code-independent coupling mechanism between two otherwise isolated bosonic modes is a crucial ingredient for realizing  universal control on these logical qubits. The isolated cavities must be sufficiently detuned from each other to ensure the coherence of each mode and the absence of undesired cross-talk. Ref.~\cite{gao2018_programmable} demonstrated how the four-wave mixing process in a Josephson junction can provide a frequency converting bilinear coupling of the form \mbox{$H_{\mathrm{int}} (t)/\hbar = g(t) (e^{i\varphi}\hat{a}\hat{b}^{\dagger} + e^{-i\varphi}\hat{a}^{\dagger}\hat{b})$}. Here, $\hat{a}, \hat{b}$ are the annihilation operations associated with each of the cavities, the time-dependent coefficient $g(t)$ is the coupling strength, and $\varphi$ is the relative phase between the two microwave drives. The coefficient $g(t)$ depends on the effective amplitudes of the drives, which satisfy the frequency matching condition \mbox{$|\omega_2 - \omega_1| = |\omega_a - \omega_b|$}. Most notably, this coupling can be programmed to implement an identity, a 50:50 beamsplitter, or a full SWAP operation between the stationary microwave fields in the cavities by simply adjusting the duration of the evolution. Such an engineered coupling provides a powerful tool for implementing programmable interferometry between cavity states~\cite{gao2018_programmable}, which is a key building block for realizing various continuous-variable information processing tasks such as boson sampling~\cite{aaronson2011_computational}, simulation of vibrational quantum dynamics of molecules \cite{huh2015_boson, sparrow2018_simulating, clements2018_approximating, wang2020_efficient}, and distributed quantum sensing~\cite{zhuang2018_distributed, zhuang2020_distributed, xia2020_demonstration, noh2020_encoding}. 

Moreover, this bilinear coupling is also a valuable resource for enacting gates on two logical elements encoded in GKP states~\cite{gottesman2001_encoding}. As mentioned in Sec.~\ref{sec:codes}, GKP encodings rely on the non-linearity of the code words and only require linear or bilinear operations for universal control \cite{baragiola2019_allgaussian}. Therefore, this engineered bilinear coupling provides a simple and effective strategy for implementing a deterministic entangling operation for the GKP code. 

For other bosonic codes, this bilinear interaction is not alone sufficient to generate a universal gate set, which requires at least one entangling gate. In this case, the exponential SWAP (eSWAP) operation can be designed to provide deterministic and code-independent entanglement~\cite{lau2016_universal}. The eSWAP operation, akin to an exchange operation between spins, implements a programmable unitary of the form:
\begin{linenomath}
\begin{equation}\label{eq:eswap}
    \hat{U}(\theta) = \cos{(\theta)} \hat{I} + \sin{(\theta)}\mathrm{SWAP},
\end{equation}
\end{linenomath}
where $\theta$ is the rotation angle on the ancilla and $\hat{I}$ is the identity operation. Intuitively, this unitary implements a weighted superposition of the identity and SWAP operations between two bosonic modes regardless of their specific encodings. The eSWAP unitary has been realized in Ref.~\cite{gao2019_entanglement} between two bosonic modes housed in 3D co-axial cavities bridged by an ancilla. In this demonstration, an additional ancilla is introduced to one of the cavity modes and the resultant dispersive coupling is used to enact a $\hat{U}_{\mathrm{CPS}}$ operation to provide the tunable rotation necessary for the eSWAP unitary. The eSWAP unitary is then enacted on several encoding schemes in the Fock, coherent, and binomial basis. The availability of such a deterministic and code-independent entangling operation is a crucial step towards universal quantum computation using bosonic logical qubits. A recent study has shown that universal control and operations on tens of bosonic qubits can be achieved in a novel architecture comprising a single transmon coupled simultaneously to a multi-mode superconducting cavity~\cite{chakram2020_seamless}.

\subsection{Implementations of QEC}\label{subsection:qec}

In general, implementations of QEC with bosonic codes suffer from an initial rise in error rates due to the presence of and need to control multi-photon states. Hence, the main experimental challenge is to achieve an enhancement in the lifetime of the encoded qubit despite this initial penalty. In the context of bosonic codes, the break-even point is defined relative to the $|0, 1\rangle$ Fock states, which are the longest-lived physical elements in the cavity without error correction. Beyond the break-even point, we can be confident that the chosen QEC protocol does not introduce more errors into the system, thus attaining an improvement in the lifetime of the logical qubit. Till date, three studies have approached~\cite{campagne-ibarcq2019_quantum, hu2019_quantum} or achieved~\cite{ofek2016_extending} the `break-even point' for QEC with bosonic codes in cQED devices without any post-selection. 

Broadly, QEC schemes fall into three categories based on how they afford protection to the logical qubit. In active QEC, error syndromes are repeatedly measured during the state evolution of the logical qubit and any errors detected are subsequently corrected based on the measurement outcomes. In autonomous QEC, errors are removed by tailored dissipation or by coupling to an auxiliary system, without repeatedly probing the logical qubit. In passive QEC, the logical information is intrinsically protected from decoherence because of specifically designed physical symmetries~\cite{terhal2015_quantum}. In this discussion, we have opted to distinguish between autonomous and passive QEC to highlight potential differences in system design. In the autonomous approach, the environment is intentionally engineered to suppress or mitigate errors. Whereas in the passive case, the system Hamiltonian itself is tailored to be robust against certain errors, which often involves constructing a unique physical element in the hardware~\cite{gyenis2019_experimental,  grimm2020_stabilization, smith2020_magnifying}.

\begin{figure}[h!]
  \centering
  \includegraphics[scale=1]{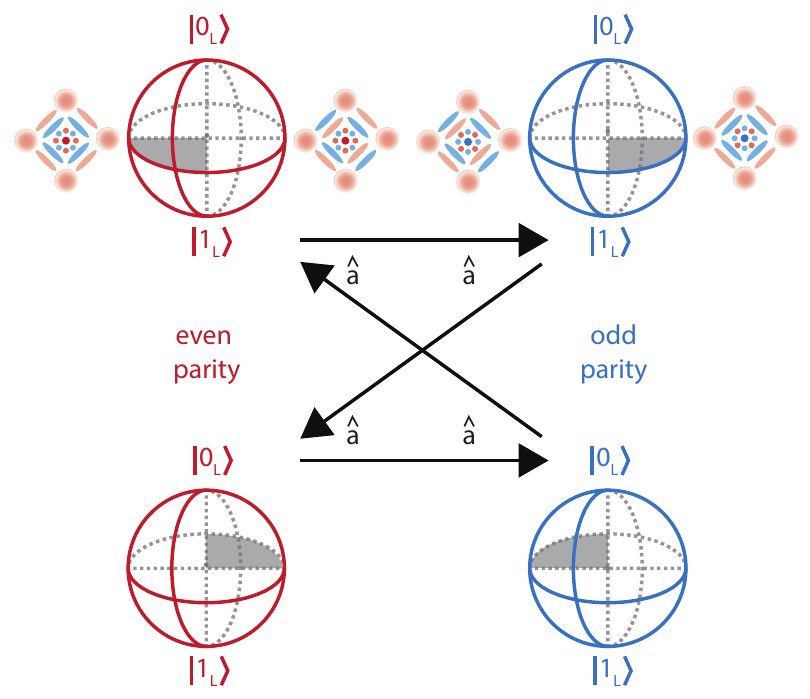}
  \caption{\textbf{Four-component cat code under photon loss.} Every photon loss event ($\hat{a}$) changes not only the parity of the basis states, but also changes the phase relationship between them. The encoded state cycles between the even (logical) and odd (error) parity subspaces, while also rotating about the Z-axis by $\pi/2$. The decoding sequence must take both these effects into account to correctly recover the logical information.}
  \label{fig:catcycle}
\end{figure}

Typically, active QEC requires robust measurements of the error syndrome and real-time feedback. In superconducting cavities, the dominant source of error is single photon loss. For encoding schemes with rotational symmetries, such as the cat and binomial codes, single photon loss results in a flip in the parity of the code words. Therefore, measuring the parity operator tells us whether a photon jump has taken place, thus allowing us to detect the error syndrome of these logical qubits~\cite{sun2014_tracking}. In Ref.~\cite{ofek2016_extending}, which corrected a four-component cat state under photon loss (Fig.~\ref{fig:catcycle}), the correct logical state was recovered by making appropriate adjustments in the decoding step based on the number of parity flips detected with real-time feedback. The corrected logical qubit showed an improved lifetime compared to both the uncorrected state and the $|0, 1\rangle$ Fock state encoding, thereby achieving the break-even point. However, this error correction strategy is not suitable for protecting bosonic qubits for timescales exceeding the intrinsic cavity lifetime as the loss of energy from the system is accounted for but not physically rectified. Thus, such energy attenuation should be physically compensated to significantly surpass the break-even point.

In contrast, Ref.~\cite{hu2019_quantum} admits a photon pumping operation to achieve QEC on a logical qubit encoded in the binomial code. The errors are detected by photon number parity measurements, as in Ref.~\cite{ofek2016_extending}. The errors occurring on the logical state are then corrected by an appropriate recovery operation as soon as they are detected. An approximated recovery operation is still required in case no errors are detected as the system evolves under the no-parity-jump operator~\cite{michael2016_new}. In this experiment, the lifetime of the logical qubit was greater than that of the uncorrected binomial code state, but was marginally below that of the $|0, 1\rangle$ Fock state.

In addition to the cat and binomial encodings, active QEC has also been recently demonstrated on a high-quality GKP state stored in a superconducting cavity~\cite{campagne-ibarcq2019_quantum}. As explained in Sec.~\ref{subsection:translation symmetric codes}, errors occurring on a moderately squeezed GKP state simply manifest as displacements of the cavity state, and are revealed and mitigated by measuring the displacement stabilizers. While mitigating errors is relatively straightforward, experimental challenges in implementing the GKP code lie in preparing finitely squeezed approximate GKP states and performing modulo quadrature measurement of stabilizers. The QEC protocol used in Ref.~\cite{campagne-ibarcq2019_quantum} suppresses all logical errors for a GKP state by alternating between two peak-sharpening and envelope-trimming rounds, each consisting of different conditional displacements on the cavity. The magnitude of these conditional displacements is dependent on the measured state of the ancilla onto which displacement stabilizers are mapped. This sharpen-and-trim technique can be generalized based on the second-order Trotter formula~\cite{royer2020_stabilization}.

An alternative strategy for QEC that does not require real-time feedback based on measurement outcomes involves engineering the dynamics of the system to correct errors autonomously. In practice, autonomous QEC can be achieved by introducing a tailored dissipation~\cite{leghtas2015_confining} such that the logical subspace is stabilized via controlled interactions with the environment. In Refs.~\cite{leghtas2015_confining, touzard2018_coherent}, the desired dissipation process subtracts photons in pairs from the bosonic mode via four-wave mixing in the Josephson junction of the ancilla. This process defines a stabilized manifold of steady states spanned by superpositions of coherent states (two-component cat states). The relative strength between the engineered two-photon dissipation and the intrinsic single photon loss of the cavity defines the separation of the two components of the cat state, which determines the extent of protection against bit-flip errors. In a more recent study, significant improvements in the suppression of bit-flip errors were demonstrated by using a tailored multi-junction ancilla to enact the two-photon dissipation process~\cite{lescanne2020_exponential}.

Similarly, an engineered four-photon dissipation achieved by an eight-wave mixing process, can protect four-component cat codes against dephasing~\cite{mundhada2019_experimental}. However, this strategy only accounts for dephasing errors and does not recover logical errors caused by photon loss. In contrast, in another autonomous QEC implementation with a truncated 4-cat state~\cite{gertler2020_protecting}, a photon was added to the cavity upon detecting a photon loss using a synthetic dissipation operator. This work effectively corrects photon loss events but does not account for dephasing errors. Similarly, another study performed autonomous QEC against photon loss with the kitten code by triggering an engineered jump operation~\cite{ma2020_error-transparent}. This operation recovered the code states whenever the system entered an error state without the need for any external probing. Although these current state-of-the-art demonstrations of autonomous QEC techniques only address one type of errors, they provide convincing evidence for the viability of the respective bosonic encoding schemes, thus paving the way for higher-order QEC protocols that can protect logical information against both photon loss and dephasing errors. 

Finally, QEC with bosonic codes can also be realized using a passive approach, for instance, by designing logical qubits with highly biased-noise channels to provide intrinsic protection without the need for probing any error syndromes. Very recently, two studies~\cite{lescanne2020_exponential, grimm2020_stabilization} verified that the bias between phase and bit-flip errors increases exponentially with $\alpha$, the size of the coherent state components of the 2-cat code. This strong asymmetry between the different types of errors can be exploited to significantly reduce the hardware overhead for fault-tolerant quantum computation~\cite{tuckett2018-ultrahigh,tuckett2019_tailoring,tuckett2020_faulttolerant}.

\section{Towards Fault-tolerant Quantum Computation with Bosonic Modes}\label{sec:FT}

A crucial consideration for a successful QEC implementation is that even the attempts to correct errors can themselves be erroneous. Thus, a key requirement for realizing `fault-tolerant' quantum computation is to carefully design error correction circuits such that faults that occur during the execution of a QEC protocol are also tolerated~\cite{gottesman2009_introduction}. A simple example of fault-tolerant circuit design is the transversal implementation of logical operations in the conventional paradigm of quantum error correction with multiple two-level systems. For instance, by implementing logical gates transversally for distance-$3$ codes that can correct any single qubit errors, one can ensure that a single fault in the circuit induces at most a single qubit error in any logical code block which remains correctable by the code~\cite{aliferis2006_quantum}.  

One key challenge in achieving fault-tolerant operations on bosonic qubits in the cat and binomial encodings is the propagation of uncorrectable errors due to ancilla decoherence~\cite{ofek2016_extending, hu2019_quantum}. The ancilla, often in the form of a transmon, is employed during state preparation, gates, measurements, and error correction on the encoded qubit via the natural dispersive coupling. Spontaneous emission of the ancilla during these processes induces dephasing on the logical qubit, thus irreversibly corrupting the encoded information. Furthermore, two-photon loss and Kerr effects also limit the effectiveness of these QEC implementations. One technique used to mitigate these imperfections is to optimize the cadence of the parity measurements~\cite{ofek2016_extending}. Frequent parity measurements allow more accurate tracking of the error syndrome but increase the likelihood of the ancilla relaxing during measurement and thereby dephasing the cavity. On the other hand, a larger interval between two measurements not only increases the probability of two-photon loss, but also results in a higher phase accumulation due to self-Kerr interactions. As the self-Kerr operator does not commute with the annihilation operator, photon jumps cause cavity dephasing. In Ref.~\cite{ofek2016_extending}, the phase accumulated due to the self-Kerr effect was first estimated and then accounted for by adjusting the phase of subsequent drive pulses. 

Several approaches have been devised to reduce the propagation of uncorrectable errors from the ancilla to the encoded information. In Ref.~\cite{rosenblum2018_fault-tolerant}, the third energy level of the ancilla is employed in the parity-mapping process to suppress the dephasing of the bosonic mode. A similar strategy can also be applied to general cavity operations, as shown in Ref.~\cite{reinhold2020_error-corrected}. In these schemes, additional levels of the ancilla as well as carefully engineered cavity-ancilla interactions are introduced to prevent incoherent evolution of the bosonic mode in the event of ancilla decay~\cite{ma2020_pathindependent}. Alternatively, error-transparent gates have been realized for binomial encodings~\cite{ma2020_error-transparent} by cleverly engineering the evolution of the bosonic qubit in both the code and error space.

For GKP codes, early discussions on fault-tolerance were focused on whether a small shift error occurring in the middle of an error correction circuit remains small throughout the entire circuit~\cite{gottesman2001_encoding,glancy2006_error}. Recently, numerous studies have approached fault-tolerant quantum computation with the GKP code by combining the GKP code with conventional error correction schemes based on multiple two-level systems that are already shown to be fault-tolerant. In performing such a concatenation of codes, additional analog information acquired during the error correction protocol can significantly boost the performance of a next-level QEC scheme based on multiple error-corrected GKP qubits~\cite{fukui2017_analog, fukui2018_tracking}. For instance, studies have shown that the fault-tolerance thresholds of the circuit-based toric code and the surface code can be increased by using the additional analog information~\cite{vuillot2019_quantum, noh2020_faulttolerant}. Another way to realize fault-tolerant quantum computing with the GKP code is to combine the GKP code with the idea of fault-tolerant measurement-based quantum computation using cluster states, for which the additional analog information also plays an important role~\cite{menicucci2014_faulttolerant, fukui2018_highthreshold, walshe2019_robust, fukui2019_highthreshold, yamasaki2020_polylog, bourassa2020_blueprint, larsen2021_faulttolerant}. From the circuit-based computing studies, we know that fault-tolerant quantum computing with the GKP code is possible if the squeezing of the GKP states is larger than $11$dB. Whereas from the measurement-based computing scenarios, a more favorable threshold value of $8$dB can be obtained by taking advantage of post-selection. However, as the measurement-based schemes use post-selection, they require a higher resource overhead. Recently, Ref. \cite{shi2019_fault-tolerant} made the phase estimation protocol to prepare GKP states in Ref.~\cite{terhal2016_encoding} fault-tolerant by using the concept of flag qubits~\cite{chao2018_quantum, chamberland2018_flag}.      

There have also been general studies on the fault-tolerance of biased-noise qubits~\cite{aliferis2008_faulttolerant, tuckett2018-ultrahigh, tuckett2019_tailoring, tuckett2020_faulttolerant}. These studies highlight the potential of achieving higher fault-tolerance thresholds and lower resource overheads by taking advantage of the noise bias. Refs.~\cite{guillaud2019_repetition,guillaud2020_error} suggested pieceable fault-tolerant~\cite{yoder2016_universal} Toffoli circuits while Ref.~\cite{puri2020_bias-preserving} offered a magic state distillation scheme tailored to biased-noise models~\cite{webster2015_reducing} as a means of achieving fault-tolerant universal quantum computation with biased-noise 2-cat qubits. A detailed roadmap for realizing biased-noise 2-cat qubits using acoustic nanomechanical resonators as well as alternative fault-tolerant universal quantum computation schemes were recently proposed~\cite{chamberland2020_building}.

\section{Outlook and Perspectives}\label{sec:future}

Continuous variables have long been recognized as promising candidates for the efficient encoding and processing of quantum information~\cite{lloyd1999_quantum}. Bosonic codes leverage the advantages afforded by continuous variable quantum devices to realize hardware-efficient QEC and pave the way towards the eventual realization of a fault-tolerant quantum computer. In this review, we have presented some of the most compelling results that attest to the viability of the cQED platform. In particular, many milestones, from reaching the break-even point to implementing logical operations between bosonic logical qubits, have thus far been demonstrated using 3D cQED devices. An important next milestone would be to demonstrate the break-even point for logical operations, i.e., to achieve gate fidelities with error-correctable bosonic qubits that are higher than the highest gate fidelity attainable with the best physical element without error correction. Achieving the break-even point in both QEC and single or two-mode logical operations forms a critical foundation upon which fault-tolerant gates and algorithms can be constructed.

We further emphasize that the techniques and methodologies developed in these 3D cQED implementations of bosonic codes are agnostic to the hardware architecture. They can therefore be readily adapted to more compact designs, such as planar or 2.5D devices, as their performance improves with further developments in materials engineering and new fabrication techniques~\cite{place2020_new}. 

Crucially, encoding logical information in bosonic codes is naturally compatible with a modular approach to scalability. In the modular architecture, each logical element can be encoded, protected, and optimized individually, and then connected to other elements via on-demand communication channels~\cite{jiang2007_distributed, kimble2008_quantum, monroe2014_large-scale}. In contrast to directly wiring up an increasing number of physical elements, modularity has the benefit of reduced cross-talk, robustness against local failure modes, and enhanced re-configurability. Encoding and manipulating individual logical qubits encoded in cavities is a crucial primitive for eventually realizing a large-scale modular quantum device. In addition, recent experiments have also demonstrated other key elements of a modular architecture, including programmable quantum communication and entanglement generation between two distant bosonic modes~\cite{axline2018_on-demand, campagne-ibarcq2018_deterministic, burkhart2020_error-detected} as well as a teleported CNOT gate~\cite{chou2018_deterministic}. 

As we continue to improve the performance and scale of these bosonic quantum systems on the hardware level, we must also expand our technological repertoire to effectively control and characterize these more complex devices. This expansion may involve, for instance, the development of more sophisticated measurement electronics and protocols as well as new theoretical frameworks to capture the signatures of multipartite entanglement. The small-scale bosonic devices developed so far provide an indispensable platform to test and refine these crucial elements.

In recent years, cQED devices have become a workhorse for the implementations of quantum error correction and quantum information processing. Alongside the developments in cQED, we are also witnessing remarkable progress with other platforms based on trapped-ions~\cite{landsman2019_two-qubit, egan2020_fault-tolerant}, silicon spins~\cite{veldhorst2015_two-qubit, watson2018_programmable}, neutral atoms~\cite{levine2019_parallel}, etc, from both academic and industrial initiatives. While the ultimate hardware to achieve fault-tolerance may not consist of millions of 3D cavities, the insights gained from and the technical achievements demonstrated by the results summarized in this article will serve as crucial building blocks for the eventual realization of a scalable and robust universal quantum computer. 

\let\oldaddcontentsline\addcontentsline% Store \addcontentsline
\renewcommand{\addcontentsline}[3]{}% Make \addcontentsline a no-op
\section*{Acknowledgments}\label{sec:acknowledgments}

The authors thank S. M. Girvin and L. Sun for fruitful discussions and comments on the manuscript. Yvonne Y. Gao acknowledges funding from the Singapore National Research Foundation Fellowship (Class of 2020). This work is also supported by the Ministry of Education in Singapore.

\let\addcontentsline\oldaddcontentsline% Restore \addcontentsline

%\newpage % to begin references on a new page

%\section*{References}
\bibliographystyle{unsrt} 
\bibliography{references.bib}        %use a bibtex bibliography file ref_lib.bib

\begin{thebibliography}{100}

\bibitem{shor1995_scheme}
P.~W. Shor.
\newblock Scheme for reducing decoherence in quantum computer memory.
\newblock {\em Phys. Rev. A}, \textbf{52}(4):R2493--R2496, 1995.
\newblock \url{https://doi.org/10.1103/PhysRevA.52.R2493}.

\bibitem{steane1996_error}
A.~M. Steane.
\newblock Error correcting codes in quantum theory.
\newblock {\em Phys. Rev. Lett.}, \textbf{77}(5):793--797, 1996.
\newblock \url{https://doi.org/10.1103/PhysRevLett.77.793}.

\bibitem{preskill1998_reliable}
J.~Preskill.
\newblock Reliable quantum computers.
\newblock {\em Proceedings of the Royal Society of London. Series A:
  Mathematical, Physical, and Engineering Sciences},
  \textbf{454}(1969):385--410, 1998.
\newblock \url{https://doi.org/10.1098/rspa.1998.0167}.

\bibitem{chiaverini2004_realization}
J.~Chiaverini, D.~Leibfried, T.~Schaetz, M.~D. Barrett, R.~B. Blakestad,
  J.~Britton, W.~M. Itano, J.~D. Jost, E.~Knill, C.~Langer, R.~Ozeri, and D.~J.
  Wineland.
\newblock Realization of quantum error correction.
\newblock {\em Nature}, \textbf{432}(7017):602--605, 2004.
\newblock \url{https://doi.org/10.1038/nature03074}.

\bibitem{reed2012_realization}
M.~D. Reed, L.~DiCarlo, S.~E. Nigg, L.~Sun, L.~Frunzio, S.~M. Girvin, and R.~J.
  Schoelkopf.
\newblock Realization of three-qubit quantum error correction with
  superconducting circuits.
\newblock {\em Nature}, \textbf{482}(7385):382--385, 2012.
\newblock \url{https://doi.org/10.1038/nature10786}.

\bibitem{taminiau2014_universal}
T.~H. Taminiau, J.~Cramer, T.~van~der Sar, V.~V. Dobrovitski, and R.~Hanson.
\newblock Universal control and error correction in multi-qubit spin registers
  in diamond.
\newblock {\em Nature Nanotech}, \textbf{9}(3):171--176, 2014.
\newblock \url{https://doi.org/10.1038/nnano.2014.2}.

\bibitem{waldherr2014_quantum}
G.~Waldherr, Y.~Wang, S.~Zaiser, M.~Jamali, T.~Schulte-Herbr{\"u}ggen, H.~Abe,
  T.~Ohshima, J.~Isoya, J.~F. Du, P.~Neumann, and J.~Wrachtrup.
\newblock Quantum error correction in a solid-state hybrid spin register.
\newblock {\em Nature}, \textbf{506}(7487):204--207, 2014.
\newblock \url{https://doi.org/10.1038/nature12919}.

\bibitem{linke2017_faulttolerant}
N.~M. Linke, M.~Gutierrez, K.~A. Landsman, C.~Figgatt, S.~Debnath, K.~R. Brown,
  and C.~Monroe.
\newblock Fault-tolerant quantum error detection.
\newblock {\em Science Advances}, \textbf{3}(10):e1701074, 2017.
\newblock \url{https://doi.org/10.1126/sciadv.1701074}.

\bibitem{andersen2020_repeated}
C.~K. Andersen, A.~Remm, S.~Lazar, S.~Krinner, N.~Lacroix, G.~J. Norris,
  M.~Gabureac, C.~Eichler, and A.~Wallraff.
\newblock Repeated quantum error detection in a surface code.
\newblock {\em Nat. Phys.}, \textbf{16}(8):204--207, 2020.
\newblock \url{https://doi.org/10.1038/s41567-020-0920-y}.

\bibitem{erhard2021_entangling}
A.~Erhard, H.~P. Nautrup, M.~Meth, L.~Postler, R.~Stricker, M.~Stadler,
  V.~Negnevitsky, M.~Ringbauer, P.~Schindler, H.~J. Briegel, R.~Blatt,
  N.~Friis, and T.~Monz.
\newblock Entangling logical qubits with lattice surgery.
\newblock {\em Nature}, \textbf{589}(7841):220--224, 2021.
\newblock \url{https://doi.org/10.1038/s41586-020-03079-6}.

\bibitem{braunstein1998_quantum}
S.~L. Braunstein.
\newblock Quantum error correction for communication with linear optics.
\newblock {\em Nature}, \textbf{394}(6688):47--49, 1998.
\newblock \url{https://doi.org/10.1038/27850}.

\bibitem{bartlett2002_universal}
S.~D. Bartlett and B.~C. Sanders.
\newblock Universal continuous-variable quantum computation: requirement of
  optical nonlinearity for photon counting.
\newblock {\em Phys. Rev. A}, \textbf{65}(4):042304, 2002.
\newblock \url{https://doi.org/10.1103/PhysRevA.65.042304}.

\bibitem{bartlett2002_quantum}
S.~D. Bartlett, H.~de~Guise, and B.~C. Sanders.
\newblock Quantum encodings in spin systems and harmonic oscillators.
\newblock {\em Phys. Rev. A}, \textbf{65}(5):052316, 2002.
\newblock \url{https://doi.org/10.1103/PhysRevA.65.052316}.

\bibitem{devoret2013_superconducting}
M.~H. Devoret and R.~J. Schoelkopf.
\newblock Superconducting circuits for quantum information: an outlook.
\newblock {\em Science}, \textbf{339}(6124):1169--1174, 2013.
\newblock \url{https://doi.org/10.1126/science.1231930}.

\bibitem{blais2020_circuit}
A.~Blais, A.~L. Grimsmo, S.~M. Girvin, and A.~Wallraff.
\newblock Circuit quantum electrodynamics.
\newblock {\em arXiv:2005.12667 [quant-ph]}, 2020.
\newblock \url{https://arxiv.org/abs/2005.12667}.

\bibitem{ofek2016_extending}
N.~Ofek, A.~Petrenko, R.~Heeres, P.~Reinhold, Z.~Leghtas, B.~Vlastakis, Y.~Liu,
  L.~Frunzio, S.~M. Girvin, L.~Jiang, M.~Mirrahimi, M.~H. Devoret, and R.~J.
  Schoelkopf.
\newblock Extending the lifetime of a quantum bit with error correction in
  superconducting circuits.
\newblock {\em Nature}, \textbf{536}(7617):441--445, 2016.
\newblock \url{https://doi.org/10.1038/nature18949}.

\bibitem{heeres2017_implementing}
R.~W. Heeres, P.~Reinhold, N.~Ofek, L.~Frunzio, L.~Jiang, M.~H. Devoret, and
  R.~J. Schoelkopf.
\newblock Implementing a universal gate set on a logical qubit encoded in an
  oscillator.
\newblock {\em Nat. Commun.}, \textbf{8}(94), 2017.
\newblock \url{https://doi.org/10.1038/s41467-017-00045-1}.

\bibitem{chou2018_deterministic}
K.~S. Chou, J.~Z. Blumoff, C.~S. Wang, P.~C. Reinhold, C.~J. Axline, Y.~Y. Gao,
  L.~Frunzio, M.~H. Devoret, L.~Jiang, and R.~J. Schoelkopf.
\newblock Deterministic teleportation of a quantum gate between two logical
  qubits.
\newblock {\em Nature}, \textbf{561}(7723):368--373, 2018.
\newblock \url{https://doi.org/10.1038/s41586-018-0470-y}.

\bibitem{gao2019_entanglement}
Y.~Y. Gao, B.~J. Lester, K.~S. Chou, L.~Frunzio, M.~H. Devoret, L.~Jiang, S.~M.
  Girvin, and R.~J. Schoelkopf.
\newblock Entanglement of bosonic modes through an engineered exchange
  interaction.
\newblock {\em Nature}, \textbf{566}(7745):509--512, 2019.
\newblock \url{https://doi.org/10.1038/s41586-019-0970-4}.

\bibitem{xu2020_demonstration}
Y.~Xu, Y.~Ma, W.~Cai, X.~Mu, W.~Dai, W.~Wang, L.~Hu, X.~Li, J.~Han, H.~Wang,
  Y.~P. Song, Z.~B. Yang, S.~B. Zheng, and L.~Sun.
\newblock Demonstration of controlled-phase gates between two error-correctable
  photonic qubits.
\newblock {\em Phys. Rev. Lett.}, \textbf{124}(12):120501, 2020.
\newblock \url{https://doi.org/10.1103/PhysRevLett.124.120501}.

\bibitem{reinhold2020_error-corrected}
P.~Reinhold, S.~Rosenblum, W.~Ma, L.~Frunzio, L.~Jiang, and R.~J. Schoelkopf.
\newblock Error-corrected gates on an encoded qubit.
\newblock {\em Nat. Phys.}, \textbf{16}(8):822--826, 2020.
\newblock \url{https://doi.org/10.1038/s41567-020-0931-8}.

\bibitem{rosenblum2018_fault-tolerant}
S.~Rosenblum, P.~Reinhold, M.~Mirrahimi, L.~Jiang, L.~Frunzio, and R.~J.
  Schoelkopf.
\newblock Fault-tolerant detection of a quantum error.
\newblock {\em Science}, \textbf{361}(6399):266--270, 2018.
\newblock \url{https://doi.org/10.1126/science.aat3996}.

\bibitem{knill1997_theory}
E.~Knill and R.~Laflamme.
\newblock Theory of quantum error-correcting codes.
\newblock {\em Phys. Rev. A}, \textbf{55}(2):900--911, 1997.
\newblock \url{https://doi.org/10.1103/PhysRevA.55.900}.

\bibitem{nielsen2000_quantum}
M.~A. Nielsen and I.~L. Chuang.
\newblock {\em Quantum Computation and Quantum Information}.
\newblock Cambridge University Press, 2000.
\newblock \url{https://doi.org/10.1017/CBO9780511976667}.

\bibitem{michael2016_new}
M.~H. Michael, M.~Silveri, R.~T. Brierley, V.~V. Albert, J.~Salmilehto,
  L.~Jiang, and S.~M. Girvin.
\newblock New class of quantum error-correcting codes for a bosonic mode.
\newblock {\em Phys. Rev. X}, \textbf{6}(3):031006, 2016.
\newblock \url{https://doi.org/10.1103/PhysRevX.6.031006}.

\bibitem{leung1997_approximate}
D.~W. Leung, M.~A. Nielsen, I.~L. Chuang, and Y.~Yamamoto.
\newblock Approximate quantum error correction can lead to better codes.
\newblock {\em Phys. Rev. A}, \textbf{56}(4):2567--2573, 1997.
\newblock \url{https://doi.org/10.1103/PhysRevA.56.2567}.

\bibitem{girvin2019_quantum}
S.~M. Girvin.
\newblock Quantum superconducting circuits and error correction.
\newblock In {\em Les Houches Summer School ``Quantum information machines"},
  2019.
\newblock \url{https://youtu.be/Na_DB5o4QNY}.

\bibitem{sheldon2016_procedure}
S.~Sheldon, E.~Magesan, J.~M. Chow, and J.~M. Gambetta.
\newblock Procedure for systematically tuning up cross-talk in the
  cross-resonance gate.
\newblock {\em Phys. Rev. A}, \textbf{93}(6):060302, 2016.
\newblock \url{https://doi.org/10.1103/PhysRevA.93.060302}.

\bibitem{rol2019_fast}
M.~A. Rol, F.~Battistel, F.~K. Malinowski, C.~C. Bultink, B.~M. Tarasinski,
  R.~Vollmer, N.~Haider, N.~Muthusubramanian, A.~Bruno, B.~M. Terhal, and
  L.~DiCarlo.
\newblock Fast, high-fidelity conditional-phase gate exploiting leakage
  interference in weakly anharmonic superconducting qubits.
\newblock {\em Phys. Rev. Lett.}, \textbf{123}(12):120502, 2019.
\newblock \url{https://doi.org/10.1103/PhysRevLett.123.120502}.

\bibitem{ouyang2014_permutationinvariant}
Y.~Ouyang.
\newblock Permutation-invariant quantum codes.
\newblock {\em Phys. Rev. A}, \textbf{90}(6):062317, 2014.
\newblock \url{https://doi.org/10.1103/PhysRevA.90.062317}.

\bibitem{ouyang2016_permutationinvariant}
Y.~Ouyang and J.~Fitzsimons.
\newblock Permutation-invariant codes encoding more than one qubit.
\newblock {\em Phys. Rev. A}, \textbf{93}(4):042340, 2016.
\newblock \url{https://doi.org/10.1103/PhysRevA.93.042340}.

\bibitem{ouyang2017_permutationinvariant}
Y.~Ouyang.
\newblock Permutation-invariant qudit codes from polynomials.
\newblock {\em Linear Algebra and its Applications}, \textbf{532}:43--59, 2017.
\newblock \url{https://doi.org/10.1016/j.laa.2017.06.031}.

\bibitem{ouyang2020_permutationinvariant}
Y.~Ouyang and R.~Chao.
\newblock Permutation-invariant constant-excitation quantum codes for amplitude
  damping.
\newblock {\em IEEE Transactions on Information Theory},
  \textbf{66}(5):2921--2933, 2020.
\newblock \url{https://doi.org/10.1109/TIT.2019.2956142}.

\bibitem{wasilewski2007_protecting}
W.~Wasilewski and K.~Banaszek.
\newblock Protecting an optical qubit against photon loss.
\newblock {\em Phys. Rev. A}, \textbf{75}(4):042316, 2007.
\newblock \url{https://doi.org/10.1103/PhysRevA.75.042316}.

\bibitem{kapit2016_hardwareefficient}
E.~Kapit.
\newblock Hardware-efficient and fully autonomous quantum error correction in
  superconducting circuits.
\newblock {\em Phys. Rev. Lett.}, \textbf{116}(15):150501, 2016.
\newblock \url{https://doi.org/10.1103/PhysRevLett.116.150501}.

\bibitem{bergmann2016_quantum}
M.~Bergmann and P.~van Loock.
\newblock Quantum error correction against photon loss using multicomponent cat
  states.
\newblock {\em Phys. Rev. A}, \textbf{94}(4):042332, 2016.
\newblock \url{https://doi.org/10.1103/PhysRevA.94.042332}.

\bibitem{kapit2018_errortransparent}
E.~Kapit.
\newblock Error-transparent quantum gates for small logical qubit
  architectures.
\newblock {\em Phys. Rev. Lett.}, \textbf{120}(5):050503, 2018.
\newblock \url{https://doi.org/10.1103/PhysRevLett.120.050503}.

\bibitem{reagor2013_reaching}
M.~Reagor, H.~Paik, G.~Catelani, L.~Sun, C.~Axline, E.~Holland, I.~M. Pop,
  N.~A. Masluk, T.~Brecht, L.~Frunzio, M.~H. Devoret, L.~Glazman, and R.~J.
  Schoelkopf.
\newblock Reaching 10 ms single photon lifetimes for superconducting aluminum
  cavities.
\newblock {\em Appl. Phys. Lett.}, \textbf{102}(19):192604, 2013.
\newblock \url{https://doi.org/10.1063/1.4807015}.

\bibitem{lescanne2020_exponential}
R.~Lescanne, M.~Villiers, T.~Peronnin, A.~Sarlette, M.~Delbecq, B.~Huard,
  T.~Kontos, M.~Mirrahimi, and Z.~Leghtas.
\newblock Exponential suppression of bit-flips in a qubit encoded in an
  oscillator.
\newblock {\em Nat. Phys.}, \textbf{16}(5):509--513, 2020.
\newblock \url{https://doi.org/10.1038/s41567-020-0824-x}.

\bibitem{grimsmo2020_quantum}
A.~L. Grimsmo, J.~Combes, and B.~Q. Baragiola.
\newblock Quantum computing with rotation-symmetric bosonic codes.
\newblock {\em Phys. Rev. X}, \textbf{10}(1):011058, 2020.
\newblock \url{https://doi.org/10.1103/PhysRevX.10.011058}.

\bibitem{ouyang2020_tradeoffs}
Y.~Ouyang and E.~Campbell.
\newblock Trade-offs on number and phase shift resilience in bosonic quantum
  codes.
\newblock {\em arXiv:2008.12576 [quant-ph]}, 2020.
\newblock \url{https://arxiv.org/abs/2008.12576}.

\bibitem{lihm2018_implementation}
J.~M. Lihm, K.~Noh, and U.~R. Fischer.
\newblock Implementation-independent sufficient condition of the
  {K}nill-{L}aflamme type for the autonomous protection of logical qudits by
  strong engineered dissipation.
\newblock {\em Phys. Rev. A}, \textbf{98}(1):012317, 2018.
\newblock \url{https://doi.org/10.1103/PhysRevA.98.012317}.

\bibitem{leghtas2013_hardware-efficient}
Z.~Leghtas, G.~Kirchmair, B.~Vlastakis, R.~J. Schoelkopf, M.~H. Devoret, and
  M.~Mirrahimi.
\newblock Hardware-efficient autonomous quantum memory protection.
\newblock {\em Phys. Rev. Lett.}, \textbf{111}(12):120501, 2013.
\newblock \url{https://doi.org/10.1103/PhysRevLett.111.120501}.

\bibitem{mirrahimi2014_dynamically}
M.~Mirrahimi, Z.~Leghtas, V.~V. Albert, S.~Touzard, R.~J. Schoelkopf, L.~Jiang,
  and M.~H. Devoret.
\newblock Dynamically protected cat-qubits: a new paradigm for universal
  quantum computation.
\newblock {\em New J. Phys.}, \textbf{16}(4):045014, 2014.
\newblock \url{https://doi.org/10.1088/1367-2630/16/4/045014}.

\bibitem{li2017_cat}
L.~Li, C.~L. Zou, V.V. Albert, S.~Muralidharan, S.~M. Girvin, and L.~Jiang.
\newblock Cat codes with optimal decoherence suppression for a lossy bosonic
  channel.
\newblock {\em Phys. Rev. Lett.}, \textbf{119}(3):030502, 2017.
\newblock \url{https://doi.org/10.1103/PhysRevLett.119.030502}.

\bibitem{albert2018_performance}
V.~V. Albert, K.~Noh, K.~Duivenvoorden, D.~J. Young, R.~T. Brierley,
  P.~Reinhold, C.~Vuillot, L.~Li, C.~Shen, S.~M. Girvin, B.~M. Terhal, and
  L.~Jiang.
\newblock Performance and structure of single-mode bosonic codes.
\newblock {\em Phys. Rev. A}, \textbf{97}(3):032346, 2018.
\newblock \url{https://doi.org/10.1103/PhysRevA.97.032346}.

\bibitem{albert2016_holonomic}
V.~V. Albert, C.~Shu, S.~Krastanov, C.~Shen, R.~B. Liu, Z.~B. Yang, R.~J.
  Schoelkopf, M.~Mirrahimi, M.~H. Devoret, and L.~Jiang.
\newblock Holonomic quantum control with continuous variable systems.
\newblock {\em Phys. Rev. Lett.}, \textbf{116}(14):140502, 2016.
\newblock \url{https://doi.org/10.1103/PhysRevLett.116.140502}.

\bibitem{albert2019_pair-cat}
V.~V. Albert, S.~O. Mundhada, A.~Grimm, S.~Touzard, M.~H. Devoret, and
  L.~Jiang.
\newblock Pair-cat codes: autonomous error-correction with low-order
  nonlinearity.
\newblock {\em Quantum Sci. Technol.}, \textbf{4}(3):035007, 2019.
\newblock \url{https://doi.org/10.1088/2058-9565/ab1e69}.

\bibitem{gottesman2001_encoding}
D.~Gottesman, A.~Kitaev, and J.~Preskill.
\newblock Encoding a qubit in an oscillator.
\newblock {\em Phys. Rev. A}, \textbf{64}(1):012310, 2001.
\newblock \url{https://doi.org/10.1103/PhysRevA.64.012310}.

\bibitem{terhal2016_encoding}
B.~M. Terhal and D.~Weigand.
\newblock Encoding a qubit into a cavity mode in circuit {QED} using phase
  estimation.
\newblock {\em Phys. Rev. A}, \textbf{93}(1):012315, 2016.
\newblock \url{https://doi.org/10.1103/PhysRevA.93.012315}.

\bibitem{matsuura2019_equivalence}
T.~Matsuura, H.~Yamasaki, and M.~Koashi.
\newblock Equivalence of approximate {G}ottesman-{K}itaev-{P}reskill codes.
\newblock {\em Phys. Rev. A}, \textbf{102}(3):032408, 2020.
\newblock \url{https://doi.org/10.1103/PhysRevA.102.032408}.

\bibitem{travaglione2002_preparing}
B.~C. Travaglione and G.~J. Milburn.
\newblock Preparing encoded states in an oscillator.
\newblock {\em Phys. Rev. A}, \textbf{66}(5):052322, 2002.
\newblock \url{https://doi.org/10.1103/PhysRevA.66.052322}.

\bibitem{pirandola2004_constructing}
S.~Pirandola, S.~Mancini, S.~Vitali, and P.~Tombesi.
\newblock Constructing finite-dimensional codes with optical continuous
  variables.
\newblock {\em EPL}, \textbf{68}(3):323, 2004.
\newblock \url{https://doi.org/10.1209/epl/i2004-10203-9}.

\bibitem{pirandola2006_generating}
S.~Pirandola, S.~Mancini, D.~Vitali, and P.~Tombesi.
\newblock Generating continuous variable quantum codewords in the near-field
  atomic lithography.
\newblock {\em J. Phys. B: At. Mol. Opt. Phys.}, \textbf{39}(4):997, 2006.
\newblock \url{https://doi.org/10.1088/0953-4075/39/4/023}.

\bibitem{vasconcelos2010_alloptical}
H.~M. Vasconcelos, L.~Sanz, and S.~Glancy.
\newblock All-optical generation of states for “{E}ncoding a qubit in an
  oscillator”.
\newblock {\em Opt. Lett.}, \textbf{35}(19):3261--3263, 2010.
\newblock \url{https://doi.org/10.1364/OL.35.003261}.

\bibitem{motes2017_encoding}
K.~R. Motes, B.~Q. Baragiola, A.~Gilchrist, and N.~C. Menicucci.
\newblock Encoding qubits into oscillators with atomic ensembles and squeezed
  light.
\newblock {\em Phys. Rev. A}, \textbf{95}(5):053819, 2017.
\newblock \url{https://doi.org/10.1103/PhysRevA.95.053819}.

\bibitem{weigand2018_generating}
D.~J. Weigand and B.~M. Terhal.
\newblock Generating grid states from {S}chr\"odinger-cat states without
  postselection.
\newblock {\em Phys. Rev. A}, \textbf{97}(2):022341, 2018.
\newblock \url{https://doi.org/10.1103/PhysRevA.97.022341}.

\bibitem{arrazola2019_machine}
J.~M. Arrazola, T.R. Bromley, J.~Izaac, C.~R. Myers, K.~Br{\'{a}}dler, and
  N.~Killoran.
\newblock Machine learning method for state preparation and gate synthesis on
  photonic quantum computers.
\newblock {\em Quantum Sci. Technol.}, \textbf{4}(2):024004, 2019.
\newblock \url{https://doi.org/10.1088/2058-9565/aaf59e}.

\bibitem{shi2019_fault-tolerant}
Y.~Shi, C.~Chamberland, and A.~Cross.
\newblock Fault-tolerant preparation of approximate {GKP} states.
\newblock {\em New J. Phys.}, \textbf{21}(9):093007, 2019.
\newblock \url{https://doi.org/10.1088/1367-2630/ab3a62}.

\bibitem{su2019_conversion}
D.~Su, C.~R. Myers, and K.~K. Sabapathy.
\newblock Conversion of {G}aussian states to non-{G}aussian states using
  photon-number-resolving detectors.
\newblock {\em Phys. Rev. A}, \textbf{100}(5):052301, 2019.
\newblock \url{https://doi.org/10.1103/PhysRevA.100.052301}.

\bibitem{eaton2019_nongaussian}
M.~Eaton, R.~Nehra, and O.~Pfister.
\newblock Non-{G}aussian and {G}ottesman-{K}itaev-{P}reskill state preparation
  by photon catalysis.
\newblock {\em New J. Phys.}, \textbf{21}(11):113034, 2019.
\newblock \url{https://doi.org/10.1088/1367-2630/ab5330}.

\bibitem{hastrup2019_measurementfree}
J.~Hastrup, K.~Park, J.~B. Brask, R.~Filip, and U.~L. Andersen.
\newblock Measurement-free preparation of grid states.
\newblock {\em npj Quantum Information}, \textbf{7} number = {1}, pages =
  {1--8}, year = {2021}, note =
  {\url{https://doi.org/10.1038/s41534-020-00353-3}},.

\bibitem{weigand2020_realizing}
D.~J. Weigand and B.~M. Terhal.
\newblock Realizing modular quadrature measurements via a tunable
  photon-pressure coupling in circuit {QED}.
\newblock {\em Phys. Rev. A}, \textbf{101}(5):053840, 2020.
\newblock \url{https://doi.org/10.1103/PhysRevA.101.053840}.

\bibitem{hastrup2020_improved}
J.~Hastrup and U.~L. Andersen.
\newblock Improved readout of qubit-coupled {G}ottesman-{K}itaev-{P}reskill
  states.
\newblock {\em arXiv:2008.10531 [quant-ph]}, 2020.
\newblock \url{https://arxiv.org/abs/2008.10531}.

\bibitem{royer2020_stabilization}
B.~Royer, S.~Singh, and S.~M. Girvin.
\newblock Stabilization of finite-energy {G}ottesman-{K}itaev-{P}reskill
  states.
\newblock {\em Phys. Rev. Lett.}, \textbf{125}(26):260509, 2020.
\newblock \url{https://doi.org/10.1103/PhysRevLett.125.260509}.

\bibitem{wan2019_memoryassisted}
K.~H. Wan, A.~Neville, and W.~S. Kolthammer.
\newblock Memory-assisted decoder for approximate
  {G}ottesman-{K}itaev-{P}reskill codes.
\newblock {\em Phys. Rev. Research}, \textbf{2}(4):043280, 2020.
\newblock \url{https://doi.org/10.1103/PhysRevResearch.2.043280}.

\bibitem{tzitrin2020_progress}
I.~Tzitrin, J.~E. Bourassa, N.~C. Menicucci, and K.~K. Sabapathy.
\newblock Progress towards practical qubit computation using approximate
  {G}ottesman-{K}itaev-{P}reskill codes.
\newblock {\em Phys. Rev. A}, \textbf{101}(3):032315, 2020.
\newblock \url{https://doi.org/10.1103/PhysRevA.101.032315}.

\bibitem{terhal2020_scalable}
B.~M. Terhal, J.~Conrad, and C.~Vuillot.
\newblock Towards scalable bosonic quantum error correction.
\newblock {\em Quantum Sci. Technol.}, \textbf{5}(4):043001, 2020.
\newblock \url{https://doi.org/10.1088/2058-9565/ab98a5}.

\bibitem{pantaleoni2020_modular}
G.~Pantaleoni, B.~Q. Baragiola, and N.~C. Menicucci.
\newblock Modular bosonic subsystem codes.
\newblock {\em Phys. Rev. Lett.}, \textbf{125}(4):040501, 2020.
\newblock \url{https://doi.org/10.1103/PhysRevLett.125.040501}.

\bibitem{walshe2020_continuousvariable}
B.~W. Walshe, B.~Q. Baragiola, R.~N. Alexander, and N.~C. Menicucci.
\newblock Continuous-variable gate teleportation and bosonic-code error
  correction.
\newblock {\em Phys. Rev. A}, \textbf{102}(6):062411, 2020.
\newblock \url{https://doi.org/10.1103/PhysRevA.102.062411}.

\bibitem{mensen2020_phasespace}
L.~J. Mensen, B.~Q. Baragiola, and N.~C. Menicucci.
\newblock Phase-space methods for representing, manipulating, and correcting
  {G}ottesman-{K}itaev-{P}reskill qubits.
\newblock {\em arXiv:2012.12488 [quant-ph]}, 2020.
\newblock \url{https://arxiv.org/abs/2012.12488}.

\bibitem{fluhmann2019_encoding}
C.~Fluhmann, T.~L. Nguyen, M.~Marinelli, V.~Negnevitsky, K.~Mehta, and J.~P.
  Home.
\newblock Encoding a qubit in a trapped-ion mechanical oscillator.
\newblock {\em Nature}, \textbf{566}(7745):513--517, 2019.
\newblock \url{https://doi.org/10.1038/s41586-019-0960-6}.

\bibitem{deNeeve2020_error}
B.~de~Neeve, T.~L. Nguyen, T.~Behrle, and J.~Home.
\newblock Error correction of a logical grid state qubit by dissipative
  pumping.
\newblock {\em arXiv:2010.09681 [quant-ph]}, 2020.
\newblock \url{https://arxiv.org/abs/2010.09681}.

\bibitem{campagne-ibarcq2019_quantum}
P.~Campagne-Ibarcq, A.~Eickbusch, S.~Touzard, E.~Zalys-Geller, N.~E. Frattini,
  V.~V. Sivak, P.~Reinhold, S.~Puri, S.~Shankar, R.~J. Schoelkopf, L.~Frunzio,
  M.~Mirrahimi, and M.~H. Devoret.
\newblock Quantum error correction of a qubit encoded in grid states of an
  oscillator.
\newblock {\em Nature}, \textbf{584}(7821):368--372, 2020.
\newblock \url{https://doi.org/10.1038/s41586-020-2603-3}.

\bibitem{noh2019_quantum}
K.~Noh, V.~V. Albert, and L.~Jiang.
\newblock Quantum capacity bounds of {G}aussian thermal loss channels and
  achievable rates with {G}ottesman-{K}itaev-{P}reskill codes.
\newblock {\em IEEE Transactions on Information Theory},
  \textbf{65}(4):2563--2582, 2019.
\newblock \url{https://doi.org/10.1109/TIT.2018.2873764}.

\bibitem{gottesman2003_secure}
D.~Gottesman and J.~Preskill.
\newblock Secure quantum key distribution using squeezed states.
\newblock In {\em Quantum Information with Continuous Variables}, pages
  317--356. Springer, 2003.
\newblock \url{https://doi.org/10.1007/978-94-015-1258-9_22}.

\bibitem{bravyi2005_universal}
S.~Bravyi and A.~Kitaev.
\newblock Universal quantum computation with ideal {C}lifford gates and noisy
  ancillas.
\newblock {\em Phys. Rev. A}, \textbf{71}(2):022316, 2005.
\newblock \url{https://doi.org/10.1103/PhysRevA.71.022316}.

\bibitem{baragiola2019_allgaussian}
B.~Q. Baragiola, G.~Pantaleoni, R.~N. Alexander, A.~Karanjai, and N.~C.
  Menicucci.
\newblock All-{G}aussian universality and fault tolerance with the
  {G}ottesman-{K}itaev-{P}reskill code.
\newblock {\em Phys. Rev. Lett.}, \textbf{123}(20):200502, 2019.
\newblock \url{https://doi.org/10.1103/PhysRevLett.123.200502}.

\bibitem{hastrup2020_cubic}
J.~Hastrup, M.~V. Larsen, J.~S. Neergaard-Nielsen, N.~C. Meniccuci, and U.~L.
  Andersen.
\newblock Cubic phase gates are not suitable for non-{C}lifford operations on
  {GKP} states.
\newblock {\em Phys. Rev. A}, \textbf{103}(3):032409, 2021.
\newblock \url{https://doi.org/10.1103/PhysRevA.103.032409}.

\bibitem{hanggli2020_enhanced}
L.~H\"anggli, M.~Heinze, and R.~K\"onig.
\newblock Enhanced noise resilience of the
  surface-{G}ottesman-{K}itaev-{P}reskill code via designed bias.
\newblock {\em Phys. Rev. A}, \textbf{102}(5):052408, 2020.
\newblock \url{https://doi.org/10.1103/PhysRevA.102.052408}.

\bibitem{harrington2001_achievable}
J.~Harrington and J.~Preskill.
\newblock Achievable rates for the {G}aussian quantum channel.
\newblock {\em Phys. Rev. A}, \textbf{64}(6):062301, 2001.
\newblock \url{https://doi.org/10.1103/PhysRevA.64.062301}.

\bibitem{rozpedek2020_quantum}
F.~Rozpedek, K.~Noh, Q.~Xu, S.~Guha, and L.~Jiang.
\newblock Quantum repeaters based on concatenated bosonic and discrete-variable
  quantum codes.
\newblock {\em arXiv:2011.15076 [quant-ph]}, 2020.
\newblock \url{https://arxiv.org/abs/2011.15076}.

\bibitem{fukui2020_alloptical}
K.~Fukui, R.~N. Alexander, and P.~van Loock.
\newblock All-optical long-distance quantum communication with
  {G}ottesman-{K}itaev-{P}reskill qubits.
\newblock {\em arXiv:2011.14876 [quant-ph]}, 2020.
\newblock \url{https://arxiv.org/abs/2011.14876}.

\bibitem{aliferis2008_faulttolerant}
P.~Aliferis and J.~Preskill.
\newblock Fault-tolerant quantum computation against biased noise.
\newblock {\em Phys. Rev. A}, \textbf{78}(5):052331, 2008.
\newblock \url{https://doi.org/10.1103/PhysRevA.78.052331}.

\bibitem{guillaud2019_repetition}
J.~Guillaud and M.~Mirrahimi.
\newblock Repetition cat qubits for fault-tolerant quantum computation.
\newblock {\em Phys. Rev. X}, \textbf{9}(4):041053, 2019.
\newblock \url{https://doi.org/10.1103/PhysRevX.9.041053}.

\bibitem{guillaud2020_error}
J.~Guillaud and M.~Mirrahimi.
\newblock Error rates and resource overheads of repetition cat qubits.
\newblock {\em arXiv:2009.10756 [quant-ph]}, 2020.
\newblock \url{https://arxiv.org/abs/2009.10756}.

\bibitem{tuckett2018-ultrahigh}
D.~K. Tuckett, S.~D. Bartlett, and S.~T. Flammia.
\newblock Ultrahigh error threshold for surface codes with biased noise.
\newblock {\em Phys. Rev. Lett.}, \textbf{120}(5):050505, 2018.
\newblock \url{https://doi.org/10.1103/PhysRevLett.120.050505}.

\bibitem{tuckett2019_tailoring}
D.~K. Tuckett, A.~S. Darmawan, C.~T. Chubb, S.~Bravyi, S.~D. Bartlett, and
  S.~T. Flammia.
\newblock Tailoring surface codes for highly biased noise.
\newblock {\em Phys. Rev. X}, \textbf{9}(4):041031, 2019.
\newblock \url{https://doi.org/10.1103/PhysRevX.9.041031}.

\bibitem{tuckett2020_faulttolerant}
D.~K. Tuckett, S.~D. Bartlett, S.~T. Flammia, and B.~J. Brown.
\newblock Fault-tolerant thresholds for the surface code in excess of $5\%$
  under biased noise.
\newblock {\em Phys. Rev. Lett.}, \textbf{124}(13):130501, 2020.
\newblock \url{https://doi.org/10.1103/PhysRevLett.124.130501}.

\bibitem{chamberland2020_building}
C.~Chamberland, K.~Noh, P.~Arrangoiz-Arriola, E.~T. Campbell, C.~T. Hann,
  J.~Iverson, H.~Putterman, T.~C. Bohdanowicz, S.~T. Flammia, A.~Keller,
  G.~Rafael, J.~Preskill, L.~Jiang, A.~H. Safavi-Naeini, O.~PAinter, and F.~G.
  S.~L. Brandão.
\newblock Building a fault-tolerant quantum computer using concatenated cat
  codes.
\newblock {\em arXiv:2012.04108 [quant-ph]}, 2020.
\newblock \url{https://arxiv.org/abs/2012.04108}.

\bibitem{cochrane1999_macroscopically}
P.~T. Cochrane, G.~J. Milburn, and W.~J. Munro.
\newblock Macroscopically distinct quantum-superposition states as a bosonic
  code for amplitude damping.
\newblock {\em Phys. Rev. A}, \textbf{59}(4):2631, 1999.
\newblock \url{https://doi.org/10.1103/PhysRevA.59.2631}.

\bibitem{jeong2002_efficient}
H.~Jeong and M.~S. Kim.
\newblock Efficient quantum computation using coherent states.
\newblock {\em Phys. Rev. A}, \textbf{65}(4):042305, 2002.
\newblock \url{https://doi.org/10.1103/PhysRevA.65.042305}.

\bibitem{ralph2003_quantum}
T.~C. Ralph, A.~Gilchrist, G.~J. Milburn, W.~J. Munro, and S.~Glancy.
\newblock Quantum computation with optical coherent states.
\newblock {\em Phys. Rev. A}, \textbf{68}(4):042319, 2003.
\newblock \url{https://doi.org/10.1103/PhysRevA.68.042319}.

\bibitem{glancy2004_transmission}
S.~Glancy, H.~M. Vasconcelos, and T.~C. Ralph.
\newblock Transmission of optical coherent-state qubits.
\newblock {\em Phys. Rev. A}, \textbf{70}(2):022317, 2004.
\newblock \url{https://doi.org/10.1103/PhysRevA.70.022317}.

\bibitem{lund2008_faulttolerant}
A.~P. Lund, T.~C. Ralph, and H.~L. Haselgrove.
\newblock Fault-tolerant linear optical quantum computing with small-amplitude
  coherent states.
\newblock {\em Phys. Rev. Lett.}, \textbf{100}(3):030503, 2008.
\newblock \url{https://doi.org/10.1103/PhysRevLett.100.030503}.

\bibitem{puri2017_engineering}
S.~Puri, S.~Boutin, and A.~Blais.
\newblock Engineering the quantum states of light in a {K}err-nonlinear
  resonator by two-photon driving.
\newblock {\em npj Quantum Inf.}, \textbf{3}(1):18, 2017.
\newblock \url{https://doi.org/10.1038/s41534-017-0019-1}.

\bibitem{puri2019_stabilized}
S.~Puri, A.~Grimm, P.~Campagne-Ibarcq, A.~Eickbusch, K.~Noh, G.~Roberts,
  L.~Jiang, M.~Mirrahimi, M.~H. Devoret, and S.~M. Girvin.
\newblock Stabilized cat in a driven nonlinear cavity: a fault-tolerant error
  syndrome detector.
\newblock {\em Phys. Rev. X}, \textbf{9}(4):041009, 2019.
\newblock \url{https://doi.org/10.1103/PhysRevX.9.041009}.

\bibitem{puri2020_bias-preserving}
S.~Puri, L.~St-Jean, J.~A. Gross, A.~Grimm, N.~E. Frattini, P.~S. Iyer,
  A.~Krishna, S.~Touzard, L.~Jiang, A.~Blais, S.~T. Flammia, and S.~M. Girvin.
\newblock Bias-preserving gates with stabilized cat qubits.
\newblock {\em Sci. Adv.}, \textbf{6}(34):5901, 2020.
\newblock \url{https://doi.org/10.1126/sciadv.aay5901}.

\bibitem{lloyd1997_capacity}
S.~Lloyd.
\newblock Capacity of the noisy quantum channel.
\newblock {\em Phys. Rev. A}, \textbf{55}(3):1613--1622, 1997.
\newblock \url{https://doi.org/10.1103/PhysRevA.55.1613}.

\bibitem{devetak2005_private}
I.~Devetak.
\newblock The private classical capacity and quantum capacity of a quantum
  channel.
\newblock {\em IEEE Transactions on Information Theory}, \textbf{51}(1):44--55,
  2005.
\newblock \url{https://doi.org/10.1109/TIT.2004.839515}.

\bibitem{holevo2001_evaluating}
A.~S. Holevo and R.~F. Werner.
\newblock Evaluating capacities of bosonic {G}aussian channels.
\newblock {\em Phys. Rev. A}, \textbf{63}(3):032312, 2001.
\newblock \url{https://doi.org/10.1103/PhysRevA.63.032312}.

\bibitem{wolf2007_quantum}
M.~M. Wolf, D.~P\'erez-Garc\'{\i}a, and G.~Giedke.
\newblock Quantum capacities of bosonic channels.
\newblock {\em Phys. Rev. Lett.}, \textbf{98}(13):130501, 2007.
\newblock \url{https://doi.org/10.1103/PhysRevLett.98.130501}.

\bibitem{wilde2012_quantum}
M.~M. Wilde, P.~Hayden, and S.~Guha.
\newblock Quantum trade-off coding for bosonic communication.
\newblock {\em Phys. Rev. A}, \textbf{86}(6):062306, 2012.
\newblock \url{https://doi.org/10.1103/PhysRevA.86.062306}.

\bibitem{wilde2018_energyconstrained}
M.~M. Wilde and H.~Qi.
\newblock Energy-constrained private and quantum capacities of quantum
  channels.
\newblock {\em IEEE Transactions on Information Theory},
  \textbf{64}(12):7802--7827, 2018.
\newblock \url{https://doi.org/10.1109/TIT.2018.2854766}.

\bibitem{arqand2020_quantum}
L.~Arqand, A.~Memarzadeh and S.~Mancini.
\newblock Quantum capacity of a bosonic dephasing channel.
\newblock {\em Phys. Rev. A}, \textbf{102}(4):042413, 2020.
\newblock \url{https://doi.org/10.1103/PhysRevA.102.042413}.

\bibitem{deleglise2008_reconstruction}
S.~Deleglise, I.~Dotsenko, C.~Sayrin, J.~Bernu, M.~Brune, J.~M. Raimond, and
  S.~Haroche.
\newblock Reconstruction of non-classical cavity field states with snapshots of
  their decoherence.
\newblock {\em Nature}, \textbf{455}(7212):510--514, 2008.
\newblock \url{https://doi.org/10.1038/nature07288}.

\bibitem{hacker2019_deterministic}
B.~Hacker, S.~Welte, S.~Daiss, A.~Shaukat, L.~Ritter, S.and~Li, and G.~Rempe.
\newblock Deterministic creation of entangled atom–light {S}chrödinger-cat
  states.
\newblock {\em Nature Photon.}, \textbf{13}(2):110--115, 2019.
\newblock \url{https://doi.org/10.1038/s41566-018-0339-5}.

\bibitem{omran2019_generation}
A.~Omran, H.~Levine, A.~Keesling, G.~Semeghini, T.~T. Wang, S.~Ebadi,
  H.~Bernien, A.~S. Zibrov, H.~Pichler, S.~Choi, J.~Cui, M.~Rossignolo,
  P.~Rembold, S.~Montangero, T.~Calarco, M.~Endres, M.~Greiner, V.~Vuletić,
  and M.~D. Lukin.
\newblock Generation and manipulation of {S}chrödinger cat states in {R}ydberg
  atom arrays.
\newblock {\em Science}, \textbf{365}(6453):570--574, 2019.
\newblock \url{https://doi.org/10.1126/science.aax9743}.

\bibitem{ourjoumtsev2006_generating}
A.~Ourjoumtsev, R.~Tualle-Brouri, J.~Laurat, and P.~Grangier.
\newblock Generating optical {S}chrödinger kittens for quantum information
  processing.
\newblock {\em Science}, \textbf{312}(5770):83--86, 2006.
\newblock \url{https://doi.org/10.1126/science.1122858}.

\bibitem{hou2016_generation}
Q.~Hou, W.~Yang, C.~Chen, and Z.~Yin.
\newblock Generation of macroscopic {S}chrödinger cat state in diamond
  mechanical resonator.
\newblock {\em Sci. Rep.}, \textbf{6}:37542, 2016.
\newblock \url{https://doi.org/10.1038/srep37542}.

\bibitem{bulutay2017_cat-state}
C.~Bulutay.
\newblock Cat-state generation and stabilization for a nuclear spin through
  electric quadrupole interaction.
\newblock {\em Phys. Rev. A}, \textbf{96}(1):012312, 2017.
\newblock \url{https://doi.org/10.1103/PhysRevA.96.012312}.

\bibitem{frunzio2005_fabrication}
L.~Frunzio, A.~Wallraff, D.~Schuster, J.~Majer, and R.~Schoelkopf.
\newblock Fabrication and characterization of superconducting circuit {QED}
  devices for quantum computation.
\newblock {\em IEEE Transactions on Applied Superconductivity},
  \textbf{15}(2):860--863, 2005.
\newblock \url{https://doi.org/10.1109/TASC.2005.850084}.

\bibitem{schoelkopf2008_wiring}
R.~J. Schoelkopf and S.~M. Girvin.
\newblock Wiring up quantum systems.
\newblock {\em Nature}, \textbf{451}(7179):664--669, 2008.
\newblock \url{https://doi.org/10.1038/451664a}.

\bibitem{wallraff2004_strong}
A.~Wallraff, D.~I. Schuster, A.~Blais, L.~Frunzio, R.~S. Huang, J.~Majer,
  S.~Kumar, S.~M. Girvin, and R.~J. Schoelkopf.
\newblock Strong coupling of a single photon to a superconducting qubit using
  circuit quantum electrodynamics.
\newblock {\em Nature}, \textbf{431}(7005):162--167, 2004.
\newblock \url{https://doi.org/10.1038/nature02851}.

\bibitem{schuster2007_resolving}
D.~I. Schuster, A.~A. Houck, J.~A. Schreier, A.~Wallraff, J.~M. Gambetta,
  A.~Blais, L.~Frunzio, J.~Majer, B.~Johnson, M.~H. Devoret, S.~M. Girvin, and
  R.~J. Schoelkopf.
\newblock Resolving photon number states in a superconducting circuit.
\newblock {\em Nature}, \textbf{445}(7127):515--518, 2007.
\newblock \url{https://doi.org/10.1038/nature05461}.

\bibitem{niemczyk2010_circuit}
T.~Niemczyk, F.~Deppe, H.~Huebl, E.~P. Menzel, F.~Hocke, M.~J. Schwarz, J.~J.
  Garcia-Ripoll, D.~Zueco, T.~Hümmer, E.~Solano, A.~Marx, and R.~Gross.
\newblock Circuit quantum electrodynamics in the ultrastrong-coupling regime.
\newblock {\em Nat. Phys.}, \textbf{6}(10):772--776, 2010.
\newblock \url{https://doi.org/10.1038/nphys1730}.

\bibitem{girvin2009_circuit}
S.~M. Girvin, M.~H. Devoret, and R.~J. Schoelkopf.
\newblock Circuit {QED} and engineering charge-based superconducting qubits.
\newblock {\em Phys. Scr.}, \textbf{2009}(T137):014012, 2009.
\newblock \url{https://doi.org/10.1088/0031-8949/2009/T137/014012}.

\bibitem{ladd2010_quantum}
T.~D. Ladd, F.~Jelezko, R.~Laflamme, Y.~Nakamura, C.~Monroe, and J.~L. O'Brien.
\newblock Quantum computers.
\newblock {\em Nature}, \textbf{464}(7285):45--53, 2010.
\newblock \url{https://doi.org/10.1038/nature08812}.

\bibitem{wendin2017_quantum}
G.~Wendin.
\newblock Quantum information processing with superconducting circuits: a
  review.
\newblock {\em Rep. Prog. Phys.}, \textbf{80}(10):106001, 2017.
\newblock \url{https://doi.org/10.1088/1361-6633/aa7e1a}.

\bibitem{krantz2019_quantum}
P.~Krantz, M.~Kjaergaard, F.~Yan, T.~P. Orlando, S.~Gustavsson, and W.~D.
  Oliver.
\newblock A quantum engineer's guide to superconducting qubits.
\newblock {\em Appl. Phys. Rev.}, \textbf{6}(2):021318, 2019.
\newblock \url{https://doi.org/10.1063/1.5089550}.

\bibitem{kjaergaard2019_superconducting}
M.~Kjaergaard, M.~E. Schwartz, J.~Braumüller, P.~Krantz, J.~I.~J. Wang,
  S.~Gustavsson, and W.~D. Oliver.
\newblock Superconducting qubits: current state of play.
\newblock {\em Annual Review of Condensed Matter Physics},
  \textbf{11}(1):369--395, 2020.
\newblock \url{https://doi.org/10.1146/annurev-conmatphys-031119-050605}.

\bibitem{blais2020_quantum}
A.~Blais, S.~M. Girvin, and W.~D. Oliver.
\newblock Quantum information processing and quantum optics with circuit
  quantum electrodynamics.
\newblock {\em Nat. Phys.}, \textbf{16}(3):247--256, 2020.
\newblock \url{https://doi.org/10.1038/s41567-020-0806-z}.

\bibitem{cai2020_bosonic}
W.~Cai, Y.~Ma, W.~Wang, C.~L. Zou, and L.~Sun.
\newblock Bosonic quantum error correction codes in superconducting quantum
  circuits.
\newblock {\em Fundamental Research}, \textbf{1}(1):50--67, 2021.
\newblock \url{https://doi.org/10.1016/j.fmre.2020.12.006}.

\bibitem{haroche2006_exploring}
S.~Haroche and J.~M. Raimond.
\newblock {\em Exploring the quantum: atoms, cavities, and photons}.
\newblock Oxford University Press, 2006.
\newblock \url{https://doi.org/10.1093/acprof:oso/9780198509141.001.0001}.

\bibitem{koch2007_charge-insensitive}
J.~Koch, M.~Y. Terri, J.~Gambetta, A.~A. Houck, D.~I. Schuster, J.~Majer,
  A.~Blais, M.~H. Devoret, S.~M. Girvin, and R.~J. Schoelkopf.
\newblock Charge-insensitive qubit design derived from the {C}ooper pair box.
\newblock {\em Phys. Rev. A}, \textbf{76}(4):042319, 2007.
\newblock \url{https://doi.org/10.1103/PhysRevA.76.042319}.

\bibitem{schreier2008_suppressing}
J.~A. Schreier, A.~A. Houck, J.~Koch, D.~I. Schuster, B.~R. Johnson, J.~M.
  Chow, J.~M. Gambetta, J.~Majer, L.~Frunzio, M.~H. Devoret, S.~M. Girvin, and
  R.~J. Schoelkopf.
\newblock Suppressing charge noise decoherence in superconducting charge
  qubits.
\newblock {\em Phys. Rev. B}, \textbf{77}(18):180502, 2008.
\newblock \url{https://doi.org/10.1103/PhysRevB.77.180502}.

\bibitem{preskill2018_quantum}
J.~Preskill.
\newblock Quantum computing in the {NISQ} era and beyond.
\newblock {\em Quantum}, \textbf{2}:79, 2018.
\newblock \url{https://doi.org/10.22331/q-2018-08-06-79}.

\bibitem{xiang2017_experimental}
X.~Fu, M.~A. Rol, C.~C. Bultink, J.~van Someren, N.~Khammassi, I.~Ashraf,
  R.~F.~L. Vermeulen, J.~C. de~Sterke, W.~J. Vlothuizen, R.~N. Schouten, C.~G.
  Almudever, L.~DiCarlo, and K.~Bertels.
\newblock An experimental microarchitecture for a superconducting quantum
  processor.
\newblock In {\em Proceedings of the 50th Annual IEEE/ACM International
  Symposium on Microarchitecture}, pages 813--825, 2017.
\newblock \url{https://doi.org/10.1145/3123939.3123952}.

\bibitem{otterbach2017_unsupervised}
J.~S. Otterbach, R.~Manenti, N.~Alidoust, A.~Bestwick, M.~Block, B.~Bloom,
  S.~Caldwell, N.~Didier, E.~Schuyler~Fried, S.~Hong, P.~Karalekas, C.~B.
  Osborn, A.~Papageorge, E.~C. Peterson, G.~Prawiroatmodjo, N.~Rubin, C.~A.
  Ryan, D.~Scarabelli, M.~Scheer, E.~A. Sete, P.~Sivarajah, R.~S. Smith,
  A.~Staley, N.~Tezak, W.~J. Zeng, A.~Hudson, B.~R. Johnson, M.~Reagor, M.~P.
  da~Silva, and C.~Rigetti.
\newblock Unsupervised machine learning on a hybrid quantum computer.
\newblock {\em arXiv:1712.05771 [quant-ph]}, 2017.
\newblock \url{https://arxiv.org/abs/1712.05771}.

\bibitem{kandala2017_hardware}
A.~Kandala, A.and~Mezzacapo, M.~Temme, K.and~Takita, M.~Brink, J.~M. Chow, and
  J.~M Gambetta.
\newblock Hardware-efficient variational quantum eigensolver for small
  molecules and quantum magnets.
\newblock {\em Nature}, \textbf{549}(7671):242--246, 2017.
\newblock \url{https://doi.org/10.1038/nature23879}.

\bibitem{arute2019_quantum}
F.~Arute, K.~Arya, R.~Babbush, D.~Bacon, J.~C. Bardin, R.~Barends, R.~Biswas,
  S.~Boixo, F.~G. S.~L. Brandao, D.~A. Buell, B.~Burkett, Y.~Chen, Z.~Chen,
  B.~Chiaro, R.~Collins, W.~Courtney, A.~Dunsworth, E.~Farhi, B.~Foxen,
  A.~Fowler, C.~Gidney, M.~Giustina, R.~Graff, K.~Guerin, S.~Habegger, M.~P.
  Harrigan, M.~J. Hartmann, A.~Ho, M.~Hoffmann, T.~Huang, T.~S. Humble, S.~V.
  Isakov, E.~Jeffrey, Z.~Jiang, D.~Kafri, K.~Kechedzhi, J.~Kelly, P.~V. Klimov,
  S.~Knysh, A.~Korotkov, F.~Kostritsa, D.~Landhuis, M.~Lindmark, E.~Lucero,
  D.~Lyakh, S.~Mandra, J.~R. McClean, M.~McEwen, A.~Megrant, X.~Mi,
  K.~Michielsen, M.~Mohseni, J.~Mutus, O.~Naaman, M.~Neeley, C.~Neill, M.~Y.
  Niu, E.~Ostby, A.~Petukhov, J.~C. Platt, C.~Quintana, E.~G. Rieffel,
  P.~Roushan, N.~C. Rubin, D.~Sank, K.~J. Satzinger, V.~Smelyanskiy, K.~J.
  Sung, M.~D. Trevithick, A.~Vainsencher, B.~Villalonga, T.~White, Z.~J. Yao,
  P.~Yeh, A.~Zalcman, H.~Neven, and J.~M. Martinis.
\newblock Quantum supremacy using a programmable superconducting processor.
\newblock {\em Nature}, \textbf{574}(7779):505--510, 2019.
\newblock \url{https://doi.org/10.1038/s41586-019-1666-5}.

\bibitem{jurcevic2020_demonstration}
P.~Jurcevic, A.~Javadi-Abhari, L.~S. Bishop, I.~Lauer, D.~F. Bogorin, M.~Brink,
  L.~Capelluto, O.~Günlük, T.~Itoko, N.~Kanazawa, A.~Kandala, K.~Keefe, G.
  A.~Krsulich, W.~Landers, E.~P. Lewandowski, D.~T. McClure, G.~Nannicini,
  A.~Narasgond, H.~M. Nayfeh, E.~Pritchett, M.~B. Rothwell, S.~Srinivasan,
  N.~Sundaresan, C.~Wang, K.~X. Wei, C.~J. Wood, J.~B. Yau, E.~J. Zhang, O.~E.
  Dial, J.~M. Chow, and J.~M. Gambetta.
\newblock Demonstration of quantum volume 64 on a superconducting quantum
  computing system.
\newblock {\em Quantum Sci. Technol.}, \textbf{6}(2):025020, 2021.
\newblock \url{https://doi.org/10.1088/2058-9565/abe519}.

\bibitem{google2020_hartree-fock}
Google~AI Quantum and Collaborators.
\newblock Hartree-{F}ock on a superconducting qubit quantum computer.
\newblock {\em Science}, \textbf{369}(6507):1084--1089, 2020.
\newblock \url{https://doi.org/10.1126/science.abb9811}.

\bibitem{lacroix2020_improving}
N.~Lacroix, C.~Hellings, C.~K. Andersen, A.~Di~Paolo, A.~Remm, S.~Lazar,
  S.~Krinner, G.~J. Norris, M.~Gabureac, J.~Heinsoo, A.~Blais, C.~Eichler, and
  A.~Wallraff.
\newblock Improving the performance of deep quantum optimization algorithms
  with continuous gate sets.
\newblock {\em PRX Quantum}, \textbf{1}(2):110304, 2020.
\newblock \url{https://doi.org/10.1103/PRXQuantum.1.020304}.

\bibitem{paik2011_observation}
H.~Paik, D.~I. Schuster, L.~S. Bishop, G.~Kirchmair, G.~Catelani, A.~P. Sears,
  B.~R. Johnson, M.~J. Reagor, L.~Frunzio, L.~I. Glazman, S.~M. Girvin, M.~H.
  Devoret, and R.~J. Schoelkopf.
\newblock Observation of high coherence in {J}osephson junction qubits measured
  in a three-dimensional circuit {QED} architecture.
\newblock {\em Phys. Rev. Lett.}, \textbf{107}(24):240501, 2011.
\newblock \url{https://doi.org/10.1103/PhysRevLett.107.240501}.

\bibitem{reagor2016_quantum}
M.~Reagor, W.~Pfaff, C.~Axline, R.~W. Heeres, N.~Ofek, K.~Sliwa, E.~Holland,
  C.~Wang, J.~Blumoff, K.~Chou, M.~J. Hatridge, L.~Frunzio, M.~H. Devoret,
  L.~Jiang, and R.~J. Schoelkopf.
\newblock Quantum memory with millisecond coherence in circuit {QED}.
\newblock {\em Phys. Rev. B}, \textbf{94}(1):014506, 2016.
\newblock \url{https://doi.org/10.1103/PhysRevB.94.014506}.

\bibitem{brecht2015_demonstration}
T.~Brecht, M.~Reagor, Y.~Chu, W.~Pfaff, C.~Wang, L.~Frunzio, M.~H. Devoret, and
  R.~J. Schoelkopf.
\newblock Demonstration of superconducting micromachined cavities.
\newblock {\em Appl. Phys. Lett.}, \textbf{107}(19):192603, 2015.
\newblock \url{https://doi.org/10.1063/1.4935541}.

\bibitem{houck2007_generating}
A.~A. Houck, D.~I. Schuster, J.~M. Gambetta, J.~A. Schreier, B.~R. Johnson,
  J.~M. Chow, L.~Frunzio, J.~Majer, M.~H. Devoret, S.~M. Girvin, and R.~J.
  Schoelkopf.
\newblock Generating single microwave photons in a circuit.
\newblock {\em Nature}, \textbf{449}(7160):328--331, 2007.
\newblock \url{https://doi.org/10.1038/nature06126}.

\bibitem{majer2007_coupling}
J.~Majer, J.~M. Chow, J.~M. Gambetta, J.~Koch, B.~R. Johnson, J.~A. Schreier,
  L.~Frunzio, D.~I. Schuster, A.~A. Houck, A.~Wallraff, A.~Blais, M.~H.
  Devoret, S.~M. Girvin, and R.~J. Schoelkopf.
\newblock Coupling superconducting qubits via a cavity bus.
\newblock {\em Nature}, \textbf{449}(7161):443--447, 2007.
\newblock \url{https://doi.org/10.1038/nature06184}.

\bibitem{sillanpaa2007_coherent}
M.~A. Sillanpää, J.~I. Park, and R.~W. Simmonds.
\newblock Coherent quantum state storage and transfer between two phase qubits
  via a resonant cavity.
\newblock {\em Nature}, \textbf{449}(7161):438--442, 2007.
\newblock \url{https://doi.org/10.1038/nature06124}.

\bibitem{dicarlo2009_demonstration}
L.~DiCarlo, J.~M. Chow, J.~M. Gambetta, L.~S. Bishop, B.~R. Johnson, D.~I.
  Schuster, J.~Majer, A.~Blais, L.~Frunzio, S.~M. Girvin, and R.~J. Schoelkopf.
\newblock Demonstration of two-qubit algorithms with a superconducting quantum
  processor.
\newblock {\em Nature}, \textbf{460}(7252):240--244, 2009.
\newblock \url{https://doi.org/10.1038/nature08121}.

\bibitem{wallraff2005_approaching}
A.~Wallraff, D.~I. Schuster, A.~Blais, L.~Frunzio, J.~Majer, M.~H. Devoret,
  S.~M. Girvin, and R.~J. Schoelkopf.
\newblock Approaching unit visibility for control of a superconducting qubit
  with dispersive readout.
\newblock {\em Phys. Rev. Lett.}, \textbf{95}(6):060501, 2005.
\newblock \url{https://doi.org/10.1103/PhysRevLett.95.060501}.

\bibitem{johansson2006_vacuum}
J.~Johansson, S.~Saito, T.~Meno, H.~Nakano, M.~Ueda, K.~Semba, and
  H.~Takayanagi.
\newblock Vacuum {R}abi oscillations in a macroscopic superconducting qubit
  oscillator system.
\newblock {\em Phys. Rev. Lett.}, \textbf{96}(12):127006, 2006.
\newblock \url{https://doi.org/10.1103/PhysRevLett.96.127006}.

\bibitem{blais2004_cavity}
A.~Blais, R.~S. Huang, A.~Wallraff, S.~M. Girvin, and R.~J. Schoelkopf.
\newblock Cavity quantum electrodynamics for superconducting electrical
  circuits: an architecture for quantum computation.
\newblock {\em Phys. Rev. A}, \textbf{69}(6):062320, 2004.
\newblock \url{https://doi.org/10.1103/PhysRevA.69.062320}.

\bibitem{krastanov2015_universal}
S.~Krastanov, V.~V. Albert, C.~Shen, C.~L. Zou, R.~W. Heeres, B.~Vlastakis,
  R.~J. Schoelkopf, and L.~Jiang.
\newblock Universal control of an oscillator with dispersive coupling to a
  qubit.
\newblock {\em Phys. Rev. A}, \textbf{92}(4):040303, 2015.
\newblock \url{https://doi.org/10.1103/PhysRevA.92.040303}.

\bibitem{brecht2017_micromachined}
T.~Brecht, Y.~Chu, C.~Axline, W.~Pfaff, J.~Z. Blumoff, K.~Chou, L.~Krayzman,
  L.~Frunzio, and R.~J. Schoelkopf.
\newblock Micromachined integrated quantum circuit containing a superconducting
  qubit.
\newblock {\em Phys. Rev. Applied}, \textbf{7}(4):044018, 2017.
\newblock \url{https://doi.org/10.1103/PhysRevApplied.7.044018}.

\bibitem{martinis2005_decoherence}
J.~M. Martinis, K.~B. Cooper, R.~McDermott, M.~Steffen, M.~Ansmann, K.~D.
  Osborn, K.~Cicak, S.~Oh, D.~P. Pappas, R.~W. Simmonds, and C.~C. Yu.
\newblock Decoherence in {J}osephson qubits from dielectric loss.
\newblock {\em Phys. Rev. Lett}, \textbf{95}(21):210503, 2005.
\newblock \url{https://doi.org/10.1103/PhysRevLett.95.210503}.

\bibitem{oconnell2008_microwave}
A.~D. O’Connell, M.~Ansmann, R.~C. Bialczak, M.~Hofheinz, N.~Katz, E.~Lucero,
  C.~McKenney, M.~Neeley, H.~Wang, E.~M. Weig, A.~N. Cleland, and J.~M.
  Martinis.
\newblock Microwave dielectric loss at single photon energies and millikelvin
  temperatures.
\newblock {\em Appl. Phys. Lett.}, \textbf{92}(11):112903, 2008.
\newblock \url{https://doi.org/10.1063/1.2898887}.

\bibitem{gao2008_experimental}
J.~Gao, M.~Daal, A.~Vayonakis, S.~Kumar, J.~Zmuidzinas, B.~Sadoulet, B.~A.
  Mazin, P.~K. Day, and H.~G. Leduc.
\newblock Experimental evidence for a surface distribution of two-level systems
  in superconducting lithographed microwave resonators.
\newblock {\em Appl. Phys. Lett.}, \textbf{92}(15):152505, 2008.
\newblock \url{https://doi.org/10.1063/1.2906373}.

\bibitem{muller2019_towards}
C.~Muller, J.~H. Cole, and J.~Lisenfeld.
\newblock Towards understanding two-level-systems in amorphous solids: insights
  from quantum circuits.
\newblock {\em Rep. Prog. Phys.}, \textbf{82}(12):124501, 2019.
\newblock \url{https://doi.org/10.1088/1361-6633/ab3a7e}.

\bibitem{sage2011_study}
J.~M. Sage, V.~Bolkhovsky, W.~D. Oliver, B.~Turek, and P.~B. Welander.
\newblock Study of loss in superconducting coplanar waveguide resonators.
\newblock {\em Journal of Applied Physics}, \textbf{109}(6):063915, 2011.
\newblock \url{https://doi.org/10.1063/1.3552890}.

\bibitem{vissers2012_reduced}
M.~R Vissers, J.~S. Kline, J.~Gao, D.~S. Wisbey, and D.~P. Pappas.
\newblock Reduced microwave loss in trenched superconducting coplanar
  waveguides.
\newblock {\em Appl. Phys. Lett.}, \textbf{100}(8):082602, 2012.
\newblock \url{https://doi.org/10.1063/1.3683552}.

\bibitem{geerlings2012_improving}
K.~Geerlings, S.~Shankar, E.~Edwards, L.~Frunzio, R.~J. Schoelkopf, and M.~H.
  Devoret.
\newblock Improving the quality factor of microwave compact resonators by
  optimizing their geometrical parameters.
\newblock {\em Appl. Phys. Lett.}, \textbf{100}(19):192601, 2012.
\newblock \url{https://doi.org/10.1063/1.4710520}.

\bibitem{calusine2018_analysis}
G.~Calusine, A.~Melville, W.~Woods, R.~Das, C.~Stull, V.~Bolkhovsky, D.~Braje,
  D.~Hover, D.~K. Kim, X.~Miloshi, D.~Rosenberg, A.~Sevi, J.~L. Yoder,
  E.~Dauler, and W.~D. Oliver.
\newblock Analysis and mitigation of interface losses in trenched
  superconducting coplanar waveguide resonators.
\newblock {\em Appl. Phys. Lett.}, \textbf{112}(6):062601, 2018.
\newblock \url{https://doi.org/10.1063/1.5006888}.

\bibitem{wang2009_improving}
H.~Wang, M.~Hofheinz, J.~Wenner, M.~Ansmann, R.~C. Bialczak, M.~Lenander,
  E.~Lucero, M.~Neeley, A.~D. O’Connell, D.~Sank, M.~Weides, A.~N. Cleland,
  and J.~M. Martinis.
\newblock Improving the coherence time of superconducting coplanar resonators.
\newblock {\em Appl. Phys. Lett.}, \textbf{95}(23):233508, 2009.
\newblock \url{https://doi.org/10.1063/1.3273372}.

\bibitem{barends2010_minimal}
R.~Barends, N.~Vercruyssen, A.~Endo, P.~J. de~Visser, T.~Zijlstra, T.~M.
  Klapwijk, P.~Diener, S.~J.~C. Yates, and J.~J.~A. Baselmans.
\newblock Minimal resonator loss for circuit quantum electrodynamics.
\newblock {\em Appl. Phys. Lett.}, \textbf{97}(2):023508, 2010.
\newblock \url{https://doi.org/10.1063/1.3458705}.

\bibitem{vissers2010_low}
M.~R. Vissers, J.~Gao, D.~S. Wisbey, D.~A. Hite, C.~C. Tsuei, A.~D. Corcoles,
  M.~Steffen, and D.~P. Pappas.
\newblock Low loss superconducting titanium nitride coplanar waveguide
  resonators.
\newblock {\em Appl. Phys. Lett.}, \textbf{97}(23):232509, 2010.
\newblock \url{https://doi.org/10.1063/1.3517252}.

\bibitem{place2020_new}
A.~P.~M. Place, L.~V.~H. Rodgers, P.~Mundada, B.~M. Smitham, M.~Fitzpatrick,
  Z.~Leng, A.~Premkumar, J.~Bryon, S.~Sussman, G.~Cheng, T.~Madhavan, H.~K.
  Babla, B.~Jaeck, A.~Gyenis, N.~Yao, R.~J. Cava, N.~P. de~Leon, and A.~A.
  Houck.
\newblock New material platform for superconducting transmon qubits with
  coherence times exceeding 0.3 milliseconds.
\newblock {\em Nature Communications}, \textbf{12}(1):1--6, 2021.
\newblock \url{https://doi.org/10.1038/s41467-021-22030-5}.

\bibitem{megrant2012_planar}
A.~Megrant, C.~Neill, R.~Barends, B.~Chiaro, Y.~Chen, L.~Feigl, J.~Kelly,
  E.~Lucero, M.~Mariantoni, P.~J.~J. O’Malley, D.~Sank, A.~Vainsencher,
  J.~Wenner, T.~C. White, Y.~Yin, J.~Zhao, C.~J. Palmstrøm, John~M. Martinis,
  and A.~N. Cleland.
\newblock Planar superconducting resonators with internal quality factors above
  one million.
\newblock {\em Appl. Phys. Lett.}, \textbf{100}(11):113510, 2012.
\newblock \url{https://doi.org/10.1063/1.3693409}.

\bibitem{sandberg2012_etch}
M.~Sandberg, M.~R. Vissers, J.~S. Kline, M.~Weides, J.~Gao, D.~S. Wisbey, and
  D.~P. Pappas.
\newblock Etch induced microwave losses in titanium nitride superconducting
  resonators.
\newblock {\em Appl. Phys. Lett.}, \textbf{100}(26):262605, 2012.
\newblock \url{https://doi.org/10.1063/1.4729623}.

\bibitem{bruno2015_reducing}
A.~Bruno, G.~de~Lange, S.~Asaad, K.~L. van~der Enden, N.~K. Langford, and
  L.~DiCarlo.
\newblock Reducing intrinsic loss in superconducting resonators by surface
  treatment and deep etching of silicon substrates.
\newblock {\em Appl. Phys. Lett.}, \textbf{106}(18):182601, 2015.
\newblock \url{https://doi.org/10.1063/1.4919761}.

\bibitem{melville2020_comparison}
A.~Melville, G.~Calusine, W.~Woods, K.~Serniak, E.~Golden, B.~M. Niedzielski,
  D.~K. Kim, A.~Sevi, J.~L. Yoder, E.~A. Dauler, and W.~D. Oliver.
\newblock Comparison of dielectric loss in titanium nitride and aluminum
  superconducting resonators.
\newblock {\em Appl. Phys. Lett.}, \textbf{117}(12):124004, 2020.
\newblock \url{https://doi.org/10.1063/5.0021950}.

\bibitem{kudra2020_high}
M.~Kudra, J.~Biznárová, F.~Roudsari, J.~J. Burnett, D.~Niepce,
  S.~Gasparinetti, B.~Wickman, and P.~Delsing.
\newblock High quality three-dimensional aluminum microwave cavities.
\newblock {\em Appl. Phys. Lett.}, \textbf{117}(7):070601, 2020.
\newblock \url{https://doi.org/10.1063/5.0016463}.

\bibitem{axline2016_architecture}
C.~Axline, M.~Reagor, R.~Heeres, P.~Reinhold, C.~Wang, K.~Shain, W.~Pfaff,
  Y.~Chu, L.~Frunzio, and R.~J. Schoelkopf.
\newblock An architecture for integrating planar and 3{D} c{QED} devices.
\newblock {\em Appl. Phys. Lett.}, \textbf{109}(4):042601, 2016.
\newblock \url{https://doi.org/10.1063/1.4959241}.

\bibitem{minev2013_planar}
Z.~K. Minev, I.~M. Pop, and M.~H. Devoret.
\newblock Planar superconducting whispering gallery mode resonators.
\newblock {\em Appl. Phys. Lett.}, \textbf{103}(14):142604, 2013.
\newblock \url{https://doi.org/10.1063/1.4824201}.

\bibitem{brecht2016_multilayer}
T.~Brecht, W.~Pfaff, C.~Wang, Y.~Chu, L.~Frunzio, M.~H. Devoret, and R.~J.
  Schoelkopf.
\newblock Multilayer microwave integrated quantum circuits for scalable quantum
  computing.
\newblock {\em npj Quantum Inf.}, \textbf{2}(1):1--4, 2016.
\newblock \url{https://doi.org/10.1038/npjqi.2016.2}.

\bibitem{minev2016_planar}
Z.~K. Minev, K.~Serniak, I.~M Pop, Z.~Leghtas, K.~Sliwa, M.~Hatridge,
  L.~Frunzio, R.~J Schoelkopf, and M.~H Devoret.
\newblock Planar multilayer circuit quantum electrodynamics.
\newblock {\em Phys. Rev. Applied}, \textbf{5}(4):044021, 2016.
\newblock \url{https://doi.org/10.1103/PhysRevApplied.5.044021}.

\bibitem{zoepfl2017_characterization}
D.~Zoepfl, P.~R. Muppalla, C.~M.~F. Schneider, S.~Kasemann, S.~Partel, and
  G.~Kirchmair.
\newblock Characterization of low loss microstrip resonators as a building
  block for circuit {QED} in a 3{D} waveguide.
\newblock {\em AIP Advances}, \textbf{7}(8):085118, 2017.
\newblock \url{https://doi.org/10.1063/1.4992070}.

\bibitem{lei2020_high}
Chan~U. Lei, L.~Krayzman, S.~Ganjam, L.~Frunzio, and R.~J. Schoelkopf.
\newblock High coherence superconducting microwave cavities with indium bump
  bonding.
\newblock {\em Appl. Phys. Lett.}, \textbf{116}(15):154002, 2020.
\newblock \url{https://doi.org/10.1063/5.0003907}.

\bibitem{hofheinz2009_synthesizing}
M.~Hofheinz, H.~Wang, M.~Ansmann, R.~C. Bialczak, E.~Lucero, M.~Neeley, A.~D.
  O'Connell, D.~Sank, J.~Wenner, J.~M. Martinis, and A.~N. Cleland.
\newblock Synthesizing arbitrary quantum states in a superconducting resonator.
\newblock {\em Nature}, \textbf{459}(7246):546--549, 2009.
\newblock \url{https://doi.org/10.1038/nature08005}.

\bibitem{leek2010_cavity}
P.~J. Leek, M.~Baur, J.~M. Fink, R.~Bianchetti, L.~Steffen, S.~Filipp, and
  A.~Wallraff.
\newblock Cavity quantum electrodynamics with separate photon storage and qubit
  readout modes.
\newblock {\em Phys. Rev. Lett.}, \textbf{104}(10):100504, 2010.
\newblock \url{https://doi.org/10.1103/PhysRevLett.104.100504}.

\bibitem{vlastakis2013_deterministically}
B.~Vlastakis, G.~Kirchmair, Z.~Leghtas, S.~E. Nigg, L.~Frunzio, S.~M. Girvin,
  M.~Mirrahimi, M.~H. Devoret, and R.~J. Schoelkopf.
\newblock Deterministically encoding quantum information using 100-photon
  {S}chrödinger cat states.
\newblock {\em Science}, \textbf{342}(6158):607--610, 2013.
\newblock \url{https://doi.org/10.1126/science.1243289}.

\bibitem{hu2019_quantum}
L.~Hu, Y.~Ma, W.~Cai, X.~Mu, Y.~Xu, W.~Wang, Y.~Wu, H.~Wang, Y.~P. Song, C.~L.
  Zou, S.~M. Girvin, L.~M. Duan, and L.~Sun.
\newblock Quantum error correction and universal gate set operation on a
  binomial bosonic logical qubit.
\newblock {\em Nat. Phys.}, \textbf{15}(5):503--508, 2019.
\newblock \url{https://doi.org/10.1038/s41567-018-0414-3}.

\bibitem{ma2020_error-transparent}
Y.~Ma, Y.~Xu, X.~Mu, W.~Cai, L.~Hu, W.~Wang, X.~Pan, H.~Wang, Y.~P. Song, C.~L.
  Zou, and L.~Sun.
\newblock Error-transparent operations on a logical qubit protected by quantum
  error correction.
\newblock {\em Nat. Phys.}, \textbf{16}(8):827--831, 2020.
\newblock \url{https://doi.org/10.1038/s41567-020-0893-x}.

\bibitem{oliver2013_materials}
W.~D Oliver and P.~B. Welander.
\newblock Materials in superconducting quantum bits.
\newblock {\em MRS Bulletin}, \textbf{38}(10):816--825, 2013.
\newblock \url{https://doi.org/10.1557/mrs.2013.229}.

\bibitem{mcrae2020_materials}
C.~R.~H. McRae, H.~Wang, J.~Gao, M.~Vissers, T.~Brecht, A.~Dunsworth,
  D.~Pappas, and J.~Mutus.
\newblock Materials loss measurements using superconducting microwave
  resonators.
\newblock {\em Rev. Sci. Instrum.}, \textbf{91}(9):091101, 2020.
\newblock \url{https://doi.org/10.1063/5.0017378}.

\bibitem{wenner2011_surface}
J.~Wenner, R.~Barends, R.~C. Bialczak, Y.~Chen, J.~Kelly, E.~Lucero,
  M.~Mariantoni, A.~Megrant, P.~J.~J. O’Malley, D.~Sank, A.~Vainsencher,
  H.~Wang, T.~C. White, Y.~Yin, J.~Zhao, A.~N. Cleland, and J.~M. Martinis.
\newblock Surface loss simulations of superconducting coplanar waveguide
  resonators.
\newblock {\em Appl. Phys. Lett.}, \textbf{99}(11):113513, 2011.
\newblock \url{https://doi.org/10.1063/1.3637047}.

\bibitem{wang2015_surface}
C.~Wang, C.~Axline, Y.~Y. Gao, T.~Brecht, Y.~Chu, L.~Frunzio, M.~H. Devoret,
  and R.~J. Schoelkopf.
\newblock Surface participation and dielectric loss in superconducting qubits.
\newblock {\em Appl. Phys. Lett.}, \textbf{107}(16):162601, 2015.
\newblock \url{https://doi.org/10.1063/1.4934486}.

\bibitem{woods2019_determining}
W.~Woods, G.~Calusine, A.~Melville, A.~Sevi, E.~Golden, D.~K. Kim,
  D.~Rosenberg, J.~L. Yoder, and W.~D. Oliver.
\newblock Determining interface dielectric losses in superconducting
  coplanar-waveguide resonators.
\newblock {\em Phys. Rev. Applied}, \textbf{12}(1):014012, 2019.
\newblock \url{https://doi.org/10.1103/PhysRevApplied.12.014012}.

\bibitem{gao2018_programmable}
Y.~Y. Gao, B.~J. Lester, Y.~Zhang, C.~Wang, S.~Rosenblum, L.~Frunzio, L.~Jiang,
  S.~M. Girvin, and R.~J. Schoelkopf.
\newblock Programmable interference between two microwave quantum memories.
\newblock {\em Phys. Rev. X}, \textbf{8}(2):021073, 2018.
\newblock \url{https://doi.org/10.1103/PhysRevX.8.021073}.

\bibitem{glauber1963_coherent}
R.~J. Glauber.
\newblock Coherent and incoherent states of the radiation field.
\newblock {\em Phys. Rev.}, \textbf{131}(6):2766, 1963.
\newblock \url{https://doi.org/10.1103/PhysRev.131.2766}.

\bibitem{frattini2017_3-wave}
N.~E. Frattini, U.~Vool, S.~Shankar, A.~Narla, K.~M. Sliwa, and M.~H. Devoret.
\newblock 3-wave mixing {J}osephson dipole element.
\newblock {\em Appl. Phys. Lett.}, \textbf{110}(22):222603, 2017.
\newblock \url{https://doi.org/10.1063/1.4984142}.

\bibitem{leghtas2013_deterministic}
Z.~Leghtas, G.~Kirchmair, B.~Vlastakis, M.~H. Devoret, R.~J. Schoelkopf, and
  M.~Mirrahimi.
\newblock Deterministic protocol for mapping a qubit to coherent state
  superpositions in a cavity.
\newblock {\em Phys. Rev. A}, \textbf{87}(4):042315, 2013.
\newblock \url{https://doi.org/10.1103/PhysRevA.87.042315}.

\bibitem{wang2016_schrodinger}
C.~Wang, Y.~Y. Gao, P.~Reinhold, R.~W. Heeres, N.~Ofek, K.~Chou, C.~Axline,
  M.~Reagor, J.~Blumoff, and K.~M. Sliwa.
\newblock A {S}chrödinger cat living in two boxes.
\newblock {\em Science}, \textbf{352}(6289):1087--1091, 2016.
\newblock \url{https://doi.org/10.1126/science.aaf2941}.

\bibitem{chow2012_universal}
J.~M. Chow, J.~M. Gambetta, A.~D. Córcoles, S.~T. Merkel, J.~A. Smolin,
  C.~Rigetti, S.~Poletto, G.~A. Keefe, M.~B. Rothwell, J.~R. Rozen, M.~B.
  Ketchen, and M.~Steffen.
\newblock Universal quantum gate set approaching fault-tolerant thresholds with
  superconducting qubits.
\newblock {\em Phys. Rev. Lett.}, \textbf{109}(6):060501, 2012.
\newblock \url{https://doi.org/10.1103/PhysRevLett.109.060501}.

\bibitem{heeres2015_cavity}
R.~W. Heeres, B.~Vlastakis, E.~Holland, S.~Krastanov, V.~V. Albert, L.~Frunzio,
  L.~Jiang, and R.~J. Schoelkopf.
\newblock Cavity state manipulation using photon-number selective phase gates.
\newblock {\em Phys. Rev. Lett.}, \textbf{115}(13):137002, 2015.
\newblock \url{https://doi.org/10.1103/PhysRevLett.115.137002}.

\bibitem{khaneja2005_optimal}
N.~Khaneja, T.~Reiss, C.~Kehlet, T.~Schulte-Herbruggen, and S.~J. Glaser.
\newblock Optimal control of coupled spin dynamics: design of {NMR} pulse
  sequences by gradient ascent algorithms.
\newblock {\em J. Magn. Reson.}, \textbf{172}(2):296--305, 2005.
\newblock \url{https://doi.org/10.1016/j.jmr.2004.11.004}.

\bibitem{defouquieres2011_second}
P.~de~Fouquieres, S.~G. Schirmer, S.~J. Glaser, and I.~Kuprov.
\newblock Second order gradient ascent pulse engineering.
\newblock {\em J. Magn. Reson.}, \textbf{212}(2):412--417, 2011.
\newblock \url{https://doi.org/10.1016/j.jmr.2011.07.023}.

\bibitem{dolde2014_high-fidelity}
F.~Dolde, V.~Bergholm, Y.~Wang, I.~Jakobi, B.~Naydenov, S.~Pezzagna, J.~Meijer,
  F.~Jelezko, P.~Neumann, T.~Schulte-Herbruggen, J.~Biamonte, and J.~Wrachtrup.
\newblock High-fidelity spin entanglement using optimal control.
\newblock {\em Nat. Commun.}, \textbf{5}(1):3371, 2014.
\newblock \url{https://doi.org/10.1038/ncomms4371}.

\bibitem{anderson2015_accurate}
B.~E. Anderson, H.~Sosa-Martinez, C.~A. Riofrío, I.~H. Deutsch, and P.~S.
  Jessen.
\newblock Accurate and robust unitary transformations of a high-dimensional
  quantum system.
\newblock {\em Phys. Rev. Lett.}, \textbf{114}(24):240401, 2015.
\newblock \url{https://doi.org/10.1103/PhysRevLett.114.240401}.

\bibitem{abdelhafez2019_gradient-based}
M.~Abdelhafez, D.~I. Schuster, and J.~Koch.
\newblock Gradient-based optimal control of open quantum systems using quantum
  trajectories and automatic differentiation.
\newblock {\em Phys. Rev. A}, \textbf{99}(5):052327, 2019.
\newblock \url{https://doi.org/10.1103/PhysRevA.99.052327}.

\bibitem{niu2019_universal}
M.~Y. Niu, S.~Boixo, V.~N. Smelyanskiy, and H.~Neven.
\newblock Universal quantum control through deep reinforcement learning.
\newblock {\em npj Quantum Inf.}, \textbf{5}(1):33, 2019.
\newblock \url{https://doi.org/10.1038/s41534-019-0141-3}.

\bibitem{zhang2019_reinforcement}
X.~M. Zhang, Z.~Wei, R.~Asad, X.~C. Yang, and X.~Wang.
\newblock When does reinforcement learning stand out in quantum control? {A}
  comparative study on state preparation.
\newblock {\em npj Quantum Inf.}, \textbf{5}(1):85, 2019.
\newblock \url{https://doi.org/10.1038/s41534-019-0201-8}.

\bibitem{bergeal2010_phase-preserving}
N.~Bergeal, F.~Schackert, M.~Metcalfe, R.~Vijay, V.~E. Manucharyan, L.~Frunzio,
  D.~E. Prober, R.~J. Schoelkopf, S.~M. Girvin, and M.~H. Devoret.
\newblock Phase-preserving amplification near the quantum limit with a
  {J}osephson ring modulator.
\newblock {\em Nature}, \textbf{465}(7294):64--68, 2010.
\newblock \url{https://doi.org/10.1038/nature09035}.

\bibitem{rosenblum2018_cnot}
S.~Rosenblum, Y.~Y. Gao, P.~Reinhold, C.~Wang, C.~J. Axline, L.~Frunzio, S.~M.
  Girvin, L.~Jiang, M.~Mirrahimi, M.~H. Devoret, and R.~J. Schoelkopf.
\newblock A {CNOT} gate between multiphoton qubits encoded in two cavities.
\newblock {\em Nat. Commun.}, \textbf{9}(1):652, 2018.
\newblock \url{https://doi.org/10.1038/s41467-018-03059-5}.

\bibitem{aaronson2011_computational}
S.~Aaronson and A.~Arkhipov.
\newblock The computational complexity of linear optics.
\newblock In {\em Proceedings of the Forty-Third Annual ACM Symposium on Theory
  of Computing}, pages 333--342, 2011.
\newblock \url{https://doi.org/10.1145/1993636.1993682}.

\bibitem{huh2015_boson}
J.~Huh, G.~Gi. Guerreschi, B.~Peropadre, J.~R. McClean, and A.~Aspuru-Guzik.
\newblock Boson sampling for molecular vibronic spectra.
\newblock {\em Nature Photon.}, \textbf{9}(9):615--620, 2015.
\newblock \url{https://doi.org/10.1038/nphoton.2015.153}.

\bibitem{sparrow2018_simulating}
C.~Sparrow, E.~Mart{\'\i}n-L{\'o}pez, N.~Maraviglia, C.~Neville, A.and~Harrold,
  J.~Carolan, Y.~N. Joglekar, T.~Hashimoto, N.~Matsuda, J.~L. O’Brien, D.~P.
  Tew, and A.~Laing.
\newblock Simulating the vibrational quantum dynamics of molecules using
  photonics.
\newblock {\em Nature}, \textbf{557}(7707):660--667, 2018.
\newblock \url{https://doi.org/10.1038/s41586-018-0152-9}.

\bibitem{clements2018_approximating}
W.~R. Clements, J.~J. Renema, A.~Eckstein, A.~A. Valido, A.~Lita, T.~Gerrits,
  S.~W. Nam, W.~S. Kolthammer, J.~Huh, and I.~A. Walmsley.
\newblock Approximating vibronic spectroscopy with imperfect quantum optics.
\newblock {\em J. Phys. B: At. Mol. Opt. Phys.}, \textbf{51}(24):245503, 2018.
\newblock \url{https://doi.org/10.1088/1361-6455/aaf031}.

\bibitem{wang2020_efficient}
C.~S. Wang, J.~C. Curtis, B.~J. Lester, Y.~Zhang, Y.~Y. Gao, J.~Freeze, V.~S.
  Batista, P.~H. Vaccaro, I.~L. Chuang, L.~Frunzio, L.~Jiang, S.~M. Girvin, and
  R.~J. Schoelkopf.
\newblock Efficient multiphoton sampling of molecular vibronic spectra on a
  superconducting bosonic processor.
\newblock {\em Phys. Rev. X}, \textbf{10}(2):021060, 2020.
\newblock \url{https://doi.org/10.1103/PhysRevX.10.021060}.

\bibitem{zhuang2018_distributed}
Q.~Zhuang, Z.~Zhang, and J.~H. Shapiro.
\newblock Distributed quantum sensing using continuous-variable multipartite
  entanglement.
\newblock {\em Phys. Rev. A}, \textbf{97}(3):032329, 2018.
\newblock \url{https://doi.org/10.1103/PhysRevA.97.032329}.

\bibitem{zhuang2020_distributed}
Q.~Zhuang, J.~Preskill, and L.~Jiang.
\newblock Distributed quantum sensing enhanced by continuous-variable error
  correction.
\newblock {\em New J. Phys.}, \textbf{22}(2):022001, 2020.
\newblock \url{https://doi.org/10.1088/1367-2630/ab7257}.

\bibitem{xia2020_demonstration}
Y.~Xia, W.~Li, W.~Clark, D.~Hart, Q.~Zhuang, and Z.~Zhang.
\newblock Demonstration of a reconfigurable entangled radio-frequency photonic
  sensor network.
\newblock {\em Phys. Rev. Lett.}, \textbf{124}(15):150502, 2020.
\newblock \url{https://doi.org/10.1103/PhysRevLett.124.150502}.

\bibitem{noh2020_encoding}
K.~Noh, S.~M. Girvin, and L.~Jiang.
\newblock Encoding an oscillator into many oscillators.
\newblock {\em Phys. Rev. Lett.}, \textbf{125}(8):080503, 2020.
\newblock \url{https://doi.org/10.1103/PhysRevLett.125.080503}.

\bibitem{lau2016_universal}
H.~K. Lau and M.~B. Plenio.
\newblock Universal quantum computing with arbitrary continuous-variable
  encoding.
\newblock {\em Phys. Rev. Lett.}, \textbf{117}(10):100501, 2016.
\newblock \url{https://doi.org/10.1103/PhysRevLett.117.100501}.

\bibitem{chakram2020_seamless}
S.~Chakram, A.~E. Oriani, R.~K. Naik, A.~V. Dixit, K.~He, A.~Agrawal, H.~Kwon,
  and D.~I. Schuster.
\newblock Seamless high-{Q} microwave cavities for multimode circuit {QED}.
\newblock {\em arXiv:2010.16382 [quant-ph]}, 2020.
\newblock \url{https://arxiv.org/abs/2010.16382}.

\bibitem{terhal2015_quantum}
B.~M. Terhal.
\newblock Quantum error correction for quantum memories.
\newblock {\em Rev. Mod. Phys.}, \textbf{87}(2):307, 2015.
\newblock \url{https://doi.org/10.1103/RevModPhys.87.307}.

\bibitem{gyenis2019_experimental}
A.~Gyenis, P.~S. Mundada, A.~Di~Paolo, T.~M. Hazard, X.~You, D.~I. Schuster,
  J.~Koch, A.~Blais, and A.~A. Houck.
\newblock Experimental realization of an intrinsically error-protected
  superconducting qubit.
\newblock {\em arXiv:1910.07542 [quant-ph]}, 2019.
\newblock \url{https://arxiv.org/abs/1910.07542}.

\bibitem{grimm2020_stabilization}
A.~Grimm, N.~E. Frattini, S.~Puri, S.~O. Mundhada, S.~Touzard, M.~Mirrahimi,
  S.~M. Girvin, S.~Shankar, and M.~H. Devoret.
\newblock Stabilization and operation of a {K}err-cat qubit.
\newblock {\em Nature}, \textbf{584}(7820):205--209, 2020.
\newblock \url{https://doi.org/10.1038/s41586-020-2587-z}.

\bibitem{smith2020_magnifying}
W.~C. Smith, M.~Villiers, A.~Marquet, J.~Palomo, M.~R. Delbecq, T.~Kontos,
  P.~Campagne-Ibarcq, B.~Doucot, and Z.~Leghtas.
\newblock Magnifying quantum phase fluctuations with {C}ooper-pair pairing.
\newblock {\em arXiv:2010.15488 [quant-ph]}, 2020.
\newblock \url{https://arxiv.org/abs/2010.15488}.

\bibitem{sun2014_tracking}
L.~Sun, A.~Petrenko, Z.~Leghtas, B.~Vlastakis, G.~Kirchmair, K.~M. Sliwa,
  A.~Narla, M.~Hatridge, S.~Shankar, J.~Blumoff, L.~Frunzio, M.~Mirrahimi,
  M.~H. Devoret, and R.~J. Schoelkopf.
\newblock Tracking photon jumps with repeated quantum non-demolition parity
  measurements.
\newblock {\em Nature}, \textbf{511}(7510):444--448, 2014.
\newblock \url{https://doi.org/10.1038/nature13436}.

\bibitem{leghtas2015_confining}
Z.~Leghtas, S.~Touzard, I.~M. Pop, A.~Kou, B.~Vlastakis, A.~Petrenko, K.~M.
  Sliwa, A.~Narla, S.~Shankar, M.~J. Hatridge, M.~Reagor, L.~Frunzio, R.~J.
  Schoelkopf, M.~Mirrahimi, and M.~H. Devoret.
\newblock Confining the state of light to a quantum manifold by engineered
  two-photon loss.
\newblock {\em Science}, \textbf{347}(6224):853--857, 2015.
\newblock \url{https://doi.org/10.1126/science.aaa2085}.

\bibitem{touzard2018_coherent}
S.~Touzard, A.~Grimm, Z.~Leghtas, S.~O. Mundhada, P.~Reinhold, C.~Axline,
  M.~Reagor, K.~Chou, J.~Blumoff, K.~M. Sliwa, S.~Shankar, L.~Frunzio, R.~J.
  Schoelkopf, M.~Mirrahimi, and M.~H. Devoret.
\newblock Coherent oscillations inside a quantum manifold stabilized by
  dissipation.
\newblock {\em Phys. Rev. X}, \textbf{8}(2):021005, 2018.
\newblock \url{https://doi.org/10.1103/PhysRevX.8.021005}.

\bibitem{mundhada2019_experimental}
S.~O. Mundhada, A.~Grimm, J.~Venkatraman, Z.~K. Minev, S.~Touzard, N.~E.
  Frattini, V.~V. Sivak, K.~Sliwa, P.~Reinhold, S.~Shankar, M.~Mirrahimi, and
  M.~H. Devoret.
\newblock Experimental implementation of a {R}aman-assisted eight-wave mixing
  process.
\newblock {\em Phys. Rev. Applied}, \textbf{12}(5):054051, 2019.
\newblock \url{https://doi.org/10.1103/PhysRevApplied.12.054051}.

\bibitem{gertler2020_protecting}
J.~M. Gertler, B.~Baker, J.~Li, S.~Shirol, J.~Koch, and C.~Wang.
\newblock Protecting a bosonic qubit with autonomous quantum error correction.
\newblock {\em Nature}, \textbf{590}(7845):243--248, 2021.
\newblock \url{https://doi.org/10.1038/s41586-021-03257-0}.

\bibitem{gottesman2009_introduction}
D.~Gottesman.
\newblock An introduction to quantum error correction and fault-tolerant
  quantum computation.
\newblock {\em arXiv:0904.2557 [quant-ph]}, 2009.
\newblock \url{https://arxiv.org/abs/0904.2557}.

\bibitem{aliferis2006_quantum}
P.~Aliferis, D.~Gottesman, and J.~Preskill.
\newblock Quantum accuracy threshold for concatenated distance-3 codes.
\newblock {\em Quantum Info. Comput.}, \textbf{6}(2):97--165, 2006.
\newblock \url{https://dl.acm.org/doi/10.5555/2011665.2011666}.

\bibitem{ma2020_pathindependent}
W.~L. Ma, M.~Zhang, Y.~Wong, K.~Noh, S.~Rosenblum, P.~Reinhold, R.~J.
  Schoelkopf, and L.~Jiang.
\newblock Path-independent quantum gates with noisy ancilla.
\newblock {\em Phys. Rev. Lett.}, \textbf{125}(11):110503, 2020.
\newblock \url{https://doi.org/10.1103/PhysRevLett.125.110503}.

\bibitem{glancy2006_error}
S.~Glancy and E.~Knill.
\newblock Error analysis for encoding a qubit in an oscillator.
\newblock {\em Phys. Rev. A}, \textbf{73}(1):012325, 2006.
\newblock \url{https://doi.org/10.1103/PhysRevA.73.012325}.

\bibitem{fukui2017_analog}
K.~Fukui, A.~Tomita, and A.~Okamoto.
\newblock Analog quantum error correction with encoding a qubit into an
  oscillator.
\newblock {\em Phys. Rev. Lett.}, \textbf{119}(18):180507, 2017.
\newblock \url{https://doi.org/10.1103/PhysRevLett.119.180507}.

\bibitem{fukui2018_tracking}
K.~Fukui, A.~Tomita, and A.~Okamoto.
\newblock Tracking quantum error correction.
\newblock {\em Phys. Rev. A}, \textbf{98}(2):022326, 2018.
\newblock \url{https://doi.org/10.1103/PhysRevA.98.022326}.

\bibitem{vuillot2019_quantum}
C.~Vuillot, H.~Asasi, Y.~Wang, L.~P. Pryadko, and B.~M. Terhal.
\newblock Quantum error correction with the toric
  {G}ottesman-{K}itaev-{P}reskill code.
\newblock {\em Phys. Rev. A}, \textbf{99}(3):032344, 2019.
\newblock \url{https://doi.org/10.1103/PhysRevA.99.032344}.

\bibitem{noh2020_faulttolerant}
K.~Noh and C.~Chamberland.
\newblock Fault-tolerant bosonic quantum error correction with the
  surface--{G}ottesman-{K}itaev-{P}reskill code.
\newblock {\em Phys. Rev. A}, \textbf{101}(1):012316, 2020.
\newblock \url{https://doi.org/10.1103/PhysRevA.101.012316}.

\bibitem{menicucci2014_faulttolerant}
N.~C. Menicucci.
\newblock Fault-tolerant measurement-based quantum computing with
  continuous-variable cluster states.
\newblock {\em Phys. Rev. Lett.}, \textbf{112}(12):120504, 2014.
\newblock \url{https://doi.org/10.1103/PhysRevLett.112.120504}.

\bibitem{fukui2018_highthreshold}
K.~Fukui, A.~Tomita, A.~Okamoto, and K.~Fujii.
\newblock High-threshold fault-tolerant quantum computation with analog quantum
  error correction.
\newblock {\em Phys. Rev. X}, \textbf{8}(2):021054, 2018.
\newblock \url{https://doi.org/10.1103/PhysRevX.8.021054}.

\bibitem{walshe2019_robust}
B.~W. Walshe, L.~J. Mensen, B.~Q. Baragiola, and N.~C. Menicucci.
\newblock Robust fault tolerance for continuous-variable cluster states with
  excess antisqueezing.
\newblock {\em Phys. Rev. A}, \textbf{100}(1):010301, 2019.
\newblock \url{https://doi.org/10.1103/PhysRevA.100.010301}.

\bibitem{fukui2019_highthreshold}
K.~Fukui.
\newblock High-threshold fault-tolerant quantum computation with the {GKP}
  qubit and realistically noisy devices.
\newblock {\em arXiv:1906.09767 [quant-ph]}, 2019.
\newblock \url{https://arxiv.org/abs/1906.09767}.

\bibitem{yamasaki2020_polylog}
H.~Yamasaki, K.~Fukui, Y.~Takeuchi, S.~Tani, and M.~Koashi.
\newblock Polylog-overhead highly fault-tolerant measurement-based quantum
  computation: all-{G}aussian implementation with
  {G}ottesman-{K}itaev-{P}reskill code.
\newblock {\em arXiv:2006.05416 [quant-ph]}, 2020.
\newblock \url{https://arxiv.org/abs/2006.05416}.

\bibitem{bourassa2020_blueprint}
J.~E. Bourassa, R.~N. Alexander, M.~Vasmer, A.~Patil, I.~Tzitrin, T.~Matsuura,
  D.~Su, B.~Q. Baragiola, S.~Guha, G.~Dauphinais, K.~K. Sabapathy, N.~C.
  Menicucci, and I.~Dhand.
\newblock Blueprint for a scalable photonic fault-tolerant quantum computer.
\newblock {\em Quantum}, \textbf{5}:392, 2021.
\newblock \url{ https://doi.org/10.22331/q-2021-02-04-392}.

\bibitem{larsen2021_faulttolerant}
M.~V. Larsen, C.~Chamberland, K.~Noh, J.~S. Neergaard-Nielsen, and U.~L.
  Andersen.
\newblock A fault-tolerant continuous-variable measurement-based quantum
  computation architecture.
\newblock {\em arXiv:2101.03014 [quant-ph]}, 2021.
\newblock \url{https://arxiv.org/abs/2101.03014}.

\bibitem{chao2018_quantum}
R.~Chao and B.~W. Reichardt.
\newblock Quantum error correction with only two extra qubits.
\newblock {\em Phys. Rev. Lett.}, \textbf{121}(5):050502, 2018.
\newblock \url{https://doi.org/10.1103/PhysRevLett.121.050502}.

\bibitem{chamberland2018_flag}
C.~Chamberland and M.~E. Beverland.
\newblock Flag fault-tolerant error correction with arbitrary distance codes.
\newblock {\em Quantum}, \textbf{2}:53, 2018.
\newblock \url{https://doi.org/10.22331/q-2018-02-08-53}.

\bibitem{yoder2016_universal}
T.~J. Yoder, R.~Takagi, and I.~L. Chuang.
\newblock Universal fault-tolerant gates on concatenated stabilizer codes.
\newblock {\em Phys. Rev. X}, \textbf{6}(3):031039, 2016.
\newblock \url{https://doi.org/10.1103/PhysRevX.6.031039}.

\bibitem{webster2015_reducing}
P.~Webster, S.~D. Bartlett, and D.~Poulin.
\newblock Reducing the overhead for quantum computation when noise is biased.
\newblock {\em Phys. Rev. A}, \textbf{92}(6):062309, 2015.
\newblock \url{https://doi.org/10.1103/PhysRevA.92.062309}.

\bibitem{lloyd1999_quantum}
S.~Lloyd and S.~L. Braunstein.
\newblock Quantum computation over continuous variables.
\newblock {\em Phys. Rev. Lett.}, \textbf{82}(8):1784--1787, 1999.
\newblock \url{https://doi.org/10.1103/PhysRevLett.82.1784}.

\bibitem{jiang2007_distributed}
L.~Jiang, J.~M. Taylor, A.~S. Sørensen, and M.~D. Lukin.
\newblock Distributed quantum computation based on small quantum registers.
\newblock {\em Phy. Rev. A}, \textbf{76}(6):062323, 2007.
\newblock \url{https://doi.org/10.1103/PhysRevA.76.062323}.

\bibitem{kimble2008_quantum}
H.~J. Kimble.
\newblock The quantum internet.
\newblock {\em Nature}, \textbf{453}(7198):1023--1030, 2008.
\newblock \url{https://doi.org/10.1038/nature07127}.

\bibitem{monroe2014_large-scale}
C.~Monroe, R.~Raussendorf, A.~Ruthven, K.~R. Brown, P.~Maunz, L.~M. Duan, and
  J.~Kim.
\newblock Large-scale modular quantum-computer architecture with atomic memory
  and photonic interconnects.
\newblock {\em Phy. Rev. A}, \textbf{89}(2):022317, 2014.
\newblock \url{https://doi.org/10.1103/PhysRevA.89.022317}.

\bibitem{axline2018_on-demand}
C.~J. Axline, L.~D. Burkhart, W.~Pfaff, M.~Zhang, K.~Chou, P.~Campagne-Ibarcq,
  P.~Reinhold, L.~Frunzio, S.~M. Girvin, L.~Jiang, M.~H. Devoret, and R.~J.
  Schoelkopf.
\newblock On-demand quantum state transfer and entanglement between remote
  microwave cavity memories.
\newblock {\em Nature Phys.}, \textbf{14}(7):705--710, 2018.
\newblock \url{https://doi.org/10.1038/s41567-018-0115-y}.

\bibitem{campagne-ibarcq2018_deterministic}
P.~Campagne-Ibarcq, E.~Zalys-Geller, A.~Narla, S.~Shankar, P.~Reinhold,
  L.~Burkhart, C.~Axline, W.~Pfaff, L.~Frunzio, R.~J. Schoelkopf, and M.~H.
  Devoret.
\newblock Deterministic remote entanglement of superconducting circuits through
  microwave two-photon transitions.
\newblock {\em Phy. Rev. Lett.}, \textbf{120}(20):200501, 2018.
\newblock \url{https://doi.org/10.1103/PhysRevLett.120.200501}.

\bibitem{burkhart2020_error-detected}
L.~D. Burkhart, J.~Teoh, Y.~Zhang, C.~J. Axline, L.~Frunzio, M.~H. Devoret,
  L.~Jiang, S.~M. Girvin, and R.~J. Schoelkopf.
\newblock Error-detected state transfer and entanglement in a superconducting
  quantum network.
\newblock {\em arXiv:2004.06168 [quant-ph]}, 2020.
\newblock \url{https://arxiv.org/abs/2004.06168}.

\bibitem{landsman2019_two-qubit}
K.~A. Landsman, Y.~Wu, P.~H. Leung, D.~Zhu, N.~M. Linke, K.~R. Brown, L.~Duan,
  and C.~Monroe.
\newblock Two-qubit entangling gates within arbitrarily long chains of trapped
  ions.
\newblock {\em Phy. Rev. A}, \textbf{100}(2):022332, 2019.
\newblock \url{https://doi.org/10.1103/PhysRevA.100.022332}.

\bibitem{egan2020_fault-tolerant}
L.~Egan, D.~M. Debroy, C.~Noel, A.~Risinger, D.~Zhu, D.~Biswas, M.~Newman,
  M.~Li, K.~R. Brown, M.~Cetina, and C.~Monroe.
\newblock Fault-tolerant operation of a quantum error-correction code.
\newblock {\em arXiv:2009.11482 [quant-ph]}, 2020.
\newblock \url{https://arxiv.org/abs/2009.11482}.

\bibitem{veldhorst2015_two-qubit}
M.~Veldhorst, C.~H. Yang, J.~C. Hwang, W.~Huang, J.~P. Dehollain, J.~T.
  Muhonen, S.~Simmons, A.~Laucht, F.~E. Hudson, K.~M. Itoh, A.~Morello, and
  A.~S. Dzurak.
\newblock A two-qubit logic gate in silicon.
\newblock {\em Nature}, \textbf{526}(7573):410--414, 2015.
\newblock \url{https://doi.org/10.1038/nature15263}.

\bibitem{watson2018_programmable}
T.~F. Watson, S.~G.~J. Philips, E.~Kawakami, D.~R. Ward, P.~Scarlino,
  M.~Veldhorst, D.~E. Savage, M.~G. Lagally, M.~Friesen, S.~N. Coppersmith,
  M.~A. Eriksson, and L.~M.~K. Vandersypen.
\newblock A programmable two-qubit quantum processor in silicon.
\newblock {\em Nature}, \textbf{555}(7698):633--637, 2018.
\newblock \url{https://doi.org/10.1038/nature25766}.

\bibitem{levine2019_parallel}
H.~Levine, A.~Keesling, G.~Semeghini, A.~Omran, T.~T. Wang, S.~Ebadi,
  H.~Bernien, M.~Greiner, V.~Vuletic, H.~Pichler, and M.~D. Lukin.
\newblock Parallel implementation of high-fidelity multiqubit gates with
  neutral atoms.
\newblock {\em Phys. Rev. Lett.}, \textbf{123}(17):170503, 2019.
\newblock \url{https://doi.org/10.1103/PhysRevLett.123.170503}.

\end{thebibliography}

\end{document}